\pdfoutput=1
\documentclass[final,1p,times]{elsarticle}

\usepackage{framed,multirow}
\usepackage[mathlines]{lineno}
\usepackage{amssymb}
\usepackage{amsmath}
\usepackage{mathrsfs} 
\usepackage{latexsym}
\biboptions{sort&compress}

\usepackage{bm}
\usepackage{subfigure}
\usepackage{tikz}
\tikzset{>=stealth}

\usepackage{url}
\usepackage{xcolor}
\definecolor{newcolor}{rgb}{.8,.349,.1}

\begin{document}


\begin{frontmatter}

\title{Application of a spectral scheme to simulate horizontally slowly varying three-dimensional ocean acoustic propagation}

\author[1]{Houwang Tu}
\ead{tuhouwang@nudt.edu.cn}
\author[1]{Yongxian Wang\corref{cor1}}
\cortext[cor1]{Corresponding author: \url{yxwang@nudt.edu.cn}}
\author[1]{Xiaolan Zhou}
\author[1]{Guojun Xu}
\author[1]{Dongbao Gao}
\author[1]{Shuqing Ma}

\address[1]{College of Meteorology and Oceanography, National University of Defense Technology, Changsha, 410073, China}


\begin{abstract}
Three-dimensional numerical models for underwater sound propagation are popular in computational ocean acoustics. For horizontally slowly varying waveguide environments, an adiabatic mode-parabolic equation hybrid theory can be used for simulation. This theory employs adiabatic modes in the vertical direction, simplifying the solution of the sound pressure to the solution of horizontal refractive index of vertical modes. The refractive equations in the horizontal direction are further solved by a ``split-step" wide-angle parabolic equation model, following the approach of the ``vertical modes and horizontal parabolic equation". Existing three-dimensional sound propagation models mostly use finite difference methods for discretization, but in recent years, the academic community has proposed new types of sound propagation models based on spectral methods. Spectral methods are numerical discretization methods based on orthogonal polynomial approximation and weighted residual principles. They offer advantages such as high computational accuracy and fast convergence. In this study, a three-dimensional adiabatic mode-parabolic equation hybrid model discretized using spectral methods is proposed. In the vertical direction, the modal functions are solved using the Chebyshev spectral method. The medium layering is handled using a domain decomposition strategy, and the leaky modes under semi-infinite boundary conditions are addressed using an eigenvalue transformation technique. In the horizontal direction, the perfectly matched layer technique is utilized to handle unbounded computational domains, and the perfectly matched layer and computational domain are segmented into multiple layers. Numerical simulations show that the Chebyshev spectral method achieves reliable results in the application of the adiabatic mode-parabolic equation hybrid model, providing a new model selection for three-dimensional sound propagation numerical simulations.
\end{abstract}

\begin{keyword}
ocean acoustics \sep spectral method \sep  adiabatic modes \sep parabolic equation \sep three-dimensional propagation
\end{keyword}

\end{frontmatter}


\section{Introduction}
With the increasing global population, human consumption of land resources has intensified. The ocean, which comprises vast amounts of resources, is becoming increasingly indispensable for human development. The exploration, development, and protection of the ocean will be the main topics of future marine research. Due to the characteristics of seawater, electromagnetic waves commonly used in spatial detection attenuate rapidly in seawater, making it difficult to propagate over long distances and complete detection activities. Sound waves, which are a type of mechanical wave, propagate quickly in seawater and can travel long distances \cite{Brekhovskikh2003}. They are currently used for underwater detection, localization and identification. The actual marine environment is complex and constantly changing. Acoustic parameters vary in both time and space, which greatly affects the propagation path of sound and consequently the detection performance of underwater equipment. Understanding the principles of sound propagation is of great importance in improving the detection capabilities of underwater equipment. The propagation of sound waves underwater follows fundamental physical laws. Under the assumption of linear acoustics, the governing equation for underwater sound propagation satisfies the linear wave equation \cite{Brekhovskikh1980}. Utilizing numerical simulation methods to simulate the propagation path and energy distribution of underwater sound waves is a common approach used for predicting sound fields. This has led to the development of the specialized subdiscipline of computational ocean acoustics \cite{Jensen2011}. Due to the complexity of the wave equation, practical numerical simulations can rarely be used to directly solve the wave equation for underwater sound propagation. Instead, the wave equation is transformed into the Helmholtz equation using Fourier transforms \cite{Orlando2023}. The numerical solution of the Helmholtz equation also requires significant computational power, and issues such as ``numerical instability" and ``numerical pollution" may arise, especially at high frequencies \cite{Kahlaf2021,Idesman2021}. Due to the computational and power limitations of underwater platforms, most sound propagation numerical models focus on solving the simplified theories of the Helmholtz equation. These simplified theories mainly include ray models, wavenumber integration models, normal mode models, and parabolic equation models, primarily focusing on two-dimensional sound propagation \cite{Etter2018}. 

Weston first introduced the concept of horizontal refraction in 1961, which is an important milestone in the development of three-dimensional sound propagation modeling \cite{Weston1961}. In recent years, with the improvement of computer performance, three-dimensional sound propagation numerical simulations that are closer to the real marine environment have gained increasing attention from the academic community \cite{Luowy2009,Ivansson2021,HeTj2020,HeTj2021a,Tuhw2022b,XiaRui2023}. Numerical simulations of three-dimensional sound propagation also face the choice between solving the Helmholtz equation directly or using simplified theories. Directly solving the Helmholtz equation provides higher accuracy, but it is computationally intensive \cite{LiCx2019}. Even with today's advanced computing capabilities, it still needs to be run on supercomputers with thousands of cores \cite{Liuw2021}. Accounting for the computational complexity and the requirement for timeliness, numerical models based on various simplified theories remain the most promising strategy for three-dimensional sound propagation simulations \cite{Tolstoy1996,XuCx2016}. Research on simplified models for three-dimensional sound propagation numerical simulations has a long history. The most natural idea is to generalize the two-dimensional simplified models to three-dimensional waveguides, including the adoption of both $N\times$2D and full 3D strategies. As a result, many classic studies on three-dimensional ray models \cite{Weston1961,HARPO,Bucker1994}, three-dimensional wavenumber integration model \cite{Schmidt1985b,OASES}, three-dimensional normal modes \cite{Chiu1990,Evans1990,Luowy2009} and three-dimensional parabolic equation models have emerged \cite{Siegmann1985,FOR3D,LinYT2013,Sturm2016}. The main difference between three-dimensional sound propagation and two-dimensional sound propagation is the presence of horizontal refraction effects. For waveguides with slowly varying terrain and acoustic parameters, horizontal refraction effects are relatively more important than mode coupling in addressing many practical problems. Winberg and Burridge developed the theory of ``vertical modes and horizontal rays" to simulate three-dimensional acoustic fields in the 1970s \cite{Henry1974,Burridge1977}. In this theory, the vertical direction utilizes the normal mode model to calculate the modal components of the sound field. Based on the modal decomposition of the sound field and by neglecting coupling effects, the modal amplitudes in three-dimensional waveguides satisfy a two-dimensional Helmholtz equation that is also known as the horizontal refraction equation (HRE). Burridge and Weinberg proposed the use of ray theory in the horizontal direction to solve the HREs. Collins developed a parabolic equation model (referred to as the mode parabolic equation, MPE) for solving the HREs in 1993 \cite{Collins1993b}. Later, Trofimov independently derived this equation \cite{Trofimov2015}. Petrov and his research group made significant contributions to the solution of the HREs and proposed analytical solutions for HREs under various conditions \cite{Petrov2019}, wide-angle MPE \cite{Petrov2020a}, MPE in curved coordinate systems \cite{Petrov2020b}, and numerical models \cite{Tyshchenko2021}. The adiabatic mode-parabolic equation (AMPE) hybrid theory of the ``vertical modes and horizontal parabolic equation" indeed provides a good simulation for three-dimensional sound propagation with horizontal variations. However, apart from the finite difference model proposed by Petrov et al. \cite{Petrov2020a,Tyshchenko2021}, there are relatively few studies on numerical solutions and validation of the AMPE hybrid model. We noticed the flourishing development of spectral methods in computational ocean acoustics, particularly in overcoming the numerical challenges of solving two-dimensional normal modes \cite{Tuhw2020b,Tuhw2023b} and parabolic equation models \cite{Tuhw2021a,Tuhw2023d}. We hope to introduce spectral methods into the numerical simulation of three-dimensional sound propagation to achieve accurate and efficient numerical sound fields.

Spectral methods are a class of numerical discretization methods that are on par with the finite difference, finite element and finite volume methods. They use orthogonal polynomial approximation functions to project the differential equations into spectral space for solution. Due to the excellent properties of orthogonal polynomials, the differential equations transformed into spectral space can be easily discretized into algebraic equation systems using the weighted residual principle. When the solution of the differential equation is sufficiently smooth, spectral methods can achieve exponential convergence rates \cite{Gottlieb1977,Jshen2006}. Spectral methods were introduced to computational ocean acoustics in 1993 but initially did not receive sufficient attention from the academic community \cite{Dzieciuch1993}. In recent years, they have experienced rapid development. The wavenumber integration model \cite{Tuhw2023c}, normal mode models \cite{Tuhw2020b,Tuhw2021b,NM-CT}, and parabolic equation model \cite{Wangyx2021a,Tuhw2023d,SMPE} based on spectral methods have successively emerged. A recent survey summarized the history of spectral methods in computational ocean acoustics and the progress made thus far \cite{Wangyx2023}. Based on the results obtained, spectral methods are particularly suitable for calculating sound propagation in regular domains. For interfaces with discontinuous acoustic parameters, the spectral accuracy can be maintained by employing a domain decomposition strategy \cite{Min2005}. In this context, we applied spectral methods to discretize the AMPE hybrid theory, developing the first fully three-dimensional acoustic propagation model based on spectral methods and separately discretizing the vertical eigen-equations and MPEs by spectral methods.

The organization of this paper is as follows. In Sec.~\ref{sec2}, we introduce the AMPE hybrid theory, derive the HRE, and then present the ``split-step" parabolic approximation for the HRE. In Sec.~\ref{sec3}, we derive the spectral discretization of the local modal equations and the spectral discretization of the MPE, including necessary numerical techniques. In Sec.~\ref{sec4}, a clear summary of the spectral algorithm is provided, along with an analysis of its parallelism. In Sec.~\ref{sec5}, three numerical experiments were constructed to validate the accuracy of the spectral scheme proposed in this paper. Finally, we conclude the paper in Sec.~\ref{sec6}.
 
\section{Physical model}
\label{sec2}
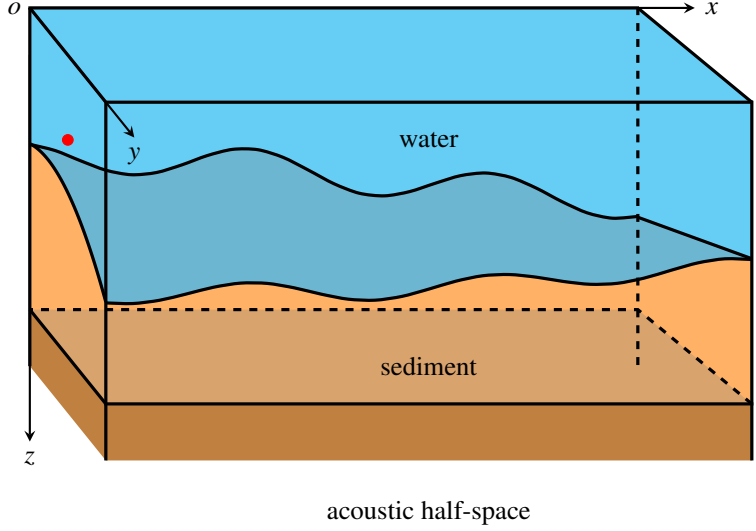
\begin{figure}
	\centering
\begin{tikzpicture}[node distance=2cm,scale = 0.5]
	\fill[cyan,opacity=0.6] (2,2)--(2,-1.6)--plot[domain=2:4,smooth] (\x ,{-0.8*\x*\x+2.7*\x-3.8)})--plot[domain=4:14*1.5,smooth] (\x ,{-5+0.5*(-1.5+0.5*sin(\x r)-2*cos(0.1*\x r)+2*sin(0.2*\x r)+cos(0.3*\x r)+0.07*\x)})--(14*1.5,-4.65)--(14*1.5,-0.5)--(18,2)--cycle;
	
	\fill[gray,opacity=0.3] (2,-1.6)--plot[domain=2:4,smooth] (\x ,{-0.8*\x*\x+2.7*\x-3.8)})--plot[domain=4:14*1.5,smooth] (\x ,{-5+0.5*(-1.5+0.5*sin(\x r)-2*cos(0.1*\x r)+2*sin(0.2*\x r)+cos(0.3*\x r)+0.07*\x)})--(12*1.5,-3.55)--plot[domain=18:2,smooth] (\x ,{-3+0.3*(-1.5+1.5*sin(\x r)+3*cos(0.1*\x r)+2*sin(0.2*\x r)+cos(0.3*\x r)+0.1*\x)})--cycle;
	
	\fill[orange,opacity=0.6] (2,-1.6)--plot[domain=2:4,smooth] (\x ,{-0.8*\x*\x+2.7*\x-3.8)})--plot[domain=4:14*1.5,smooth] (\x ,{-5+0.5*(-1.5+0.5*sin(\x r)-2*cos(0.1*\x r)+2*sin(0.2*\x r)+cos(0.3*\x r)+0.07*\x)})--(14*1.5,-8.5)--(4,-8.5)--(2,-6)--cycle;
	\fill[brown] (2,-6)--(4,-8.5)--(14*1.5,-8.5)--(14*1.5,-10)--(4,-10)--(2,-7.5)--cycle;	
	\fill[gray,opacity=0.3] (2,-6)--(4,-8.5)--(14*1.5,-8.5)--(12*1.5,-6)--cycle;
	
	\draw[thick, ->](2,2)--(19.5,2) node[right]{$x$};
	\draw[thick, ->](2,0.75)--(2,-9.5) node[below]{$z$};       
	\draw[very thick](1.96,2)--(12.02*1.5,2);
	\draw[very thick](4,-0.5)--(14.02*1.5,-0.5);
	\draw[very thick](2,2)--(4,-0.5);
	\draw[thick, ->](2,2)--(4.75,-1.4375) node[below]{$y$}; 
	\draw[very thick](12*1.5,2)--(14*1.5,-0.5);
	\draw[dashed,very thick](12*1.5,2.02)--(12*1.5,-7.5);
	\draw[very thick](2,2)--(2,-7.5);
	\draw[very thick](4,-0.5)--(4,-10);
	\draw[very thick](14*1.5,-0.5+0.04)--(14*1.5,-10.0);
	
	\draw [very thick]  plot[domain=2:18,smooth] (\x ,{-3+0.3*(-1.5+1.5*sin(\x r)+3*cos(0.1*\x r)+2*sin(0.2*\x r)+cos(0.3*\x r)+0.1*\x)});		
	\draw [very thick]  plot[domain=4:14*1.5,smooth] (\x ,{-5+0.5*(-1.5+0.5*sin(\x r)-2*cos(0.1*\x r)+2*sin(0.2*\x r)+cos(0.3*\x r)+0.07*\x)});		
	\draw [very thick]  plot[domain=2:4,smooth] (\x ,{-0.8*\x*\x+2.7*\x-3.8)});		
	\draw[very thick](18,-3.55)--(14*1.5,-4.65);
	
	\draw[very thick](2,-6)--(4,-8.5); 
	\draw[very thick](4-0.04,-8.5)--(14*1.5+0.04,-8.5);   
	\draw[dashed,very thick](14*1.5,-8.5)--(12*1.5,-6)--(2,-6); 
	
	\filldraw [red] (3,-1.5) circle [radius=4pt];
	\node at (1.6,2){$o$};
	\node at (9*1.5-1,-1.5){water};
	\node at (9*1.5-1,-7.5){sediment};
	\node at (9*1.5-1,-11.4){acoustic half-space};
\end{tikzpicture}
\caption{Schematic diagram of a three-dimensional waveguide.}
	\label{Figure1}
\end{figure}

\subsection{Horizontal refraction equations}
The three-dimensional acoustic propagation is governed by the Helmholtz equation, which can be expressed as follows:
\begin{equation}
	\rho \nabla \cdot\left(\frac{1}{\rho} \nabla p\right)+k^2(\mathbf{r}) p=-\delta\left(\mathbf{r}_\mathrm{s}\right).
\end{equation}
In particular, for the marine environment depicted in Fig.~\ref{Figure1}, the three-dimensional Helmholtz equation can be simplified as follows:
\begin{equation}
	\label{eq.2}
	\frac{\partial^2p}{\partial x^2}+\frac{\partial^2p}{\partial y^2}+\rho(z) \frac{\partial}{\partial z}\left(\frac{1}{\rho(z)} \frac{\partial p}{\partial z}\right)+{k^2(x,y,z)} p=-\delta\left(x\right) \delta\left(y-y_{\mathrm{s}}\right) \delta\left(z-z_{\mathrm{s}}\right),
\end{equation}
where $x\in[0,x_{\max}]$, $y\in[0,y_{\max}]$, $z\in[0,H]$, $p\equiv p(x,y,z)$ and $\rho\equiv\rho(z)$. It is assumed that the density is only a function of the depth. The solution for acoustic pressure can be expressed using the ``modes superposition" as follows:
\begin{equation}
	\label{eq.3}
	p(x, y, z)=\sum_m \Phi_m(x,y) \Psi_m(x,y,z),
\end{equation}
where $\{\Psi_m(x,y,z)\}$ represents the local modes, and $\Phi_m(x,y)$ is referred to as the horizontal refractive index of the $m$-th modes. The vertical modes satisfy the eigen-equation:
\begin{subequations}
	\label{eq.4}
	\begin{gather}
		\label{eq.4a}
		\rho(z) \frac{\mathrm{d}}{\mathrm{d} z}\left[\frac{1}{\rho(z)} \frac{\mathrm{d} \Psi(z)}{\mathrm{d} z}\right]+\left[k^{2}(z)-\kappa^{2}\right]\Psi(z)=0,\\
		k(z)=(1+\mathrm{i}\eta\alpha)\omega/c(z),\quad \eta=(40\pi \lg{\mathrm{e}})^{-1}.
	\end{gather}
\end{subequations}
This is a Sturm--Liouville problem with an infinite set of eigensolutions $\{\kappa_m,\Psi_m(z)\}_{m=1}^\infty$, and the eigenmodes satisfy the orthogonality defined as follows:
\begin{equation}
	\label{eq.5}
	\int \frac{{\Psi_m(z)}{\Psi_n(z)}}{\rho(z)}\mathrm {d} z=\delta_{mn},
	\quad m,n = 1, 2, \ldots.
\end{equation}

Substituting Eq.~\eqref{eq.3} into the aforementioned three-dimensional Helmholtz equation and applying the following operator yields:
\begin{equation*}
\int\left(\cdot\right) \frac{\Psi_n(x,y,z)}{\rho(z)} \mathrm{d} z.
\end{equation*}	
Considering the orthogonality and completeness of the local modes, it is straightforward to obtain the following:	
\begin{equation}
	\frac{\partial^2 \Phi_n}{\partial x^2}+\frac{\partial^2 \Phi_n}{\partial y^2}+\kappa_{n}^2(x, y) \Phi_n+\sum_m A_{m n} \Phi_m+\sum_m 2 B_{m n} \frac{\partial \Phi_m}{\partial x}+\sum_m 2 C_{m n} \frac{\partial \Phi_m}{\partial y}=-\frac{\Psi_n\left(0,y_{\mathrm{s}},z_{\mathrm{s}}\right)\delta\left(x\right)\delta\left(y-y_{\mathrm{s}}\right)}{\rho\left(z_{\mathrm{s}}\right)},
\end{equation}
where
\begin{equation}
	A_{m n}=\int\left(\frac{\partial^2}{\partial x^2}+\frac{\partial^2}{\partial y^2}\right) \Psi_m \frac{\Psi_n}{\rho} \mathrm{d} z, \quad	B_{m n}=-B_{n m}=\int \frac{\partial \Psi_m}{\partial x} \frac{\Psi_n}{\rho} \mathrm{d} z, \quad C_{m n}=-C_{n m}=\int \frac{\partial \Psi_m}{\partial y} \frac{\Psi_n}{\rho} \mathrm{d} z.
\end{equation}
$A_{m n}$, $B_{m n}$, and $C_{m n}$ are terms representing mode coupling, where $\kappa_n(x,y)$ denotes the horizontal wavenumber of the $n$-th mode. When the horizontal fluctuations of ocean acoustic parameters are small, the contributions of the coupling matrices $A_{m n}$, $B_{m n}$, and $C_{m n}$ can be ignored based on adiabatic mode theory \cite{Burridge1977}. In this case, the following HRE can be derived:
\begin{equation}
	\label{eq.8}
	\frac{\partial^2 \Phi_n}{\partial x^2}+\frac{\partial^2 \Phi_n}{\partial y^2}+\kappa_{n}^2(x, y) \Phi_n=-\frac{\Psi_n\left(0,y_{\mathrm{s}},z_{\mathrm{s}}\right)\delta\left(x\right)\delta\left(y-y_{\mathrm{s}}\right)}{\rho\left(z_{\mathrm{s}}\right)}.
\end{equation}

The condition under which the adiabatic approximation can be applied is:

\begin{equation}
	\left|\frac{2S_{m,n}\sqrt{\kappa_{m}\kappa_{n}}}{\kappa_{m}^2-\kappa_{n}^2}\right| \ll 1,\quad S_{m,n}=\int \Psi_m(x,y,z)\nabla\Psi_n(x,y,z)\mathrm{d}z.
\end{equation}

The HRE \eqref{eq.8} is a two-dimensional Helmholtz equation. It is equivalent to eliminating one-dimensional variables by using local normal modes compared to Eq.~\eqref{eq.2}. In the HRE, the equivalent wavenumber $\kappa_{n}(x,y)$ is determined by the eigenvalues of the local modes, thus each mode corresponds to its own HRE. In practical computations, it is common to select modes of finite order, i.e., $m,n\leq M$. To simplify the equations, we omit the subscript $n$ when introducing the solutions of the HREs.

\subsection{Mode parabolic equation}
\label{sec2.2}
Many models and theories in computational ocean acoustics are used to solve two-dimensional Helmholtz equations. Theoretically, rays, normal modes, and parabolic models can all be utilized for solving HREs. Among them, the parabolic equation model strikes a good balance between efficiency and accuracy.

\subsubsection{Classic parabolic equation (PE)}
Referring to the derivation of the parabolic equation model, and neglecting the non-homogeneous term on the right-hand side, if we select the $x$-axis as the principal direction of propagation and omit the subscript $n$, Eq.~\eqref{eq.8} can be naturally decomposed into the following:
\begin{equation}
	\left(\frac{\partial}{\partial x}-\mathrm{i}\sqrt{\frac{\partial^2}{\partial y^2}+\kappa^2}\right)\left(\frac{\partial}{\partial x}+\mathrm{i}\sqrt{\frac{\partial^2}{\partial y^2}+\kappa^2}\right) \Phi +\left[\frac{\partial}{\partial x},\mathrm{i}\sqrt{\frac{\partial^2}{\partial y^2}+\kappa^2}\right]\Phi = 0.
\end{equation}
The square brackets in the above equation denote the commutative operator, expressed in the following form:
\[
\left[A,B\right]\Phi=AB\Phi-BA\Phi.
\]

In environments where horizontal fluctuations are not significant, the outcomes of the commutative operator are small and can be disregarded. Consequently, the above equation simplifies to:
\begin{equation}
	\left(\frac{\partial}{\partial x}-\mathrm{i}\sqrt{\frac{\partial^2}{\partial y^2}+\kappa^2}\right)\left(\frac{\partial}{\partial x}+\mathrm{i}\sqrt{\frac{\partial^2}{\partial y^2}+\kappa^2}\right)\Phi = 0.
\end{equation}
By disregarding the backscattered waves, we can obtain the equation for forward propagation:
\begin{equation}
	\label{eq.12}
	\frac{\partial \Phi}{\partial x}=\left(\mathrm{i}\sqrt{\frac{\partial^2}{\partial y^2}+\kappa^2}\right)\Phi.
\end{equation}
The principal oscillation can be eliminated from $\Phi$ by substituting it with the following:
\begin{equation}
	\label{eq.13}
	\Phi(x,y)=\exp{\left(\mathrm{i}\bar{\kappa}x\right)}\phi(x,y),
\end{equation}	
where $\bar{\kappa}$ is referred to as the reference wavenumber and is typically taken as the value of $\kappa_{\mathrm{s}}$ ($\kappa$ at the sound source).

By substituting the aforementioned equation into Eq.~\eqref{eq.12}, we can derive the governing equation of $\phi(x,y)$:
\begin{subequations}
	\label{eq.14}
	\begin{gather}
		\frac{\partial \phi}{\partial x}=\mathrm{i}\kappa_{\mathrm{s}}\left(\sqrt{1+\mathcal{Y}}-1\right)\phi,\\
		\kappa_{\mathrm{s}}^2\mathcal{Y}=\frac{\partial^2}{\partial y^2}+\kappa^2-\kappa_{\mathrm{s}}^2.
	\end{gather}
\end{subequations}
Based on the theory of first-order ordinary differential equations, the above equation exhibits a step-wise parabolic solution in the following form:
\begin{equation}
	\label{eq.15}
	\phi(x+\Delta x,y)=\exp\left[\mathrm{i}\kappa_{\mathrm{s}}\Delta x\left(\sqrt{1+\mathcal{Y}}-1\right)\right]\phi(x,y).
\end{equation}

However, the MPE \eqref{eq.15} cannot be directly advanced due to the presence of the linear operator $\mathcal{Y}$, which includes the second-order derivative of $y$ within the exponential radical term. The most widely adopted approach is to apply a rational approximation technique, specifically the Pad\'e series expansion, to the exponential radical operator. This approach, also known as the wide-angle PE algorithm \cite{Collins1993a}, combines the efficiency of split-step methods with the accuracy of Pad\'e approximation. In the subsequent steps, we will employ this method.

\subsubsection{Rational approximation of the $\mathcal{Y}$ operator}
Rational function approximation involves using the ratio of two algebraic polynomials to approximate a given function.
\begin{equation}
	f(X) \approx R_{i,j}(X) = \frac{S_i(X)}{Q_j(X)}.
\end{equation}
Here, $S_i(X)$ and $Q_j(X)$ are $i$-order and $j$-order polynomials of $X$, respectively. The Pad\'e approximation aims to find $R_{i,j}(X)$ such that
\[
f^{(m)}(0) = R_{i,j}^{(m)}(0) , \quad \textrm{for} \  m = 0,1,\ldots,i+j.
\]
This method extends the concept of Taylor series by representing the approximation in rational form. The coefficients can be calculated by solving the system of linear equations derived from equating the truncated Taylor series of $f(X)$ and rational function $R_{i,j}(X)$.

In the rational approximation of operators in MPEs, it is often effective to choose $i=j$ to obtain satisfactory results in most cases.
\begin{equation}
	f(X) \approx R_{n,n}(X) =\frac{1+\alpha_1 X+\ldots+\alpha_{n-1} X^{n-1}+\alpha_n X^n}{1+\beta_1 X+\ldots+\beta_{n-1} X^{n-1}+\beta_n X^n} =\prod_{j=1}^n \frac{1+c_j X}{1+b_j X},
\end{equation}
where $\{b_j\}$ and $\{c_j\}$ can be calculated by root-finding algorithms. 
\begin{equation}
	\label{eq.18}
	R_{n,n}(X)=\prod_{j=1}^n \frac{1+c_j X}{1+b_j X}=1+\sum_{j=1}^n \frac{a_j X}{1+b_j X} =d_0+\sum_{j=1}^n \frac{d_j}{1+b_j X},			
\end{equation}
The expression mentioned above, \eqref{eq.18}, is more convenient for numerical implementation. The coefficients $\{a_j\}$ and $\{d_j\}$ can be calculated using the following formula \cite{Antoine2010}:
\[
a_j=(c_j-b_j) \prod_{i \neq j} \frac{c_i-b_j}{b_i-b_j}, \quad  d_j=-\frac{a_j}{b_j}, \quad d_0=1-\sum_{j=1}^n d_j.
\]
Based on the Pad\'e approximation theory discussed above, the exponential radical operator in the parabolic solution \eqref{eq.15} of the MPE can be approximated as follows:
\begin{equation}
	\label{eq.19}
	\exp\left[\mathrm{i}\kappa_{\mathrm{s}}\Delta x\left(\sqrt{1+\mathcal{Y}}-1\right)\right] \approx R_{n,n}(\mathcal{Y})= \left(d_0+\sum_{j=1}^n \frac{d_j}{1+b_j \mathcal{Y}}\right).
\end{equation}
Therefore, obtaining the stepwise MPE solution is straightforward:
\begin{equation}
	\label{eq.20}
	\phi(x+\Delta x)=\left(d_0+\sum_{j=1}^n \frac{d_j }{1+b_j \mathcal{Y}}\right) \phi(x).
\end{equation}
The rational approximation within each step iteration can be completed by following these two steps:
\begin{subequations}
	\label{eq.21}
	\begin{gather}
		(1+b_j\mathcal{Y})W_j=d_j\phi(x),\quad j=1,2,\ldots,n,\\
		\phi(x+\Delta x)=d_0\phi(x)+\sum_{j=1}^{n}W_j,
	\end{gather}
\end{subequations}
where $\{W_j\}$ is the intermediate function.

Importantly, in many cases, parabolic models that utilize weighted rational approximations of $R_{n,n}(X)$ and $R_{n-1,n}(X)$ are able to achieve more accurate and stable approximations of the $X$ operator.
\begin{subequations}
	\begin{gather}
		R_{n-1,n}(X) =\frac{1+\alpha_1 X+\cdots+\alpha_{n-1} X^{n-1}}{1+\beta_1 X+\cdots+\beta_{n-1} X^{n-1}+\beta_n X^n},\\
		f(X) \approx \theta R_{n,n}(X) +(1-\theta)R_{n-1,n}(X),\quad \theta\in[0,1].
	\end{gather}
\end{subequations}
The Pad\'e approximation in Eq.~\eqref{eq.19} can also be formulated using the aforementioned weighted approach.

\subsection{Perfectly matched layer}
In Eq.~\eqref{eq.2}, there are no other barriers in the horizontal direction that impede the propagation of sound waves, except for the sea surface and seabed. The boundary conditions at the sea surface and seabed are accounted for by solving local modes. The unbounded domain implies that sound energy can freely penetrate through the planes of $y=0$, $y={\max}$, and $x=x_{\max}$ to infinity, without any reflected waves impacting the sound field within the defined domain. When solving for the MPEs in the context of HREs, the primary propagation direction is set to align with the radiation boundary conditions by default. However, numerically safeguarding the free boundary in the $y$-direction is crucial.

\begin{figure}
	\centering
			\begin{tikzpicture}[node distance=2cm,scale = 0.8]		
		\shade [left color=black,right color=black,opacity=0.3](-0.5,3) rectangle (10.5,4.5);
		\shade [left color=green,right color=cyan,opacity=0.3](-0.5,3) rectangle (10.5,-1.5);
		\shade [left color=black,right color=black,opacity=0.3](-0.5,-3) rectangle (10.5,-1.5);
		
		\draw[very thick,->] (-0.02-0.5,4.5)--(11,4.5) node[right] {$x$};
		\draw[very thick,->] (0-0.5,4.5)--(0-0.5,-3.5) node[below] {$y$};		
		
		\filldraw [blue] (-0.5,-1.5) circle [radius=1.5pt];	
		\filldraw [blue] (-0.5,-1.39) circle [radius=1.5pt];
		\filldraw [blue] (-0.5,-1.08) circle [radius=1.5pt];
		\filldraw [blue] (-0.5,-0.60) circle [radius=1.5pt];
		\filldraw [blue] (-0.5,-0.05) circle [radius=1.5pt];
		\filldraw [blue] (-0.5,0.50) circle [radius=1.5pt];
		\filldraw [blue] (-0.5,0.97) circle [radius=1.5pt];			
		\filldraw [blue] (-0.5,1.29) circle [radius=1.5pt];
		\filldraw [blue] (-0.5,1.40) circle [radius=1.5pt];
		\filldraw [blue] (-0.5,1.51) circle [radius=1.5pt];	
		\filldraw [blue] (-0.5,1.80) circle [radius=1.5pt];	
		\filldraw [blue] (-0.5,2.20) circle [radius=1.5pt];	
		\filldraw [blue] (-0.5,2.60) circle [radius=1.5pt];	
		\filldraw [blue] (-0.5,2.89) circle [radius=1.5pt];		
		\filldraw [blue] (-0.5,3) circle [radius=1.5pt];			
		\filldraw [blue] (-0.5,3.22) circle [radius=1.5pt];
		\filldraw [blue] (-0.5,3.75) circle [radius=1.5pt];
		\filldraw [blue] (-0.5,4.28) circle [radius=1.5pt];
		\filldraw [blue] (-0.5,4.50) circle [radius=1.5pt];			
		\filldraw [blue] (-0.5,-1.72) circle [radius=1.5pt];
		\filldraw [blue] (-0.5,-2.25) circle [radius=1.5pt];
		\filldraw [blue] (-0.5,-2.78) circle [radius=1.5pt];
		\filldraw [blue] (-0.5,-3.0) circle [radius=1.5pt];
		
		\filldraw [blue] (0.5,-1.5) circle [radius=1.5pt];
		\filldraw [blue] (0.5,-1.39) circle [radius=1.5pt];
		\filldraw [blue] (0.5,-1.08) circle [radius=1.5pt];
		\filldraw [blue] (0.5,-0.60) circle [radius=1.5pt];
		\filldraw [blue] (0.5,-0.05) circle [radius=1.5pt];
		\filldraw [blue] (0.5,0.50) circle [radius=1.5pt];
		\filldraw [blue] (0.5,0.97) circle [radius=1.5pt];			
		\filldraw [blue] (0.5,1.29) circle [radius=1.5pt];
		\filldraw [blue] (0.5,1.40) circle [radius=1.5pt];
		\filldraw [blue] (0.5,1.51) circle [radius=1.5pt];	
		\filldraw [blue] (0.5,1.80) circle [radius=1.5pt];	
		\filldraw [blue] (0.5,2.20) circle [radius=1.5pt];	
		\filldraw [blue] (0.5,2.60) circle [radius=1.5pt];	
		\filldraw [blue] (0.5,2.89) circle [radius=1.5pt];
		\filldraw [blue] (0.5,3) circle [radius=1.5pt];			
		\filldraw [blue] (0.5,3.22) circle [radius=1.5pt];
		\filldraw [blue] (0.5,3.75) circle [radius=1.5pt];
		\filldraw [blue] (0.5,4.28) circle [radius=1.5pt];
		\filldraw [blue] (0.5,4.50) circle [radius=1.5pt];			
		\filldraw [blue] (0.5,-1.72) circle [radius=1.5pt];
		\filldraw [blue] (0.5,-2.25) circle [radius=1.5pt];
		\filldraw [blue] (0.5,-2.78) circle [radius=1.5pt];
		\filldraw [blue] (0.5,-3.0) circle [radius=1.5pt];
		
		\filldraw [blue] (1.5,-1.5) circle [radius=1.5pt];
		\filldraw [blue] (1.5,-1.39) circle [radius=1.5pt];
		\filldraw [blue] (1.5,-1.08) circle [radius=1.5pt];
		\filldraw [blue] (1.5,-0.60) circle [radius=1.5pt];
		\filldraw [blue] (1.5,-0.05) circle [radius=1.5pt];
		\filldraw [blue] (1.5,0.50) circle [radius=1.5pt];
		\filldraw [blue] (1.5,0.97) circle [radius=1.5pt];			
		\filldraw [blue] (1.5,1.29) circle [radius=1.5pt];
		\filldraw [blue] (1.5,1.40) circle [radius=1.5pt];
		\filldraw [blue] (1.5,1.51) circle [radius=1.5pt];	
		\filldraw [blue] (1.5,1.80) circle [radius=1.5pt];	
		\filldraw [blue] (1.5,2.20) circle [radius=1.5pt];	
		\filldraw [blue] (1.5,2.60) circle [radius=1.5pt];	
		\filldraw [blue] (1.5,2.89) circle [radius=1.5pt];
		\filldraw [blue] (1.5,3) circle [radius=1.5pt];			
		\filldraw [blue] (1.5,3.22) circle [radius=1.5pt];
		\filldraw [blue] (1.5,3.75) circle [radius=1.5pt];
		\filldraw [blue] (1.5,4.28) circle [radius=1.5pt];
		\filldraw [blue] (1.5,4.50) circle [radius=1.5pt];			
		\filldraw [blue] (1.5,-1.72) circle [radius=1.5pt];
		\filldraw [blue] (1.5,-2.25) circle [radius=1.5pt];
		\filldraw [blue] (1.5,-2.78) circle [radius=1.5pt];
		\filldraw [blue] (1.5,-3.0) circle [radius=1.5pt];
		
		\filldraw [blue] (2.5,-1.5) circle [radius=1.5pt];
		\filldraw [blue] (2.5,-1.39) circle [radius=1.5pt];
		\filldraw [blue] (2.5,-1.08) circle [radius=1.5pt];
		\filldraw [blue] (2.5,-0.60) circle [radius=1.5pt];
		\filldraw [blue] (2.5,-0.05) circle [radius=1.5pt];
		\filldraw [blue] (2.5,0.50) circle [radius=1.5pt];
		\filldraw [blue] (2.5,0.97) circle [radius=1.5pt];			
		\filldraw [blue] (2.5,1.29) circle [radius=1.5pt];
		\filldraw [blue] (2.5,1.40) circle [radius=1.5pt];
		\filldraw [blue] (2.5,1.51) circle [radius=1.5pt];	
		\filldraw [blue] (2.5,1.80) circle [radius=1.5pt];	
		\filldraw [blue] (2.5,2.20) circle [radius=1.5pt];	
		\filldraw [blue] (2.5,2.60) circle [radius=1.5pt];	
		\filldraw [blue] (2.5,2.89) circle [radius=1.5pt];
		\filldraw [blue] (2.5,3) circle [radius=1.5pt];			
		\filldraw [blue] (2.5,3.22) circle [radius=1.5pt];
		\filldraw [blue] (2.5,3.75) circle [radius=1.5pt];
		\filldraw [blue] (2.5,4.28) circle [radius=1.5pt];
		\filldraw [blue] (2.5,4.50) circle [radius=1.5pt];			
		\filldraw [blue] (2.5,-1.72) circle [radius=1.5pt];
		\filldraw [blue] (2.5,-2.25) circle [radius=1.5pt];
		\filldraw [blue] (2.5,-2.78) circle [radius=1.5pt];
		\filldraw [blue] (2.5,-3.0) circle [radius=1.5pt];
		
		\filldraw [blue] (3.5,-1.5) circle [radius=1.5pt];
		\filldraw [blue] (3.5,-1.39) circle [radius=1.5pt];
		\filldraw [blue] (3.5,-1.08) circle [radius=1.5pt];
		\filldraw [blue] (3.5,-0.60) circle [radius=1.5pt];
		\filldraw [blue] (3.5,-0.05) circle [radius=1.5pt];
		\filldraw [blue] (3.5,0.50) circle [radius=1.5pt];
		\filldraw [blue] (3.5,0.97) circle [radius=1.5pt];			
		\filldraw [blue] (3.5,1.29) circle [radius=1.5pt];
		\filldraw [blue] (3.5,1.40) circle [radius=1.5pt];
		\filldraw [blue] (3.5,1.51) circle [radius=1.5pt];	
		\filldraw [blue] (3.5,1.80) circle [radius=1.5pt];	
		\filldraw [blue] (3.5,2.20) circle [radius=1.5pt];	
		\filldraw [blue] (3.5,2.60) circle [radius=1.5pt];	
		\filldraw [blue] (3.5,2.89) circle [radius=1.5pt];
		\filldraw [blue] (3.5,3) circle [radius=1.5pt];			
		\filldraw [blue] (3.5,3.22) circle [radius=1.5pt];
		\filldraw [blue] (3.5,3.75) circle [radius=1.5pt];
		\filldraw [blue] (3.5,4.28) circle [radius=1.5pt];
		\filldraw [blue] (3.5,4.50) circle [radius=1.5pt];			
		\filldraw [blue] (3.5,-1.72) circle [radius=1.5pt];
		\filldraw [blue] (3.5,-2.25) circle [radius=1.5pt];
		\filldraw [blue] (3.5,-2.78) circle [radius=1.5pt];
		\filldraw [blue] (3.5,-3.0) circle [radius=1.5pt];
		
		\filldraw [blue] (4.5,-1.5) circle [radius=1.5pt];
		\filldraw [blue] (4.5,-1.39) circle [radius=1.5pt];
		\filldraw [blue] (4.5,-1.08) circle [radius=1.5pt];
		\filldraw [blue] (4.5,-0.60) circle [radius=1.5pt];
		\filldraw [blue] (4.5,-0.05) circle [radius=1.5pt];
		\filldraw [blue] (4.5,0.50) circle [radius=1.5pt];
		\filldraw [blue] (4.5,0.97) circle [radius=1.5pt];			
		\filldraw [blue] (4.5,1.29) circle [radius=1.5pt];
		\filldraw [blue] (4.5,1.40) circle [radius=1.5pt];
		\filldraw [blue] (4.5,1.51) circle [radius=1.5pt];	
		\filldraw [blue] (4.5,1.80) circle [radius=1.5pt];	
		\filldraw [blue] (4.5,2.20) circle [radius=1.5pt];	
		\filldraw [blue] (4.5,2.60) circle [radius=1.5pt];	
		\filldraw [blue] (4.5,2.89) circle [radius=1.5pt];
		\filldraw [blue] (4.5,3) circle [radius=1.5pt];			
		\filldraw [blue] (4.5,3.22) circle [radius=1.5pt];
		\filldraw [blue] (4.5,3.75) circle [radius=1.5pt];
		\filldraw [blue] (4.5,4.28) circle [radius=1.5pt];
		\filldraw [blue] (4.5,4.50) circle [radius=1.5pt];			
		\filldraw [blue] (4.5,-1.72) circle [radius=1.5pt];
		\filldraw [blue] (4.5,-2.25) circle [radius=1.5pt];
		\filldraw [blue] (4.5,-2.78) circle [radius=1.5pt];
		\filldraw [blue] (4.5,-3.0) circle [radius=1.5pt];
		
		\filldraw [blue] (5.5,-1.5) circle [radius=1.5pt];
		\filldraw [blue] (5.5,-1.39) circle [radius=1.5pt];
		\filldraw [blue] (5.5,-1.08) circle [radius=1.5pt];
		\filldraw [blue] (5.5,-0.60) circle [radius=1.5pt];
		\filldraw [blue] (5.5,-0.05) circle [radius=1.5pt];
		\filldraw [blue] (5.5,0.50) circle [radius=1.5pt];
		\filldraw [blue] (5.5,0.97) circle [radius=1.5pt];			
		\filldraw [blue] (5.5,1.29) circle [radius=1.5pt];
		\filldraw [blue] (5.5,1.40) circle [radius=1.5pt];
		\filldraw [blue] (5.5,1.51) circle [radius=1.5pt];	
		\filldraw [blue] (5.5,1.80) circle [radius=1.5pt];	
		\filldraw [blue] (5.5,2.20) circle [radius=1.5pt];	
		\filldraw [blue] (5.5,2.60) circle [radius=1.5pt];	
		\filldraw [blue] (5.5,2.89) circle [radius=1.5pt];
		\filldraw [blue] (5.5,3) circle [radius=1.5pt];			
		\filldraw [blue] (5.5,3.22) circle [radius=1.5pt];
		\filldraw [blue] (5.5,3.75) circle [radius=1.5pt];
		\filldraw [blue] (5.5,4.28) circle [radius=1.5pt];
		\filldraw [blue] (5.5,4.50) circle [radius=1.5pt];			
		\filldraw [blue] (5.5,-1.72) circle [radius=1.5pt];
		\filldraw [blue] (5.5,-2.25) circle [radius=1.5pt];
		\filldraw [blue] (5.5,-2.78) circle [radius=1.5pt];
		\filldraw [blue] (5.5,-3.0) circle [radius=1.5pt];
		
		\filldraw [blue] (6.5,-1.5) circle [radius=1.5pt];
		\filldraw [blue] (6.5,-1.39) circle [radius=1.5pt];
		\filldraw [blue] (6.5,-1.08) circle [radius=1.5pt];
		\filldraw [blue] (6.5,-0.60) circle [radius=1.5pt];
		\filldraw [blue] (6.5,-0.05) circle [radius=1.5pt];
		\filldraw [blue] (6.5,0.50) circle [radius=1.5pt];
		\filldraw [blue] (6.5,0.97) circle [radius=1.5pt];			
		\filldraw [blue] (6.5,1.29) circle [radius=1.5pt];
		\filldraw [blue] (6.5,1.40) circle [radius=1.5pt];
		\filldraw [blue] (6.5,1.51) circle [radius=1.5pt];	
		\filldraw [blue] (6.5,1.80) circle [radius=1.5pt];	
		\filldraw [blue] (6.5,2.20) circle [radius=1.5pt];	
		\filldraw [blue] (6.5,2.60) circle [radius=1.5pt];	
		\filldraw [blue] (6.5,2.89) circle [radius=1.5pt];
		\filldraw [blue] (6.5,3) circle [radius=1.5pt];			
		\filldraw [blue] (6.5,3.22) circle [radius=1.5pt];
		\filldraw [blue] (6.5,3.75) circle [radius=1.5pt];
		\filldraw [blue] (6.5,4.28) circle [radius=1.5pt];
		\filldraw [blue] (6.5,4.50) circle [radius=1.5pt];			
		\filldraw [blue] (6.5,-1.72) circle [radius=1.5pt];
		\filldraw [blue] (6.5,-2.25) circle [radius=1.5pt];
		\filldraw [blue] (6.5,-2.78) circle [radius=1.5pt];
		\filldraw [blue] (6.5,-3.0) circle [radius=1.5pt];
		
		\filldraw [blue] (7.5,-1.5) circle [radius=1.5pt];
		\filldraw [blue] (7.5,-1.39) circle [radius=1.5pt];
		\filldraw [blue] (7.5,-1.08) circle [radius=1.5pt];
		\filldraw [blue] (7.5,-0.60) circle [radius=1.5pt];
		\filldraw [blue] (7.5,-0.05) circle [radius=1.5pt];
		\filldraw [blue] (7.5,0.50) circle [radius=1.5pt];
		\filldraw [blue] (7.5,0.97) circle [radius=1.5pt];			
		\filldraw [blue] (7.5,1.29) circle [radius=1.5pt];
		\filldraw [blue] (7.5,1.40) circle [radius=1.5pt];
		\filldraw [blue] (7.5,1.51) circle [radius=1.5pt];	
		\filldraw [blue] (7.5,1.80) circle [radius=1.5pt];	
		\filldraw [blue] (7.5,2.20) circle [radius=1.5pt];	
		\filldraw [blue] (7.5,2.60) circle [radius=1.5pt];	
		\filldraw [blue] (7.5,2.89) circle [radius=1.5pt];
		\filldraw [blue] (7.5,3) circle [radius=1.5pt];			
		\filldraw [blue] (7.5,3.22) circle [radius=1.5pt];
		\filldraw [blue] (7.5,3.75) circle [radius=1.5pt];
		\filldraw [blue] (7.5,4.28) circle [radius=1.5pt];
		\filldraw [blue] (7.5,4.50) circle [radius=1.5pt];			
		\filldraw [blue] (7.5,-1.72) circle [radius=1.5pt];
		\filldraw [blue] (7.5,-2.25) circle [radius=1.5pt];
		\filldraw [blue] (7.5,-2.78) circle [radius=1.5pt];
		\filldraw [blue] (7.5,-3.0) circle [radius=1.5pt];
		
		\filldraw [blue] (8.5,-1.5) circle [radius=1.5pt];
		\filldraw [blue] (8.5,-1.39) circle [radius=1.5pt];
		\filldraw [blue] (8.5,-1.08) circle [radius=1.5pt];
		\filldraw [blue] (8.5,-0.60) circle [radius=1.5pt];
		\filldraw [blue] (8.5,-0.05) circle [radius=1.5pt];
		\filldraw [blue] (8.5,0.50) circle [radius=1.5pt];
		\filldraw [blue] (8.5,0.97) circle [radius=1.5pt];			
		\filldraw [blue] (8.5,1.29) circle [radius=1.5pt];
		\filldraw [blue] (8.5,1.40) circle [radius=1.5pt];
		\filldraw [blue] (8.5,1.51) circle [radius=1.5pt];	
		\filldraw [blue] (8.5,1.80) circle [radius=1.5pt];	
		\filldraw [blue] (8.5,2.20) circle [radius=1.5pt];	
		\filldraw [blue] (8.5,2.60) circle [radius=1.5pt];	
		\filldraw [blue] (8.5,2.89) circle [radius=1.5pt];
		\filldraw [blue] (8.5,3) circle [radius=1.5pt];			
		\filldraw [blue] (8.5,3.22) circle [radius=1.5pt];
		\filldraw [blue] (8.5,3.75) circle [radius=1.5pt];
		\filldraw [blue] (8.5,4.28) circle [radius=1.5pt];
		\filldraw [blue] (8.5,4.50) circle [radius=1.5pt];			
		\filldraw [blue] (8.5,-1.72) circle [radius=1.5pt];
		\filldraw [blue] (8.5,-2.25) circle [radius=1.5pt];
		\filldraw [blue] (8.5,-2.78) circle [radius=1.5pt];
		\filldraw [blue] (8.5,-3.0) circle [radius=1.5pt];
		
		\filldraw [blue] (9.5,-1.5) circle [radius=1.5pt];
		\filldraw [blue] (9.5,-1.39) circle [radius=1.5pt];
		\filldraw [blue] (9.5,-1.08) circle [radius=1.5pt];
		\filldraw [blue] (9.5,-0.60) circle [radius=1.5pt];
		\filldraw [blue] (9.5,-0.05) circle [radius=1.5pt];
		\filldraw [blue] (9.5,0.50) circle [radius=1.5pt];
		\filldraw [blue] (9.5,0.97) circle [radius=1.5pt];			
		\filldraw [blue] (9.5,1.29) circle [radius=1.5pt];
		\filldraw [blue] (9.5,1.40) circle [radius=1.5pt];
		\filldraw [blue] (9.5,1.51) circle [radius=1.5pt];	
		\filldraw [blue] (9.5,1.80) circle [radius=1.5pt];	
		\filldraw [blue] (9.5,2.20) circle [radius=1.5pt];	
		\filldraw [blue] (9.5,2.60) circle [radius=1.5pt];	
		\filldraw [blue] (9.5,2.89) circle [radius=1.5pt];
		\filldraw [blue] (9.5,3) circle [radius=1.5pt];			
		\filldraw [blue] (9.5,3.22) circle [radius=1.5pt];
		\filldraw [blue] (9.5,3.75) circle [radius=1.5pt];
		\filldraw [blue] (9.5,4.28) circle [radius=1.5pt];
		\filldraw [blue] (9.5,4.50) circle [radius=1.5pt];			
		\filldraw [blue] (9.5,-1.72) circle [radius=1.5pt];
		\filldraw [blue] (9.5,-2.25) circle [radius=1.5pt];
		\filldraw [blue] (9.5,-2.78) circle [radius=1.5pt];
		\filldraw [blue] (9.5,-3.0) circle [radius=1.5pt];
		
		\filldraw [blue] (10.5,-1.5) circle [radius=1.5pt];
		\filldraw [blue] (10.5,-1.39) circle [radius=1.5pt];
		\filldraw [blue] (10.5,-1.08) circle [radius=1.5pt];
		\filldraw [blue] (10.5,-0.60) circle [radius=1.5pt];
		\filldraw [blue] (10.5,-0.05) circle [radius=1.5pt];
		\filldraw [blue] (10.5,0.50) circle [radius=1.5pt];
		\filldraw [blue] (10.5,0.97) circle [radius=1.5pt];			
		\filldraw [blue] (10.5,1.29) circle [radius=1.5pt];
		\filldraw [blue] (10.5,1.40) circle [radius=1.5pt];
		\filldraw [blue] (10.5,1.51) circle [radius=1.5pt];	
		\filldraw [blue] (10.5,1.80) circle [radius=1.5pt];	
		\filldraw [blue] (10.5,2.20) circle [radius=1.5pt];	
		\filldraw [blue] (10.5,2.60) circle [radius=1.5pt];	
		\filldraw [blue] (10.5,2.89) circle [radius=1.5pt];
		\filldraw [blue] (10.5,3) circle [radius=1.5pt];			
		\filldraw [blue] (10.5,3.22) circle [radius=1.5pt];
		\filldraw [blue] (10.5,3.75) circle [radius=1.5pt];
		\filldraw [blue] (10.5,4.28) circle [radius=1.5pt];
		\filldraw [blue] (10.5,4.50) circle [radius=1.5pt];			
		\filldraw [blue] (10.5,-1.72) circle [radius=1.5pt];
		\filldraw [blue] (10.5,-2.25) circle [radius=1.5pt];
		\filldraw [blue] (10.5,-2.78) circle [radius=1.5pt];
		\filldraw [blue] (10.5,-3.0) circle [radius=1.5pt];	
		
		\draw[thick,dashed] (-0.5,1.4)--(10.5,1.4);	
		\filldraw [red] (0-0.5,1.4) circle [radius=2.5 pt]; 
		\node at (-0.5,1.6)[right]{source}; 
		\node at (-0.6-1.2,1.4)[right]{$y_\mathrm{s}+\epsilon$};
		\node at (0-0.5,4.6)[left]{0};
		\node at (-0.6-0.4,3.0)[right]{$\epsilon$};
		\node at (-0.6-1.6,-1.5)[right]{$y_{\max}+\epsilon$};
		\node at (-0.6-1.9,-3.0)[right]{$y_{\max}+2\epsilon$};
		\node at (5,3.625){Perfectly matched layer};
		\node at (5,0.5){$\kappa_{n}(x,y)$};
		\node at (5,-2.275){Perfectly matched layer};
		\node at (10,4.75)[right]{$x_{\max}$};
	\end{tikzpicture}
	\caption{Schematic diagram of the computational domain of the MPEs. The blue dots represent the Chebyshev--Gauss--Lobatto points used for spectral discretization, and the black dotted line represents a virtual interface set at the sound source.}
	\label{Figure2}
\end{figure}
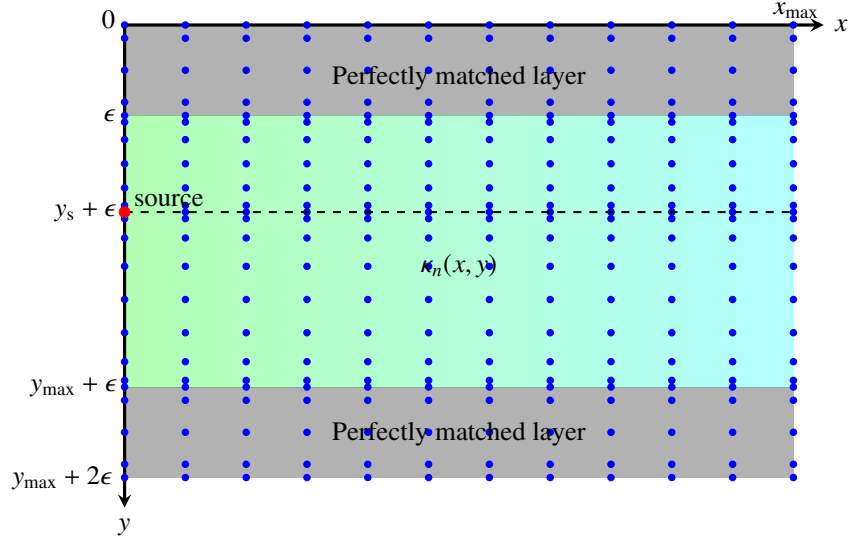

The absorption layer is a commonly employed technique to imitate free boundaries. The fundamental concept behind this approach is to introduce an artificial layer with a high absorption coefficient to effectively absorb waves that penetrate the boundary, thereby ensuring their minimal impact on the region of interest. As illustrated in Fig.~\ref{Figure2}, we establish an absorption layer with a thickness of $\epsilon$ at the top and bottom of the $y$-axis. Traditional artificial absorption layers often require a substantial thickness to guarantee the desired absorption effect. However, this can result in a larger computational domain and occupy numerous grid points, subsequently increasing computational costs for regions that are not of immediate concern.

Currently, the most widely adopted technique is the perfectly matched layer (PML), which offers superior absorption capabilities with thinner layers while minimizing computational costs. Originally developed for simulating Maxwell's equations in computational electromagnetics \cite{Berenger1994}, the PML technique was adapted for solving parabolic equations in the early 21st century \cite{Yevick2000, LuYy2007, Levy2001}. The key component of the PML technique lies in complex coordinate transformation, which converts the $y$-axis coordinates into $\tilde{y}$ coordinates incorporating an absorption parameter $\sigma(y)$. This transformation enables effective absorption within the PML.

\begin{equation}
	\label{eq.23}
	\tilde{y}=y+\mathrm{i}\int_0^y \sigma(\tau) \mathrm{d}\tau,\quad \begin{cases}
		\sigma(y)=0, &y\in[\epsilon,y_{\max}],\\
		\sigma(y)>0, &y\in[0,\epsilon]\cup[y_{\max}+\epsilon,y_{\max}+2\epsilon].
	\end{cases}
\end{equation}
After performing the complex coordinate transformation, the $\mathcal{Y}$ operator in the PML takes on the following form:
\begin{equation}
	\kappa_{\mathrm{s}}^2 \mathcal{Y}=\frac{1}{1+\mathrm{i} \sigma(y)} \frac{\partial}{\partial y} \frac{1}{1+\mathrm{i} \sigma(y)} \frac{\partial}{\partial y}+\kappa^2-\kappa_{\mathrm{s}}^2.
\end{equation}
By carefully choosing the absorption parameter $\sigma(y)$, efficient absorption can be attained. A comprehensive analysis of sound wave absorption using PML is provided in Ref.~\cite{Tuhw2023d}. When solving MPEs, the PML parameters can be determined as follows:
\begin{equation*}
\sigma(y)=\sigma_0(y-y_{\max})^3/\epsilon^3.
\end{equation*}

\subsection{Starter}
Eqs.~\eqref{eq.20} to \eqref{eq.21} necessitate an initial condition $\phi(0,y)$ for their forward advancement, commonly known as the ``starter'' in the parabolic model. In the context of the underwater acoustic parabolic model, starters are typically designed to generate an approximately excited field as if it were produced by a point source \cite{Collins2019}. We outline the initialization process for the two-dimensional underwater acoustic parabolic model and introduce three types of starters.

\subsubsection{Greene starter}
An analytical sound source offers an efficient way to generate a starter with minimal computational effort. By ensuring that the source aperture aligns with the angular limitations in the parabolic model, a stable analytical starter can be obtained. One example of such a source is Greene's sound source, which exhibits excellent performance over a wide range of angles. This source can be described by a weighted Gaussian expression:
\begin{equation}
	\phi(0, y)=\frac{\Psi\left(0,y_{\mathrm{s}},z_\mathrm{s}\right)}{2 \sqrt{\pi}}\left[1.4467-0.8402 \kappa_{\mathrm{s}}^2 (y-y_\mathrm{s})^2\right] \exp\left[{-\frac{\kappa_{\mathrm{s}}^2 (y-y_\mathrm{s})^2}{1.5256}}\right].
\end{equation}
It can efficiently induce wave excitation within a tensor angle of less than 30°, presenting efficiency advantages over wide-angle starters \cite{Greene1984}.

\subsubsection{Ray-based starter}
For the unbounded domains under consideration, the ray model proves to be especially well suited for calculating the initial field of MPEs, eliminating the need to account for sound wave reflections from boundaries. In 2020, Petrov et al. introduced a wide-angle parabolic model starter utilizing ray tracing techniques \cite{Petrov2020a}. A succinct overview of this approach follows.

Referring to the two-dimensional ray model in the $(x,z)$ coordinate system, the horizontal refractive index $\Phi(x,y)$ can be expressed in the following manner:
\begin{equation}
	\Phi(x,y)=Q(x,y) \mathrm{e}^{\mathrm{i} \kappa_{\mathrm{s}} S(x, y)}+o\left(1 / \kappa_{\mathrm{s}} \right),
\end{equation}
where $Q$ and $S$ denote the amplitude and phase of $\Phi$, respectively. The above equation implies the assumption of $x<\lambda$ (where $\lambda$ represents the equivalent wavelength), suggesting that the medium properties remain independent of $x$ within short distances. $S$ and $Q$ satisfy the following eikonal and transport equations, respectively.
\begin{subequations}
	\begin{gather}
		\left(\nabla S\right)^2=n(x, y),\quad n(x, y)=\kappa(x, y) / \kappa_{\mathrm{s}},\\
		2\nabla Q \cdot \nabla S + Q \nabla^2 S=0.
	\end{gather}
\end{subequations}
The following curvilinear coordinate system for the rays are introduced as follows:

\begin{equation}
	\begin{aligned}
		\frac{\mathrm{d} x}{\mathrm{d} \ell}=\frac{\xi}{n} , &\quad \frac{\mathrm{d} \xi}{\mathrm{d} \ell}=\frac{\partial n}{\partial x}, \\
		\frac{\mathrm{d} y}{\mathrm{d} \ell}=\frac{\eta}{n}, &\quad \frac{\mathrm{d} \eta}{\mathrm{d} \ell}=\frac{\partial n}{\partial y}.
	\end{aligned}
\end{equation}
where the parameter $\ell$ represents the arc length along the ray. In Cartesian coordinates $(x,y)$, the ray equations can be expressed in the aforementioned first-order form. The ray possesses a take-off angle $\alpha$ and initiates from the source position $(0,y_\mathrm{s})$ as its initial conditions.	
\begin{equation}
	\begin{aligned}
		x(0)=0, &\quad \xi(0)=\cos \alpha, \\
		y(0)=y_\mathrm{s}, &\quad \eta(0)=\sin \alpha.
	\end{aligned}		
\end{equation}
According to classical ray theory, solving for the following equations is not difficult:
\begin{subequations}
	\begin{gather}
		S(\ell)=S(0)+\int_0^{\ell} n(\ell) \mathrm{d} \ell,\\
		Q(\ell)=\frac{Q_0}{n(\ell)}\sqrt{\frac{\cos\alpha}{\partial y(\ell,\alpha)/\partial \alpha}},\quad Q_0=\frac{\mathrm{e}^{\mathrm{i}\pi/4}}{\sqrt{8\pi\kappa_{\mathrm{s}}}}.
	\end{gather}
\end{subequations}
Since we calculate the initial field within a very short segment where the medium properties are independent of $x$, we can further assume that the medium within this segment is homogeneous ($n(\ell)=1$). This assumption may not be entirely reasonable in a two-dimensional ray model, as the sound speed exhibits distinct vertical distribution characteristics. However, the sound speed variation is slow in the $y$ direction. In this case, the amplitude and phase of the ray-based starter can be simplified as follows:
\begin{equation}
	x(\ell)=\ell\cos\alpha,\quad y(\ell)=\ell\sin\alpha,\quad S(\ell)=\ell,\quad Q(\ell)=\frac{Q_0}{\sqrt{x^2+(y-y_\mathrm{s}^2)}}.
\end{equation}

The desired aperture for the ray-based starter can be specified by appropriately defining interval values for $\alpha$.

\subsubsection{Self-starter}
The application of the self-starter technique in parabolic models has a long-standing history, originally proposed by Collins \cite{Collins1992,Robert1997,Collins1999b}. This technique has been implemented in RAM \cite{RAM}, showcasing its practicality and effectiveness. The self-starter is computed by solving a boundary value problem (BVP) that contains the PE operator, which is more efficient than the traditional normal-mode starter. Following Collins' approach, a self-starter applicable to MPEs can be constructed by reference \cite{Petrov2020a}.

Starting from Eq.~\eqref{eq.14}, we assume that over a short distance, the equivalent wavenumber $\kappa$ is $x$-independent, i.e., $\kappa(x,y)=\kappa(y)$. We let $\{f_\nu(y)\}$ be a complete system of eigenfunctions for the operator $\left(\frac{\partial^2}{\partial y^2}+\kappa^2\right)$ on the interval $(-\infty,\infty)$, with corresponding eigenvalues $\lambda_\nu$. When $x>0$, the solution to the HRE \eqref{eq.8} is given by:
\begin{equation}
	\phi(x, y)=\frac{\mathrm{i} \Psi(z_\mathrm{s})}{2} \int \frac{1}{\lambda_\nu} f_\nu(0) f_\nu(y) \mathrm{e}^{\mathrm{i} \lambda_\nu x} \mathrm{d} \nu.
\end{equation}
When $x=0$, the above equation degenerates to:
\begin{equation}
	\phi_0(y) \equiv \phi(0, y)=\frac{\mathrm{i} \Psi(z_\mathrm{s})}{2} \int \frac{1}{\lambda_\nu} f_\nu(0) f_\nu(y) \mathrm{d} \nu.
\end{equation}
The above equation represents the solution to the one-dimensional BVP of
\begin{equation}
	\label{eq.34}
	\left(\sqrt{\frac{\partial^2}{\partial y^2}+\kappa^2}\right) \phi_0=\frac{\mathrm{i} \Psi(z_\mathrm{s})}{2} \delta(y-y_{\mathrm{s}}),
\end{equation}
where the boundary condition at infinity is the radiation condition. This BVP cannot be directly solved numerically when $k=k(y)$. However, Collins' indirect method \cite{Collins1992} can be applied in this situation. The first step of this method is to solve the following auxiliary BVP:
\begin{equation}
	\label{eq.35}
	(1+\mathcal{Y}) \zeta=\frac{\mathrm{i} \Psi(z_\mathrm{s})}{2 \kappa_{\mathrm{s}}^2} \delta(y-y_{\mathrm{s}}).
\end{equation}
By imposing matching conditions for the Dirac delta function $\delta(y-y_{\mathrm{s}})$ at $y=0$, the equation can be easily solved numerically. The solution to the above equation can be represented by the eigenfunction $\{f_\nu\}$ as follows:	
\begin{equation}
	\zeta_0(y)=\frac{\mathrm{i} \Psi(z_\mathrm{s})}{2} \int \frac{1}{\lambda_\nu^2} f_\nu(0) f_\nu(y) \mathrm{d} \nu \delta(y).
\end{equation}	
Specifically, in the case of a constant wavenumber $\kappa(y)=\kappa_\mathrm{s}$, $\zeta_0$ has the following analytical solution:
\begin{equation}
	\label{eq.37}
	\zeta_0(y)=\frac{\Psi(z_\mathrm{s})}{4} \mathrm{e}^{\mathrm{i} \kappa_{\mathrm{s}}|y-y_{\mathrm{s}}|}.
\end{equation}
From Eqs.~\eqref{eq.34} and \eqref{eq.35}, it can be observed that the self-starter $\phi_0(y)$ can be calculated using the following formula:
\begin{equation}
	\phi_0(y)=\kappa_{\mathrm{s}} \sqrt{1+\mathcal{Y}} \zeta_{0}(y).
\end{equation}
Similarly, the square operator in the above equation can also be approximated using the Pad\'e series.

Notably, Eq.~\eqref{eq.37} cannot be directly applied to PML as PML involves complex coordinate transformation \eqref{eq.23}. Taking the second layer of PML as an example, the modification for $\zeta_0$ should be as follows:
\begin{equation}
	\zeta_0(y)=\frac{\Psi(z_\mathrm{s})}{4} \mathrm{e}^{\mathrm{i} \kappa_{\mathrm{s}} |y-y_{\mathrm{s}}|} \exp\left(-\kappa_{\mathrm{s}} \int_{y_{\max}}^y \sigma(y) \mathrm{d} y\right).
\end{equation}

\section{Spectral discretization}
\label{sec3}
\subsection{Spectral method}
Spectral methods are a numerical discretization technique used to solve differential equations by discretizing continuous differential equations into algebraic equations, similar to finite difference and finite element methods \cite{Orszag1972,Gottlieb1977}. In spectral methods, the function $u(t)$ is approximated using orthogonal polynomials $\{\varphi_j(t)\}$ to obtain the expansion coefficients $\{\hat{u}_j\}$ of $u(t)$. This process, known as spectral expansion or transformation, yields the spectral coefficients $\{\hat{u}_j\}$ \cite{Canuto1982}.

Clearly, achieving precise spectral approximation necessitates an infinite series, as shown in Eq.~\eqref{eq.40}. However, the favorable properties of orthogonal polynomials render the spectral series progressively less significant as the number of terms increases. This convergence offers theoretical assurance for truncation in spectral approximation \cite{Canuto1988,Canuto2006}.	
\begin{equation}
	\label{eq.40}
	u(t)=\sum_{j=0}^{\infty}\hat{u}_j\varphi_j(t) \approx u_N(t)=\sum_{j=0}^{N}\hat{u}_j\varphi_j(t).
\end{equation}

Finite order numerical truncation can introduce errors in spectral approximation, resulting in deviations from the original differential equation and giving rise to a residual term $R_N(t)$. The concept of discretizing a system of linear equations using spectral methods is based on the principle of weighted residuals. This involves selecting a suitable set of weight functions $\{w_j(t)\}$, multiplying them with the residuals $R_N(t)$, and integrating them over the defined domain. This process, known as weighted residuals, leads to a system of algebraic equations involving the spectral coefficients $\{\hat{u}_j\}$, with the aim of minimizing the weighted residuals to zero \cite{Jshen2006,Jshen2011}. More specifically, in classical Galerkin- and Tau-type spectral methods, the weight function is chosen to be the basis function itself, $w_j(t)=\varphi_j(t)$.
\begin{equation}
	\int_{\Omega} R_N(t)\varphi_j(t) \mathrm{d}x=0,\quad j=0,1,2,\ldots,N.
\end{equation}
The weighted residual method utilizes the orthogonality property of the weight function to generate a series of algebraic equations involving the spectral coefficients. Detailed discussions on this topic can be found in monographs dedicated to spectral methods \cite{Guoby1998}.

Achieving accurate computation of the integral in the equation above is crucial in spectral methods. In most cases, evaluating the integral of the weighted residual cannot be performed analytically and requires numerical methods. Gaussian quadrature, known for its $(2N+1)$ order accuracy (where $N$ represents the number of nodes), is the preferred technique for numerical integration in spectral methods. Gaussian quadrature involves special Gauss nodes and weights, which are typically tailored to the choice of basis functions (orthogonal polynomials) $\{\varphi_j(t)\}$. In this paper, we employ the Chebyshev spectral method that utilizes Chebyshev polynomials $\{T_j(t)\}$ as the basis functions, i.e., $\varphi_j(t)=T_j(t)$. Notably, the term ``Chebyshev" here specifically refers to the first kind of Chebyshev polynomial. These polynomials are defined on the interval $t\in[-1,1]$, exhibit symmetry, and are well-suited for spectral approximation in bounded domains. Chebyshev polynomials possess excellent numerical properties and can be efficiently utilized for numerical integration, differentiation, and other calculations \cite{Boyd2001,Mason2002}. Leveraging Chebyshev polynomials as the basis functions ensures high accuracy and rapid convergence.

\begin{itemize}
	\item
	Numerical integration
	
	When conducting numerical integration with Chebyshev polynomials, the Gauss-Chebyshev quadrature method can be employed. This method utilizes carefully selected nodes and weights to attain highly accurate numerical integration. For bounded regions that include the endpoints, the most effective choice of nodes is the Gauss-Chebyshev-Lobatto (GCL) nodes. The GCL nodes and weights can be defined as follows:
	\begin{equation}
		t_j=\cos\left(\frac{j\pi}{N}\right), \quad j = 0, 1, 2, \ldots, N,\quad \omega_{j}=\left\{\begin{array}{ll}
			\frac{\pi}{2N}, & j=0, N, \\
			\frac{\pi}{N}, & \mathrm{otherwise}.
		\end{array}\right.
	\end{equation}
	In the above definition, $N+1$ denotes the total number of nodes, and $t_j$ represents the $j$-th GCL node. Essentially, this definition divides a semicircle into $N+1$ equal parts through equidistant central angles. The projection of each division point is taken on the diameter axis, as shown in Fig.~\ref{Figure3}. By utilizing these nodes in conjunction with suitable weights, precise numerical integration can be achieved when working with the Chebyshev spectral method \cite{Canuto2006}.
	\begin{figure}
		\centering
		\includegraphics[width=0.75\textwidth]{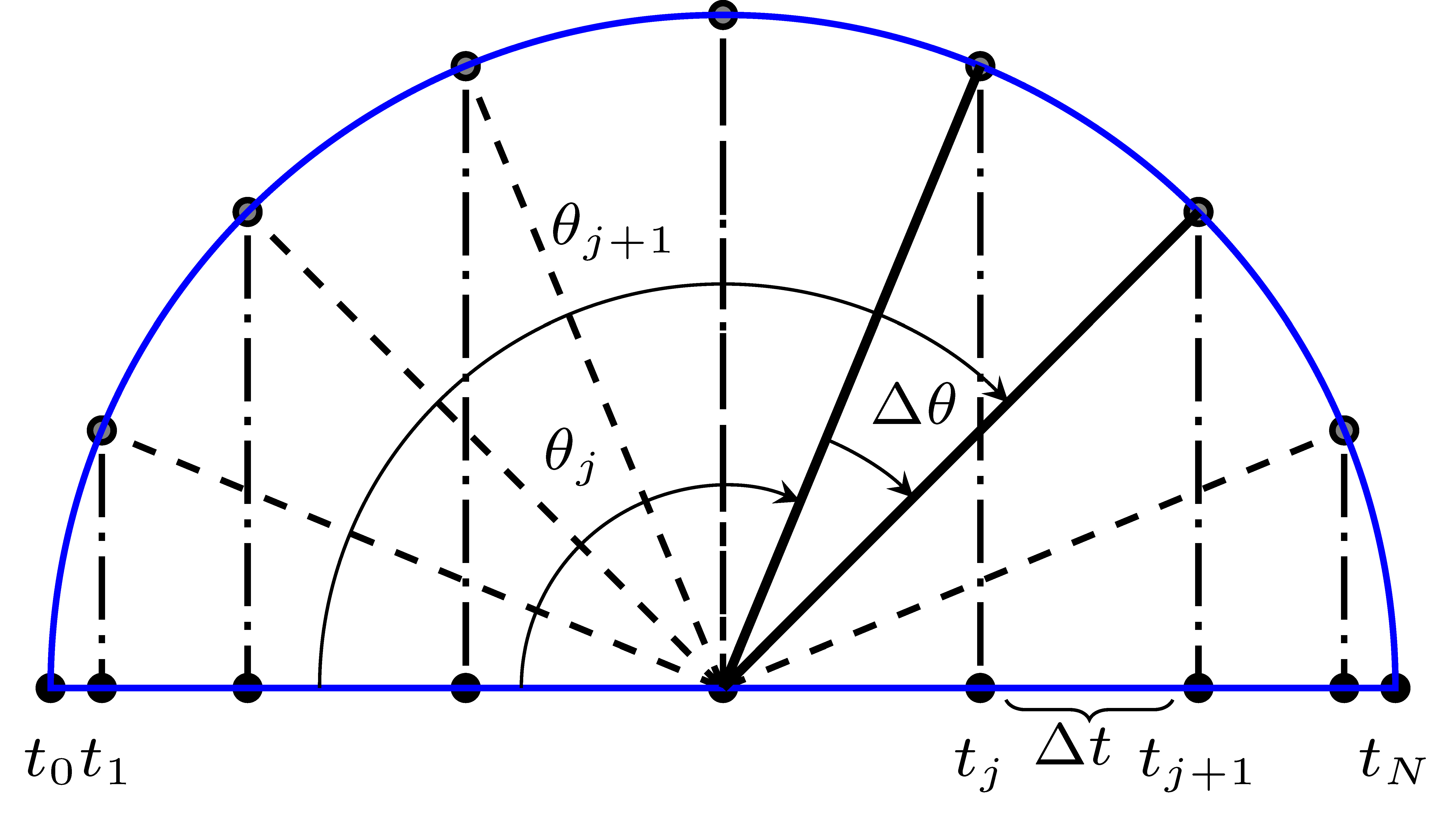}
		\caption{The GCL nodes used in $(N+1)$-oder spectral approximation.}
		\label{Figure3}
	\end{figure}
	
	\item
	Numerical differentiation
	
	If the function $u(t)$ can be represented by an expansion using Chebyshev polynomials $\{\hat{u}_j\}$, then the derivative of $u(t)$, denoted as $u'(t)$, can also be expanded using Chebyshev polynomials. The corresponding spectral coefficients of $u'(t)$ are denoted as $\{\hat{u}_j'\}$. There exists a numerical relationship between $\{\hat{u}_j'\}$ and $\{\hat{u}_j\}$, given by the following expression:
	\begin{equation}
		\label{eq.43}
		\hat{u}'_i \approx \frac{2}{c_i}
		\sum_{\substack{j=i+1,\\ 
				j+i=\mathrm{odd}
		}}^{N} j \hat{u}_j,\quad c_i=\begin{cases}
			2,  & i=0,\\
			1,  & i>0,
		\end{cases} \quad \bm{\hat{u}}' \approx \bm{D}_N \bm{\hat{u}} .
	\end{equation}
	This relationship allows us to compute the spectral coefficients of the derivative $\{\hat{u}_j'\}$ based on the known spectral coefficients $\{\hat{u}_j\}$ \cite{Jshen2006}.
	
	\item
	Convolution
	
	The spectral coefficients $\{\hat{w}_j\}_{j=0}^N$ of $w(t) = u(t)v(t)$ and the spectral coefficients $\{\hat{u}_j\}_{j=0}^N$ of $u(t)$ have the following relationship:
	
	\begin{equation}
		\label{eq.44}
		\hat{w}_j \approx 
		\frac{1}{2} \sum_{m+i=j}^{N} \hat{u}_m\hat{v}_i +
		\frac{1}{2} \sum_{|m-i|=j}^{N} \hat{u}_m\hat{v}_i,\quad \bm{\hat{w}} \approx \bm{C}_v \bm{\hat{u}}.
	\end{equation}	
\end{itemize}

The above equations serve as a fundamental connection between operations carried out in physical space and Chebyshev spectral space. These relationships are particularly useful for determining the spectral coefficients of the solution to a differential equation in spectral space. By employing an inverse spectral transformation, we can effectively obtain an equivalent solution to the given differential equation \cite{Kopriva2009}.

Comprehensive mathematical analysis regarding the stability and convergence of spectral methods can be found in dedicated textbooks on the computational theory of spectral methods \cite{Jshen2006}. These books typically offer thorough explanations of the theoretical underpinnings, mathematical principles, and accompanying analytical proofs associated with spectral methods. Some of the topics covered include approximation errors, the impact of chosen basis functions, stability criteria, and convergence theorems.

\subsection{Spectrally discretized vertical modes}
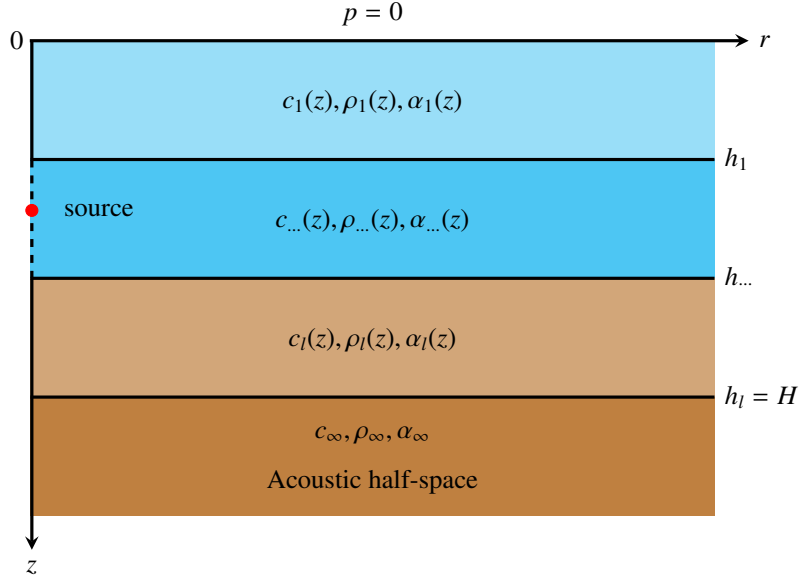
\begin{figure}
	\centering
			\begin{tikzpicture}[node distance=2cm,scale=0.9]
		\fill[cyan,opacity=0.4] (2,0)--(12,0)--(12,-1.75)--(2,-1.75)--cycle;
		\fill[cyan,opacity=0.7] (2,-1.75)--(12,-1.75)--(12,-3.5)--(2,-3.5)--cycle;
		\fill[brown,opacity=0.7] (2,-3.5)--(12,-3.5)--(12,-5.25)--(2,-5.25)--cycle;
		\fill[brown] (2,-5.25)--(12,-5.25)--(12,-7)--(2,-7)--cycle;		
		\node at (1.8,0){$0$};
		\draw[very thick, ->](1.99,0)--(12.5,0) node[right]{$r$};
		\draw[very thick, ->](2.02,-5.25)--(2.02,-7.5) node[below]{$z$};
		\draw[very thick](2.02,0)--(2.02,-1.75);
		\draw[dashed, very thick](2.02,-1.75)--(2.02,-3.5);
		\draw[very thick](2.02,-3.5)--(2.02,-5.25);		
		\draw[very thick](2,-1.75)--(12,-1.75) node[right]{$h_1$};
		\draw[very thick](2,-3.5)--(12,-3.5) node[right]{$h_{\cdots}$};
		\draw[very thick](2,-5.25)--(12,-5.25) node[right]{$h_l=H$};	
		\filldraw [red] (2.02,-2.5) circle [radius=2.5pt];
		\node at (3,-2.5){source};
		\node at (7,-0.9){$c_1(z),\rho_1(z),\alpha_1(z)$};
		\node at (7,-2.65){$c_{\dots}(z),\rho_{\dots}(z),\alpha_{\dots}(z)$};
		\node at (7,-4.4){$c_l(z),\rho_l(z),\alpha_l(z)$};
		\node at (7,-5.8){$c_\infty,\rho_\infty,\alpha_\infty$};
		\node at (7,-6.5){Acoustic half-space};
		\node at (7,0.4){$p=0$};
	\end{tikzpicture}	
	\caption{Schematic diagram of the waveguide with any number of layers.}
	\label{Figure4}
\end{figure}

The key to solving local modes lies in Eq.~\eqref{eq.4a}. Prior to performing spectral discretization, it is crucial to scale the problem domain to $t\in[-1,1]$ to align with the domain of $\{T_j(t)\}$.
\begin{equation}
	\rho(t)\frac{4}{\Delta z^2}\frac{\mathrm{d}}{\mathrm{d} t}\left[\frac{1}{\rho(t)} \frac{\mathrm{d} \Psi(t)}{\mathrm{d} t}\right]+\left[k^{2}(t)-\kappa^{2}\right]\Psi(t)=0,
\end{equation}
where $\Delta z$ represents the depth of the waveguide. Following the Chebyshev spectral discretization process, as outlined in Eqs.~\eqref{eq.43} and \eqref{eq.44}, Eq.~\eqref{eq.4a} transforms into the following matrix eigenvalue problem:
\begin{equation}
	\bm{A}\bm{\hat{\Psi}}=\kappa^{2}\bm{\hat{\Psi}},\quad \bm{A}=\left[\frac{4}{\Delta z^2}\bm{C}_{\rho}\bm{D}_N\bm{C}_{\left(\frac{1}{\rho}\right)}\bm{D}_N+\bm{C}_{k^{2}}\right] .
\end{equation}
When the waveguide consists of multiple layers with varying material properties, as depicted in Fig.~\ref{Figure4}, it is necessary for the modal equation mentioned above to hold within each individual layer.
However, it is difficult for a single set of basis functions to effectively capture the discontinuity of acoustic parameters across interfaces. To overcome this challenge, the domain decomposition strategy is employed, which involves utilizing separate sets of basis functions in each layer and performing independent Chebyshev spectral discretization within each layer \cite{Min2005}. The resulting discretized modal equations are then organized in the following manner:
\begin{equation}
	\label{eq.47}
	\left[\begin{array}{cccc}
		\bm{A}_1&\bm{0}&\bm{0}&\bm{0}\\
		\bm{0}&\bm{A}_2&\bm{0}&\bm{0}\\
		\bm{0}&\bm{0}&\ddots&\bm{0}\\
		\bm{0}&\bm{0}&\bm{0}&\bm{A}_l\\
	\end{array}\right]
	\left[\begin{array}{c}
		\bm{\hat{\Psi}}_1\\
		\bm{\hat{\Psi}}_2\\
		\vdots\\
		\bm{\hat{\Psi}}_l\\
	\end{array}
	\right]=\kappa^2\left[\begin{array}{c}
		\bm{\hat{\Psi}}_1\\
		\bm{\hat{\Psi}}_2\\
		\vdots\\
		\bm{\hat{\Psi}}_l\\
	\end{array}
	\right].
\end{equation}
Solving the aforementioned equation requires the application of boundary conditions and interface conditions, the continuities of sound pressure and normal particle velocity. A total of $2l$ boundary conditions is implemented. The method of imposing boundary conditions is a distinctive aspect of spectral methods. While the classical Galerkin method achieves this by preselecting basis functions that comply with the boundary conditions, this approach lacks flexibility. In this paper, the Tau-type spectral method is employed. The Chebyshev--Tau spectral method transforms the boundary conditions into spectral space and solves them together with the algebraic equations resulting from the spectral discretization \cite{Lanczos1938}. At this stage, the algebraic system becomes overdetermined. However, the Tau method preserves the well-posedness of the linear system by neglecting the last few algebraic equations derived from the spectral discretization. In particular, in Eq.~\eqref{eq.47}, by substituting the boundary conditions and interface conditions for the last two rows of each subblock, it becomes possible to accurately allocate all $2l$ conditions.

For ideal seafloors that are either perfectly soft or rigid, the above model is sufficient. However, for waveguides with an acoustic half-space, the bottom boundary must fulfill the condition:
\begin{equation}
	\label{eq.48}
	\Psi(H)+\frac{\rho_\infty}{\rho_l(H)\gamma_\infty} \Psi'(H)=0,\quad \gamma_{\infty}=\sqrt{\kappa^{2}-k_\infty^2},\quad k_\infty=(1+\mathrm{i}\eta\alpha_\infty)\omega/c_\infty.
\end{equation}
When applying the bottom boundary condition \eqref{eq.48}, the directly discretized modal equations encounter difficulties since they involve the unknown eigenvalue $\kappa$. As a result, the discretized eigenvalue system can only be solved using root-finding algorithms. To overcome this issue, an eigenvalue transformation technique developed by Sabatini and Cristini is employed \cite{Sabatini2019}. By substituting $k_{z,\infty}=\sqrt{k_\infty^{2}-\kappa^{2}}$, Eqs.~\eqref{eq.4} and \eqref{eq.48} can be transformed into:
\begin{subequations}
	\begin{gather}
		\rho(z) \frac{\mathrm{d}}{\mathrm{d} z}\left[\frac{1}{\rho(z)} \frac{\mathrm{d} \Psi}{\mathrm{d}z}\right]+\left[k^{2}(z)-k_{\infty}^{2}+k_{z,\infty}^{2}\right] \Psi=0, \\	
		\frac{\mathrm{i}\rho_\infty}{\rho_b(H)}\left.\frac{\mathrm{d} \Psi(z)}{\mathrm{d} z}\right|_{z=H}+k_{z,\infty} \Psi(H)=0.
	\end{gather}
\end{subequations}
Moreover, $\kappa$ is indirectly obtained by solving for the vertical wavenumber $k_{z,\infty}$ in the above equation. 

\subsection{Spectrally discretized MPEs}
The MPE is further simplified using a Pad\'e series approximation into Eqs.~\eqref{eq.20} and \eqref{eq.21}. In Eq.~\eqref{eq.21}, the most crucial part is the discretization of the $\mathcal{Y}$ operator. In the Chebyshev spectral method, the $\mathcal{Y}$ operator is spectrally discretized as follows:
\begin{equation}
	\label{eq.50}
	\bm{Y}=\frac{4}{\kappa_{\mathrm{s}}^2\Delta y^2}\bm{D}_N^2+\bm{C}_{\left(\frac{\kappa^2}{\kappa_{\mathrm{s}}^2}\right)}-\bm{I}.
\end{equation}
where $\bm{I}$ is the identity matrix. Therefore, Eq.~\eqref{eq.20} can be discretized in the Chebyshev spectral space as follows:
\begin{equation}
	\label{eq.51}
	\bm{\hat{\phi}}(x+\Delta x)=\left(d_0\bm{I}+\sum_{q=1}^p \frac{d_q\bm{I}}{\bm{I}+b_q \bm{Y}}\right) \bm{\hat{\phi}}(x).
\end{equation}
Eq.~\eqref{eq.21} becomes:
\begin{subequations}
	\label{eq.52}
	\begin{gather}
		(\bm{I}+b_{j}\bm{Y})\bm{W}_{j}=d_{j} \bm{\hat{\phi}}(x),\quad j=1,2,\ldots,n,\\
		\bm{\hat{\phi}}(x+\Delta x)=d_0\bm{\hat{\phi}}(x)+\sum_{j=1}^{n}\bm{W}_j.
	\end{gather}
\end{subequations}

Since the parameters in the PML are continuous with the acoustic parameters in the computational domain, discretizing the $\mathcal{Y}$ operator does not introduce additional numerical errors. Therefore, Eqs.~\eqref{eq.50} to \eqref{eq.52} are suitable for numerical solutions of the MPE in Fig.~\ref{Figure2}. However, it should be noted that the GCL nodes used in spectral discretization have the characteristic of being ``dense at both endpoints and sparse in the middle". This can cause the GCL nodes to concentrate heavily in the two PMLs, potentially leading to a loss of numerical accuracy in the computational domain. In such cases, a domain decomposition strategy can be considered \cite{Min2005}, where the PML and computational domain are separated into four layers. The computational domain is divided into two layers by a virtual interface set at the sound source to ensure that an adequate number of GCL nodes are arranged near the sound source. Each layer undergoes spectral discretization independently, and they are subsequently solved uniformly, similar to solving local modes in the previous subsection. This division into four layers has a noticeable effect. By allowing different numbers of spectral truncation orders in the PML and computational domain, nodes are distributed more effectively after layering. As a result, the variations in the acoustic profiles are captured more effectively.	

\begin{figure}
	\centering
	\includegraphics[width=0.75\textwidth]{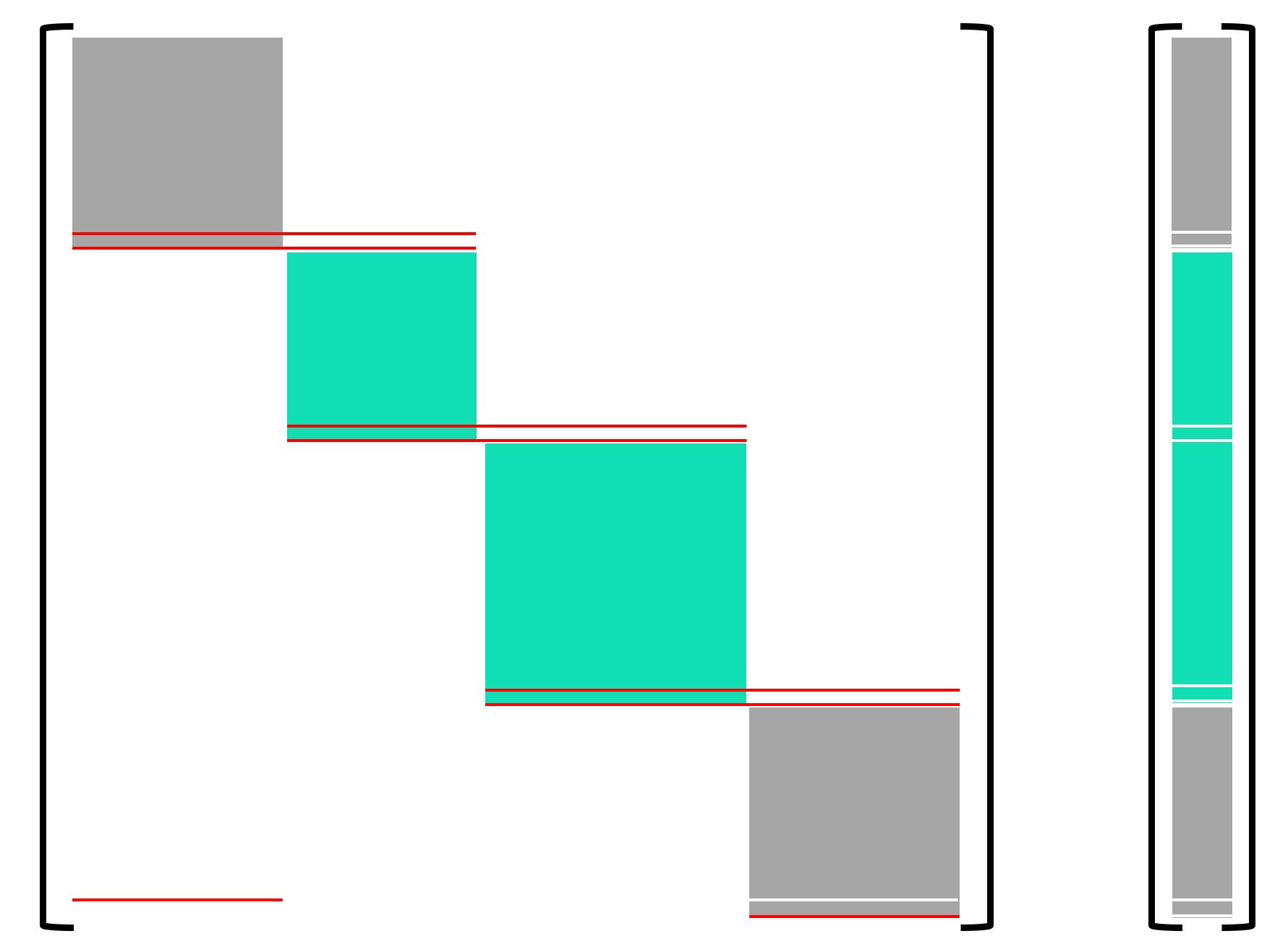}
	\caption{The shape features of the spectrally discretized global matrix and vector; white lines and areas represent 0 elements, while red lines represent rows that have been replaced by boundary conditions.}
	\label{Figure5}
\end{figure}

Based on the approach described in Eq.~\eqref{eq.50}, the spectral discretization of the $\mathcal{Y}$ operator in the four layers yields the following assembled new global matrix and vector:
\begin{equation}
	\label{eq.53}
	\bm{Y}=\left[\begin{array}{cccc}
		\bm{Y}_1&\bm{0}&\bm{0}&\bm{0}\\
		\bm{0}&\bm{Y}_2&\bm{0}&\bm{0}\\
		\bm{0}&\bm{0}&\bm{Y}_3&\bm{0}\\
		\bm{0}&\bm{0}&\bm{0}&\bm{Y}_4
	\end{array}\right],\quad
	\bm{\hat{\phi}}=\left[\begin{array}{c}
		\bm{\hat{\phi}}_1\\
		\bm{\hat{\phi}}_2\\
		\bm{\hat{\phi}}_3\\
		\bm{\hat{\phi}}_4\\
	\end{array}
	\right].
\end{equation}
After adopting a domain decomposition strategy, it is necessary to explicitly impose boundary conditions at the three interfaces, naturally considering the zeroth and first-order continuity of $\phi$:
\begin{subequations}
	\begin{gather}
		\phi_1(\epsilon)=\phi_2(\epsilon),\quad \left.\frac{\partial \phi_1}{\partial y} \right|_{y=\epsilon}= \left.\frac{\partial \phi_2}{\partial y}\right|_{y=\epsilon},\\
		\phi_2(y_\mathrm{s}+\epsilon)=\phi_3(y_\mathrm{s}+\epsilon),\quad \left.\frac{\partial \phi_2}{\partial y}\right|_{y=y_\mathrm{s}+\epsilon} = \left.\frac{\partial \phi_3}{\partial y}\right|_{y=y_\mathrm{s}+\epsilon},\\
		\phi_3(y_{\max}+\epsilon)=\phi_4(y_{\max}+\epsilon),\quad \left.\frac{\partial \phi_3}{\partial y}\right|_{y=y_{\max}+\epsilon} = \left.\frac{\partial \phi_4}{\partial y}\right|_{y=y_{\max}+\epsilon},		
	\end{gather}
\end{subequations}
where the subscript of $\phi$ represents its layer number. Following the approach of the Tau method, we apply Chebyshev spectral discretization to the six conditions at the three interfaces and perform spectral discretization of the pressure-release conditions on the outer sides of the two PMLs ($y=0$ and $y=y_{\max}+2\epsilon$):
\begin{subequations}
	\begin{gather}			\bm{s}_1\bm{\hat{\phi}}_1-\bm{q}_2\bm{\hat{\phi}}_2=0,\\
		\frac{1}{\epsilon}\bm{s}_1\bm{D}_{N_1}\bm{\hat{\phi}}_1-\frac{1}{y_{\mathrm{s}}}\bm{q}_2\bm{D}_{N_2}\bm{\hat{\phi}}_2=0,\\
		\bm{s}_2\bm{\hat{\phi}}_2-\bm{q}_3\bm{\hat{\phi}}_3=0,\\
		\frac{1}{y_{\mathrm{s}}}\bm{s}_2\bm{D}_{N_2}\bm{\hat{\phi}}_2-\frac{1}{y_{\max}-y_{\mathrm{s}}}\bm{q}_3\bm{D}_{N_3}\bm{\hat{\phi}}_3=0,\\		
		\bm{s}_3\bm{\hat{\phi}}_3-\bm{q}_4\bm{\hat{\phi}}_4=0,\\
		\frac{1}{y_{\max}-y_{\mathrm{s}}}\bm{s}_3\bm{D}_{N_3}\bm{\hat{\phi}}_3-\frac{1}{\epsilon}\bm{q}_4\bm{D}_{N_1}\bm{\hat{\phi}}_4=0,\\		
		\bm{q}_1\bm{\hat{\phi}}_1=0,\\
		\bm{s}_4\bm{\hat{\phi}}_4=0,
	\end{gather} 
\end{subequations}
where 
\[
\bm{s}_\ell=[s_0, s_1, s_2, \ldots, s_{N_\ell}], \quad s_i=T_i(-1)=(-1)^i,
\]
\[
\bm{q}_\ell=[q_0, q_1, q_2, \ldots, q_{N_\ell}],\quad q_i = T_i(+1) = 1.
\]
By applying the eight conditions mentioned above to the four layers, with two conditions allocated for each layer, the global matrix and vector in Eq.~\eqref{eq.53} will have the distribution characteristics depicted in Fig.~\ref{Figure5}.	

Therefore, following the approach described in Eqs.~\eqref{eq.51} and \eqref{eq.52}, we can solve for $\bm{\hat{\phi}}(x+\Delta x)$. When stepping forward, if $\kappa(x)$ changes, we discretize the $\mathcal{Y}$ operator again and step forward using the updated $\bm{Y}$ matrix. Finally, the overall solution to the MPE can be obtained through an inverse Chebyshev transformation [Eq.~\eqref{eq.40}].

\section{Numerical implementation}
\label{sec4}
\subsection{Numerical algorithm}
After introducing the theory of the three-dimensional AMPE model and the spectral discretization approach, we can summarize the overall spectral algorithm as follows:
\begin{enumerate}
	\item
	Set up the original parameters of the three-dimensional waveguide.
	
	Specifically, it involves (1) terrain parameters and acoustic parameter profiles; (2) source frequency and location; (3) spectral truncation orders for discretizing the eigen-equation and MPEs; (4) type of starter, number of terms in the Pad\'e series $n$, and step forward $\Delta x$; (5) thickness of PMLs $\epsilon$, absorption coefficient $\sigma_0$ in the PMLs, and spectral truncation orders for PMLs; and (6) spatial resolution of the output sound field, $\Delta y$, and $\Delta z$.
	
	\item
	Calculate several important intermediate parameters for the solving process.
	
	Specifically, it involves: (1) the number of retained modes $M$; (2) the positions of the GCL nodes in the four layers along the $y$-axis (see Fig.~\ref{Figure2}); and (3) the coefficients $\{b_j\}_{j=1}^n$ and $\{d_j\}_{j=1}^n$ in Eq.~\eqref{eq.21} of the Pad\'e approximation.
	
	\item
	To solve the local eigenpairs $(\kappa_m,\Psi_m)$ corresponding to each GCL point on the $xoy$ plane. 
	
	Moreover, $\kappa_m(x,y)$ and $\Psi_m(x,y)$ can be calculated on the $xoy$ plane using equidistant grids. Then, interpolation is used to obtain $\kappa_m(x,y)$ and $\Psi_m(x,y)$ at the GCL nodes. However, the directly calculated $\kappa_m(x,y)$ and $\Psi_m(x,y)$ at the GCL nodes will certainly be more accurate than interpolation.
	
	\item
	For each mode $m$: Calculate the starter of MPE and perform spectral transformation on it for step forward, $\bm{\hat{\phi}}(x_0)$.
	
	\item
	Perform Chebyshev spectral discretization of the $\mathcal{Y}$ operators and boundary conditions. Assemble the global matrix and vector, as depicted in Fig.~\ref{Figure5}.
	
	\item
	Solve the resulting linear system obtained in Eq.~\eqref{eq.52}, then step forward to the next distance, repeatedly obtaining the spectral coefficients of the horizontal refractive index $\bm{\hat{\phi}}(x)$ on the entire $xoy$ plane.
	
	\item
	By performing the inverse Chebyshev transformation [Eq.~\eqref{eq.40}] on $\bm{\hat{\phi}}(x)$ and substituting the result into Eq.~\eqref{eq.13}, we can obtain the horizontal refractive index $\Phi(x,y)$ on the $xoy$ plane.
	
	\item
	Then, by utilizing Eq.~\eqref{eq.3}, we can effectively synthesize the three-dimensional sound field throughout the entire space.
	
\end{enumerate}

Overall, the spectral algorithm provides a reliable approach for simulating three-dimensional acoustic waveguides. Considering the efficiency issue, which is of great concern in numerical simulations of three-dimensional sound propagation, we will now analyze the parallelism of the spectral algorithm to discover its potential for acceleration on increasingly common high-performance computers.

\subsection{Computational Parallelism}	
\begin{figure}
	\centering
	\includegraphics[width=\textwidth]{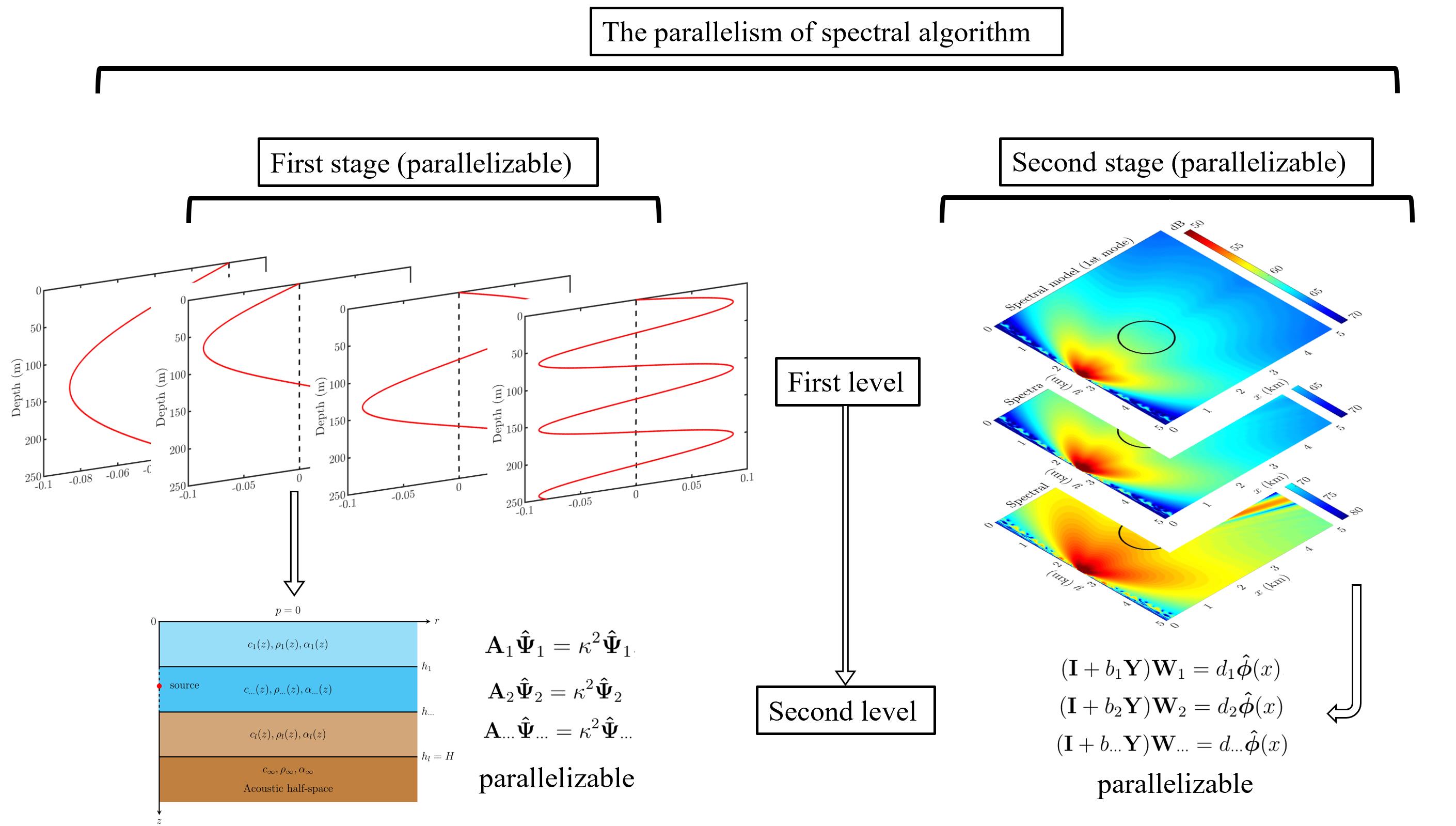}
	\caption{The schematic diagram illustrates the parallelism of the spectral algorithm.}
	\label{Figure6}
\end{figure}

The computation of the aforementioned spectral algorithm can be primarily divided into two stages. The first stage involves solving local modes, which essentially means solving the modal equations discretized by the Chebyshev spectral method. The second stage involves solving the MPEs discretized by the spectral method. The local eigen-equations need to be solved first to obtain the local modes and equivalent wavenumber $\kappa(x,y)$, which are then used as inputs to solve the MPEs. The computational workload of the spectral model is mainly concentrated in these two stages, as illustrated in Fig.~\ref{Figure6}.

In the case of undulating seafloor topography and complex and variable acoustic parameters, the maximum number of modal equations to be solved in the first stage is $nx\times ny$ ($nx$ is related to the sum of spectral truncation orders). However, if the environmental parameters in the ocean change gradually, we can also interpolate the eigenpairs of adjacent points to obtain modal data with higher resolution. This approach can reduce the computational workload and improve the efficiency. Since there are no data dependencies among the modes and no explicit temporal ordering requirements in the calculation of local modes during the third step, the computation of local modes for different points can be naturally parallelized. This means that local modes for multiple points can be calculated simultaneously, thereby improving computational efficiency. We refer to this as the first level of parallelism. When solving for local modes at a specific point, the layers of the medium can be naturally parallelized due to the lack of data dependencies during spectral discretization. However, this parallelism disappears once the spectral discretization of each layer is completed, and they are assembled into the algebraic system in Eq.~\eqref{eq.47}. Nevertheless, parallelism can once again be employed during the inverse Chebyshev transformation after obtaining the spectral coefficients of the submodes for each layer. We refer to this parallelism between the layers of the medium during the calculation of local modes as the second level of parallelism.

In the second stage, each of the $M$ modes corresponds to an HRE, so there is natural parallelism at the first level in Steps four to seven of the spectral algorithm. In the sixth step of the algorithm, a parabolic model using the Pad\'e approximation is used, which requires solving $n$ linear systems of equations at each forward step. Since the summation form of the Pad\'e series is utilized in this paper, the solution of these $n$ linear systems of equations can be parallelized.

\section{Numerical experiments}
\label{sec5}	
We implemented the above spectral algorithm as a numerical model. Next, we will test the developed three-dimensional spectral model through three numerical experiments, which represent analytical examples, quasi-three-dimensional waveguides, and fully three-dimensional waveguides. These studies are of great importance for understanding and predicting the propagation behavior of sound waves in real marine environments. In the following examples, we set the PML thickness $\epsilon$ of the MPE to 500 m, the absorption coefficient $\sigma_0$ to 5, and the number of terms in the Pad\'e approximation to 6.

\subsection{Analytical example: ideal fluid waveguide}
The ideal fluid waveguide is the simplest form of a three-dimensional waveguide, consisting of a homogeneous layer of water with boundaries at the sea surface and seabed. In this particular example, both the sea surface and seabed are assumed to be perfectly free boundaries. The depth of the seawater is $H=100$ m, and the sound speed and density are set to 1500 m/s and 1 g/cm$^3$, respectively. The source frequency is 20 Hz, and it is located at coordinates (0, 1500, 36) m. This waveguide can only excite two modes in the vertical direction, i.e., $M=2$.

\begin{figure}
	\centering
	\subfigure[]{\label{Figure7a}\includegraphics[width=0.49\textwidth]{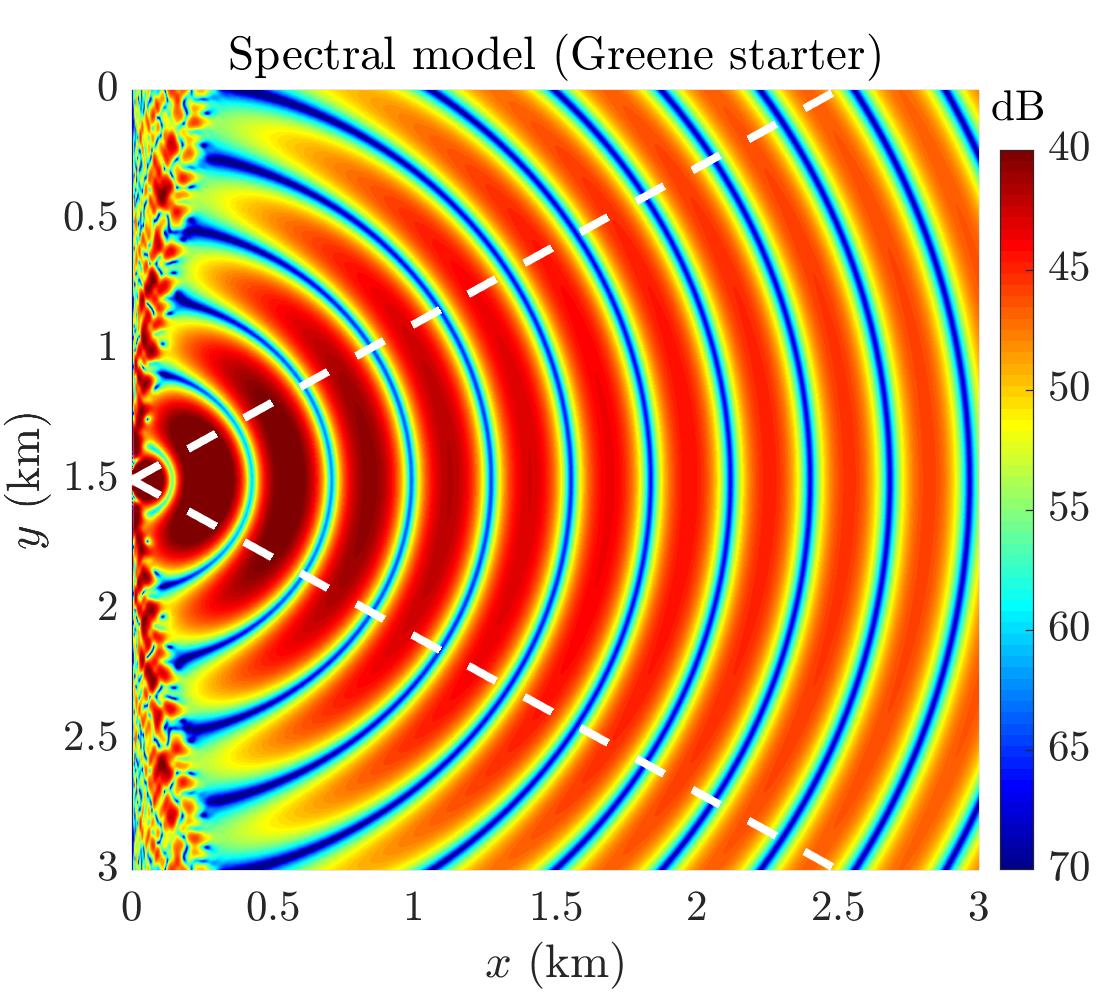}}
	\subfigure[]{\label{Figure7b}\includegraphics[width=0.49\textwidth]{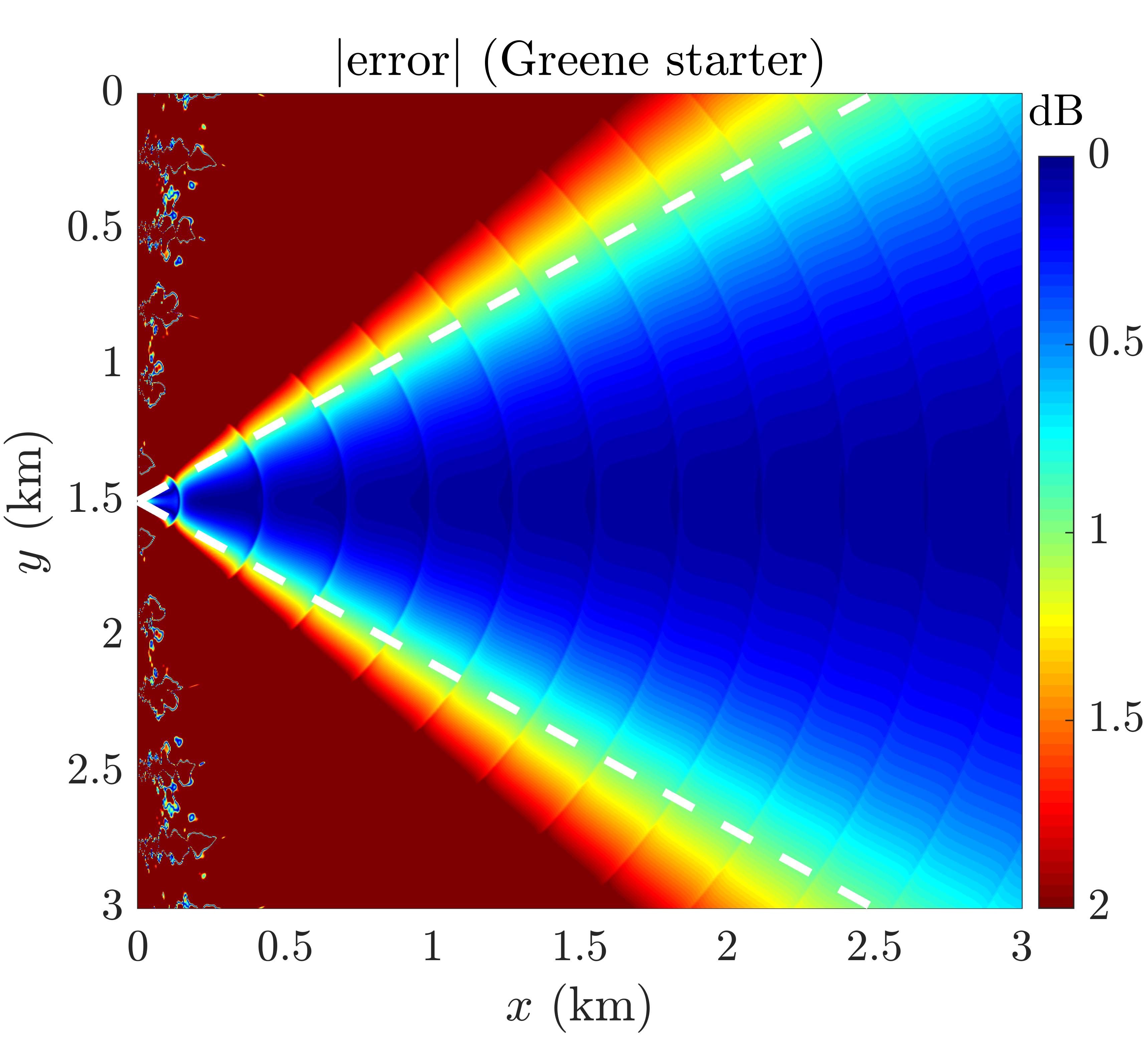}}\\
	\subfigure[]{\label{Figure7c}\includegraphics[width=0.49\textwidth]{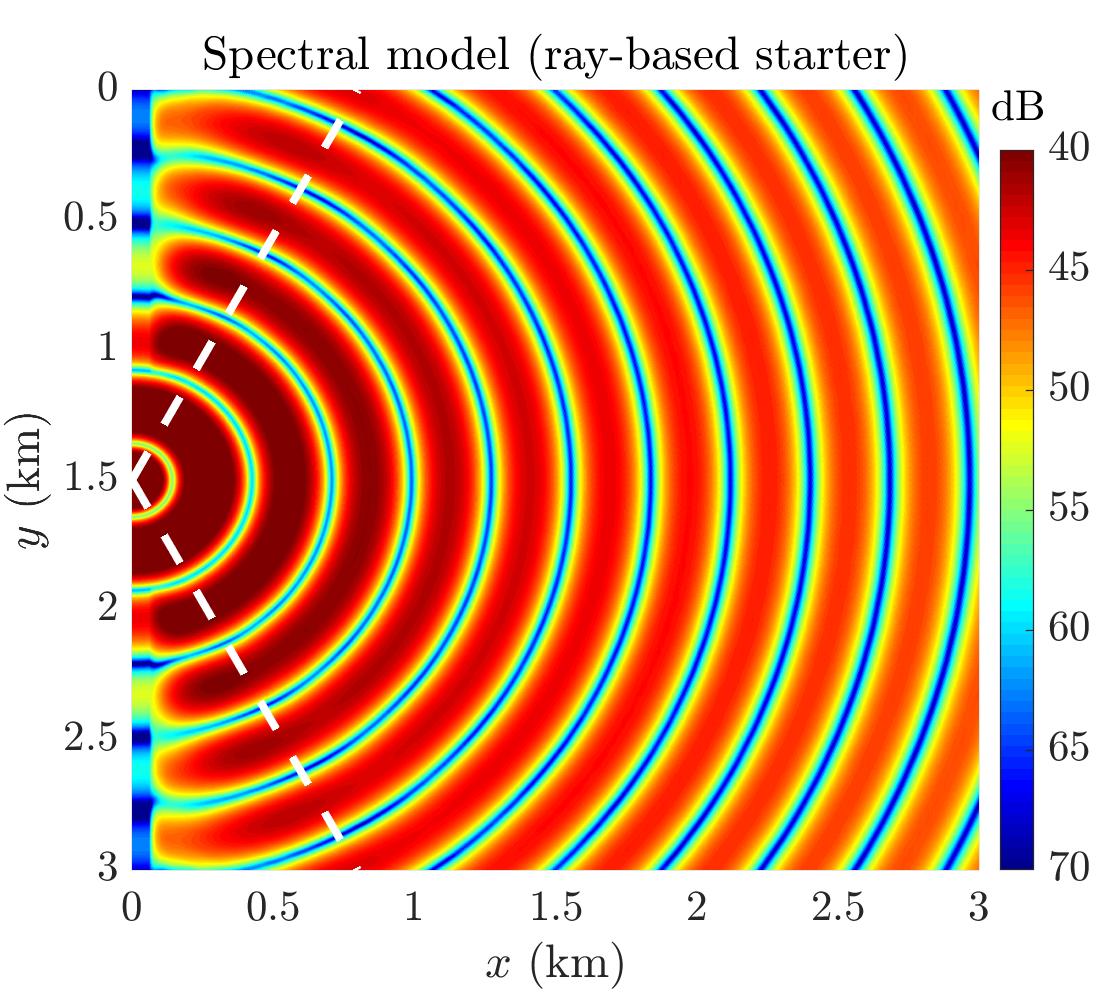}}
	\subfigure[]{\label{Figure7d}\includegraphics[width=0.49\textwidth]{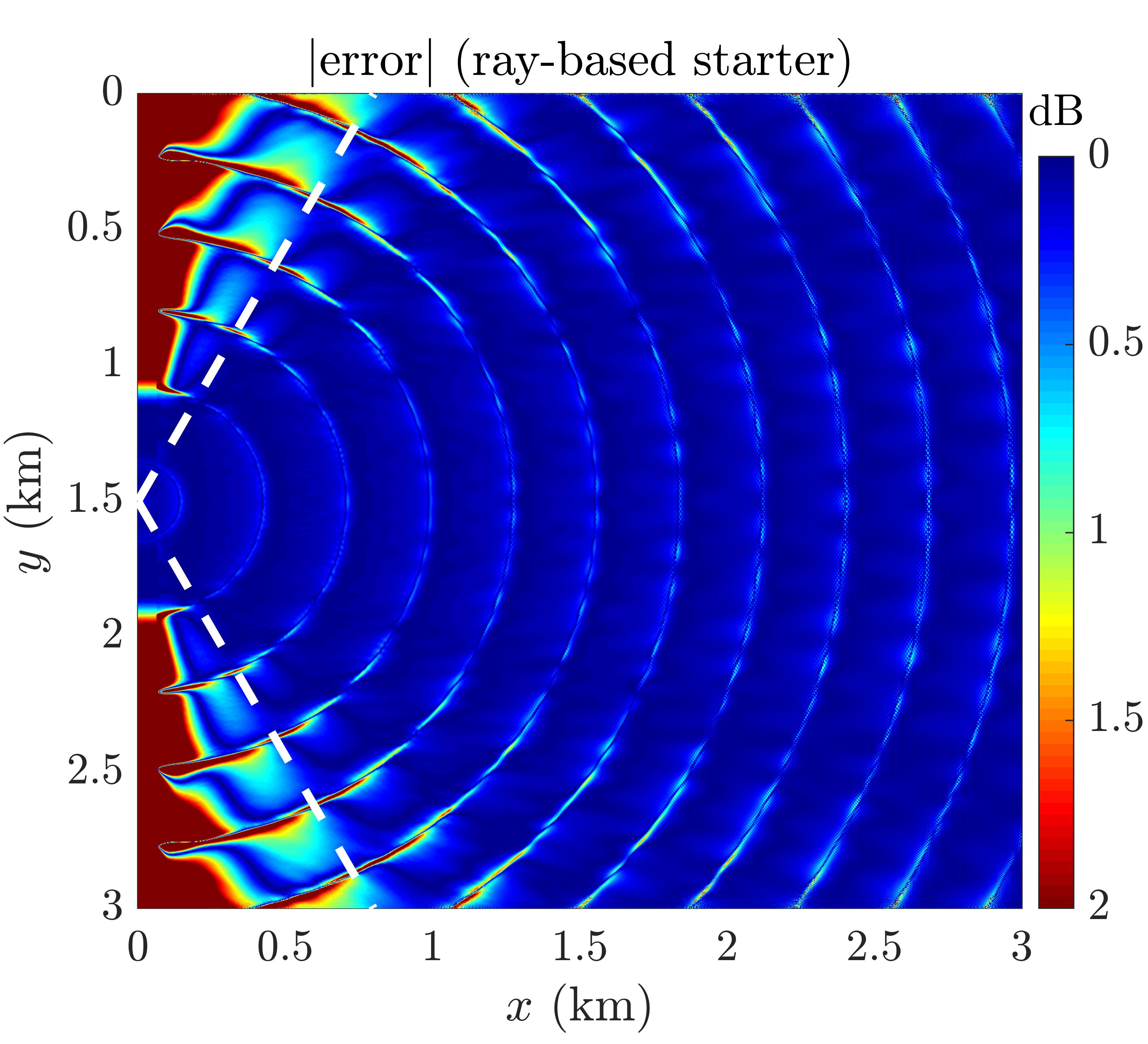}}\\
	\subfigure[]{\label{Figure7e}\includegraphics[width=0.49\textwidth]{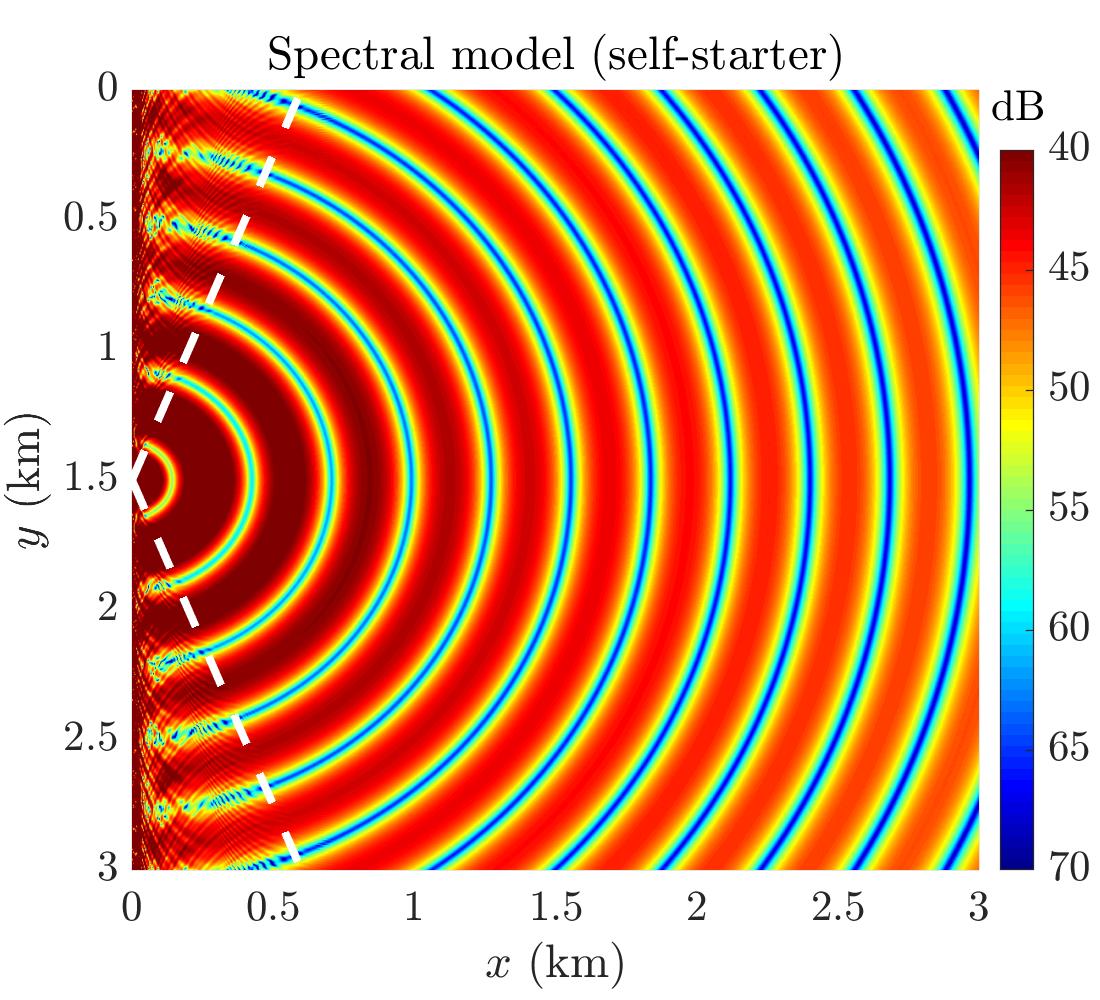}}
	\subfigure[]{\label{Figure7f}\includegraphics[width=0.49\textwidth]{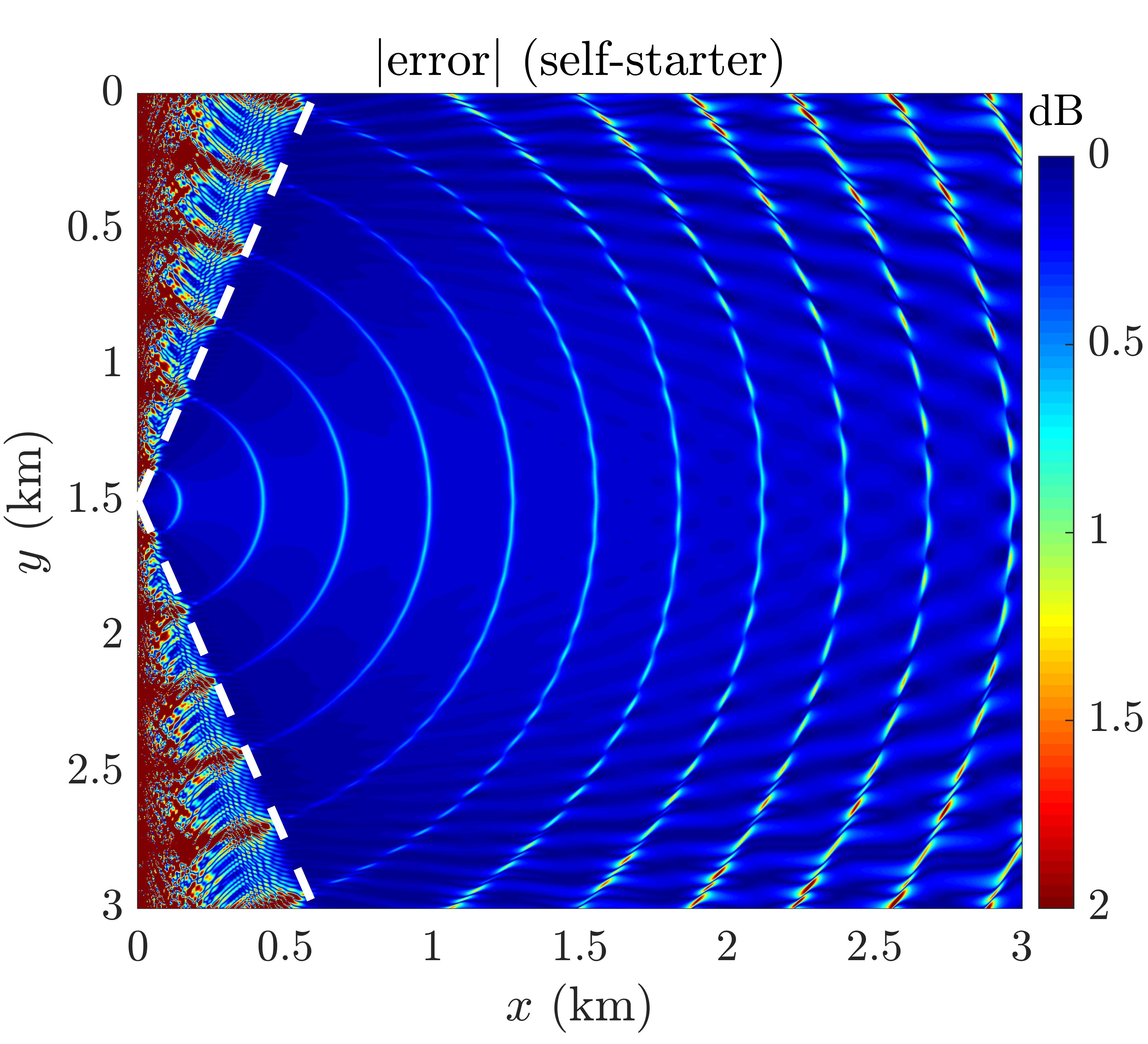}}
	\caption{TL field [(a), (c), (e)] and absolute error [(b), (d), (f)] slices of the three-dimensional ideal fluid waveguide at the $z_\mathrm{r}=36$ m plane calculated using the Greene starter (a)--(b), ray-based starter (c)--(d) and self-starter (e)--(f). The white dashed lines represent the range of angles within which the average error of the TL is less than 1 dB.}
	\label{Figure7}
\end{figure}

The three-dimensional ideal fluid waveguide has an exact solution in the following analytical form, which can be conveniently used for comparison in our model.
\begin{subequations}
	\begin{gather}
		p(x,y,z)=\frac{\mathrm{i}}{2H}\sum_{m=1}^{M}\sin(k_{z,m}z_\mathrm{s})\sin(k_{z,m}z)\mathcal{H}_0^{(1)}\left(\kappa_m\sqrt{x^2+(y-y_\mathrm{s})^2}\right),\\
		k_{z,m}=\frac{m\pi}{H},\quad \kappa_{m}=\sqrt{k^2-k_{z,m}^2}.
	\end{gather}		
\end{subequations}
To compare the wide-angle performance of the starter, we fixed the number of truncation terms in the Pad\'e series at 6 and set the spectral truncation order of the PMLs to 100. The spectral truncation order for both layers of the computational domain was set to 250. In Fig.~\ref{Figure7a}, \ref{Figure7c} and \ref{Figure7e}, we present the horizontal transmission loss (TL) slices computed under three different starters. From the sound field slices at a depth of $z_\mathrm{r}=36$ m, it is evident that all three starters have achieved the right computational results. Fig.~\ref{Figure7b}, \ref{Figure7d} and \ref{Figure7f} illustrate the absolute error of the three starters for the spectral model. We defined reliable results as those where the absolute error of TL is less than 1 dB along the polar axis with the source as the pole. The white dashed line indicates the critical angle for reliable results. In terms of the beam angle, the self-starter has the best wide-angle performance (approximately 70$^\circ$), followed by the ray-based starter (approximately 60$^\circ$), and the least favorable performance is exhibited by the Greene starter (approximately 30$^\circ$). From the theoretical perspective of starters, the self-starter can further improve its wide-angle performance by increasing the number of terms in the Pad\'e approximation, but this will result in a linear increase in computational complexity. The ray-based starter can enhance its wide-angle capability by increasing the take-off angle without incurring significant additional computational cost. The wide-angle capability of the Greene starter is difficult to improve, but it can still be a good choice in situations where near-field requirements are not stringent.

Next, we discuss the computational costs of different starters. Table \ref{tab1} provides the runtimes for a three-dimensional ideal fluid waveguide with different number of truncation orders. Here, $N$ represents the sum of the truncation orders set in the two layers of the computational domain. It is evident from the table that, at the same spectral truncation order, the simulation times for ideal fluid waveguides with Greene starter and ray-based starter configurations are comparable, and both are significantly faster than the self-starter. Since the ray-based starter has better wide-angle capability and is not slower than the Greene starter, using the ray-based starter in practical simulations offers better cost-effectiveness.
\begin{table}[htbp]
	\centering
	\caption{\label{tab1} Different spectral truncation orders result in varying simulation runtimes for the three starters in ideal fluid waveguide. (unit: seconds).}
	\begin{tabular}{crrr}
		\hline
		Trucation order & Greene starter& Ray-based starter& Self-starter \\
		\hline
		$N$=200 & 51.670  &50.762  &123.774\\
		$N$=300 & 74.644  &73.809  &177.687\\
		$N$=400 & 109.057 &110.920 &241.765\\
		$N$=500 & 146.288 &147.033 &342.657\\				
		\hline
	\end{tabular}
\end{table}

\begin{figure}
	\centering
	\subfigure[]{\includegraphics[width=0.49\textwidth]{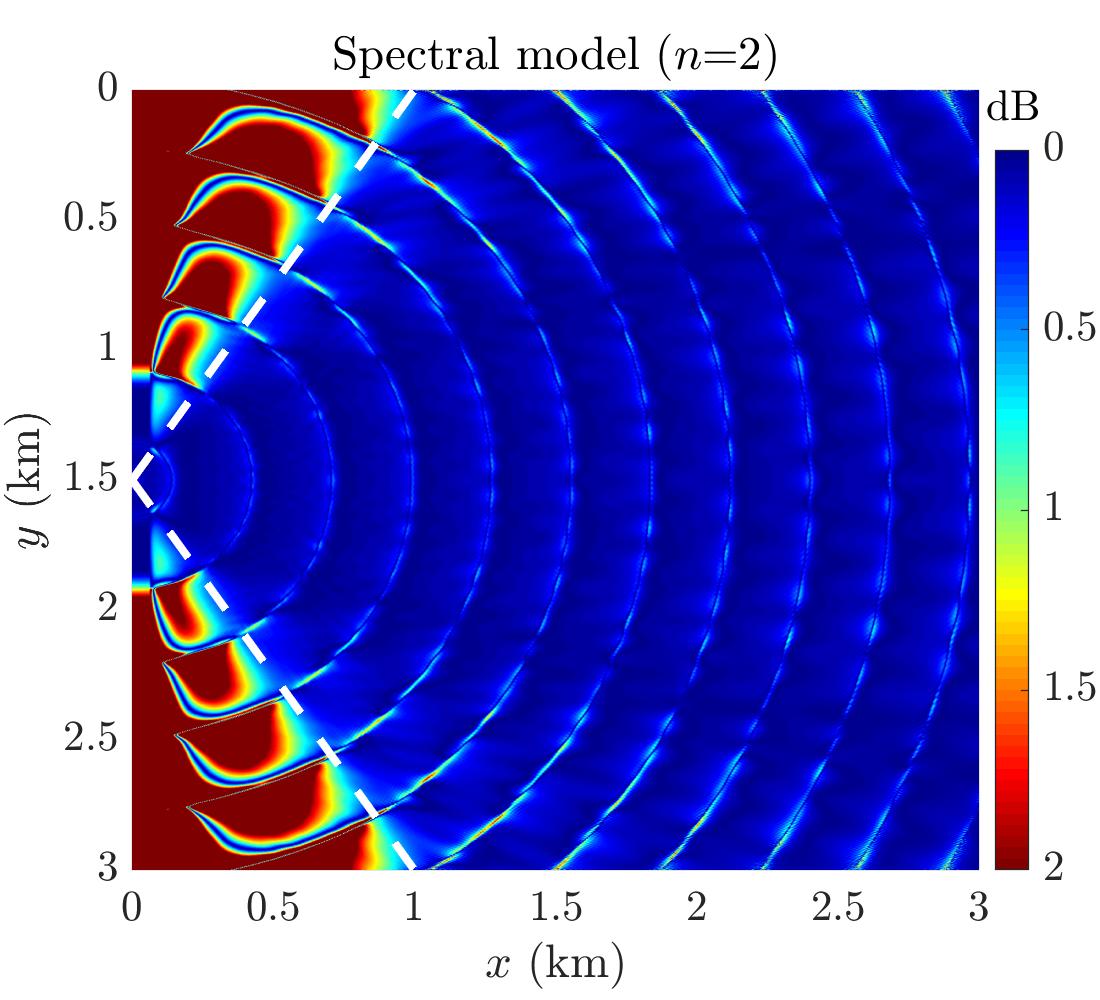}}
	\subfigure[]{\includegraphics[width=0.49\textwidth]{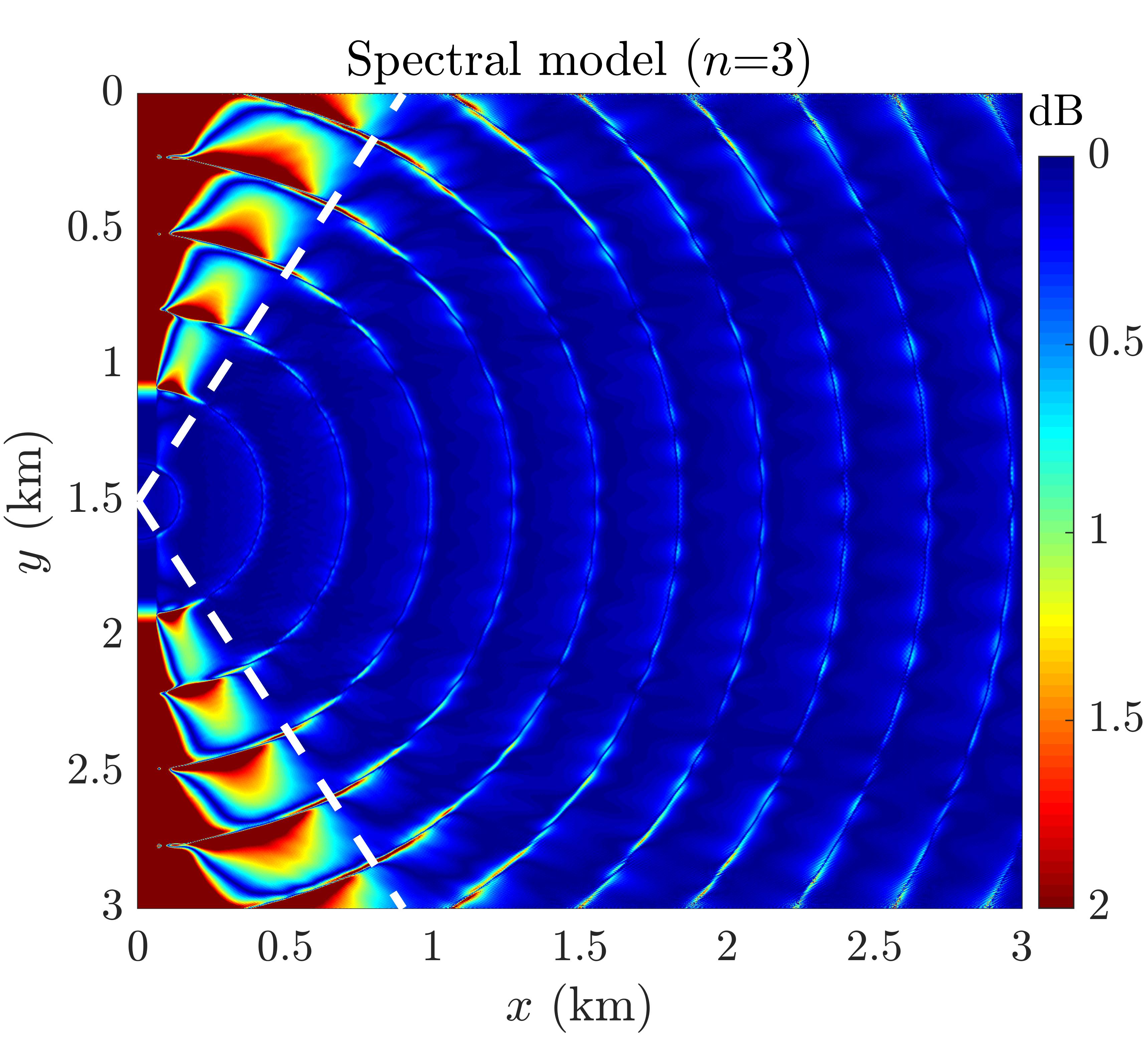}}\\
	\subfigure[]{\includegraphics[width=0.49\textwidth]{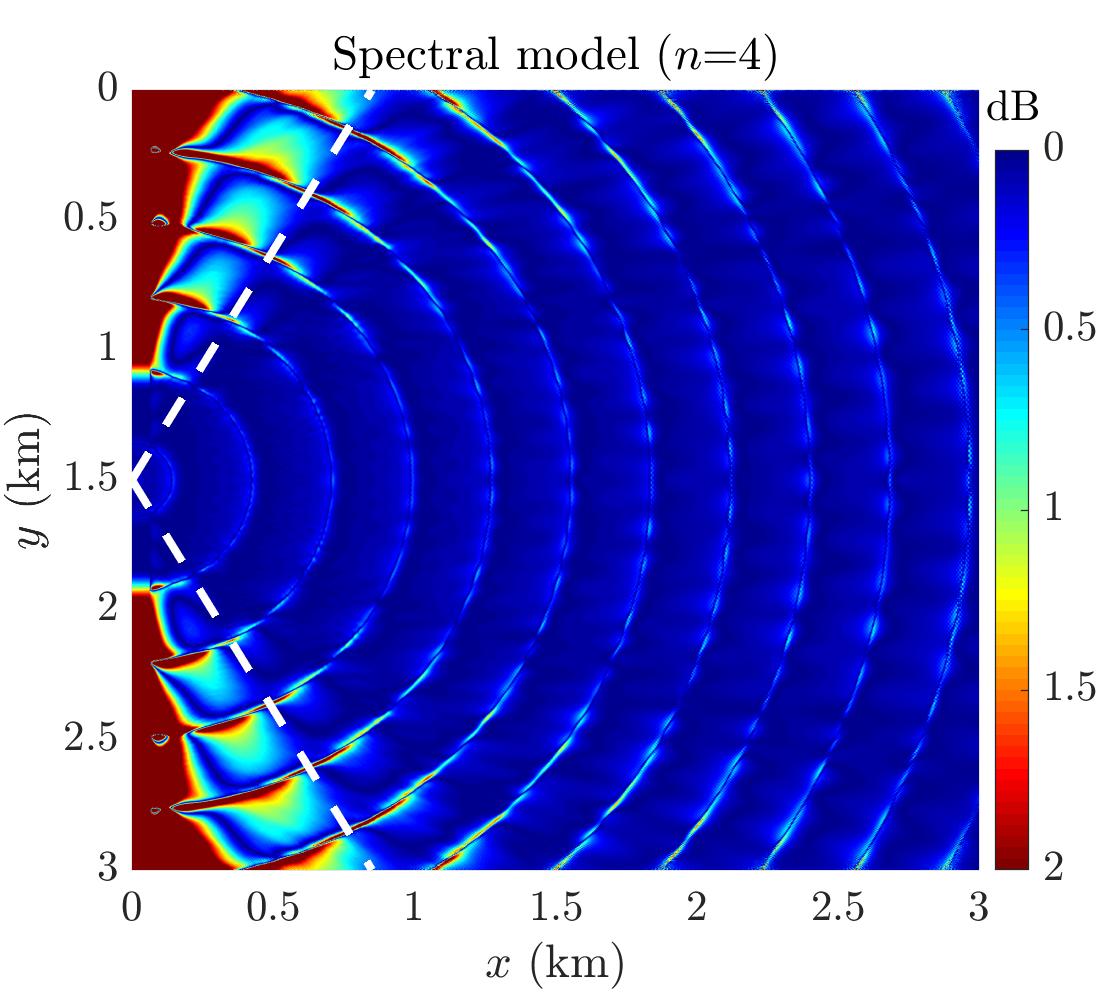}}
	\subfigure[]{\includegraphics[width=0.49\textwidth]{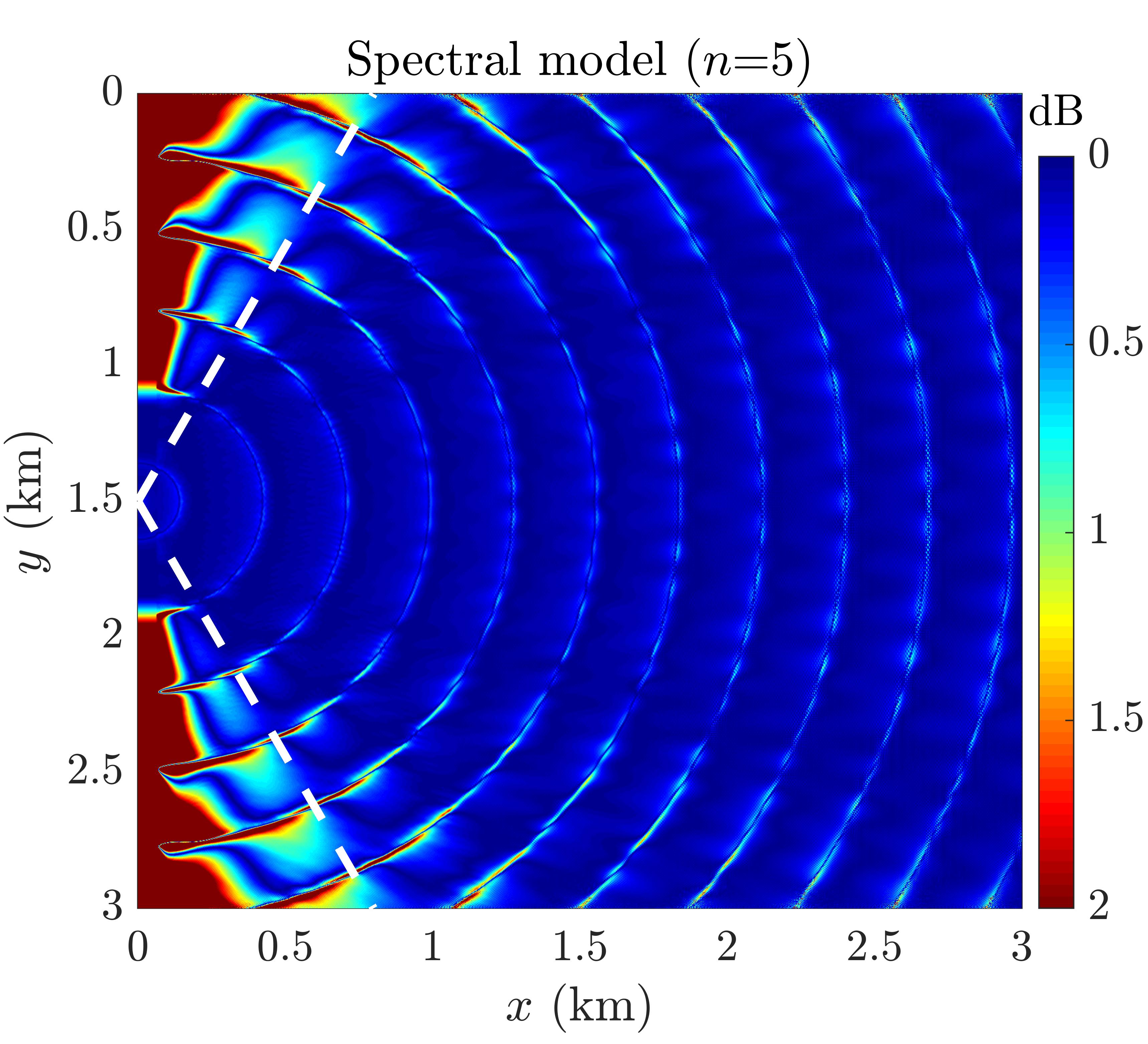}}
	\caption{The TL field slices of the three-dimensional ideal fluid waveguide on the $z_\mathrm{r} = 36$ m plane, computed using the spectral model in this paper, under different orders of the Pad\'e approximation.}
	\label{Figure8}
\end{figure}

Apart from the starter, the accuracy of the Pad\'e approximation is also a key factor that affects the success of the simulation. Next, we will study the influence of different orders of Pad\'e approximation on the accuracy of the solution based on the spectral model using a ray-based starter. To ensure that the grazing angle of the ray-based starter is sufficiently large (so that the order of the Pad\'e approximation becomes the only factor affecting the propagation angles), we set the take-off angle $\alpha$ to 80°. Fig.~\ref{Figure8} depicts the slices of the absolute error of the acoustic field at a depth of $z_\mathrm{r} = 36$ m, computed using the spectral model in this paper under Pad\'e approximations of orders 2 to 5. It can be compared together with Fig.~\ref{Figure7d}, which has a Pad\'e approximation order of 6. Clearly, as the number of terms in the Pad\'e series increases, both the near-field accuracy and the trustworthy opening angle of the acoustic field gradually increase. Even at $n=2$, reliable solutions can be obtained in the far field. A significant improvement in the opening angle capability can be observed at approximately $n=5$, while there is limited improvement in the opening angle capability at $n=6$. This indicates that beyond a Pad\'e series order of 5, the influence of the number of terms on the precision of the solution becomes negligible.

\subsection{Shallow sea with an underwater ridge}
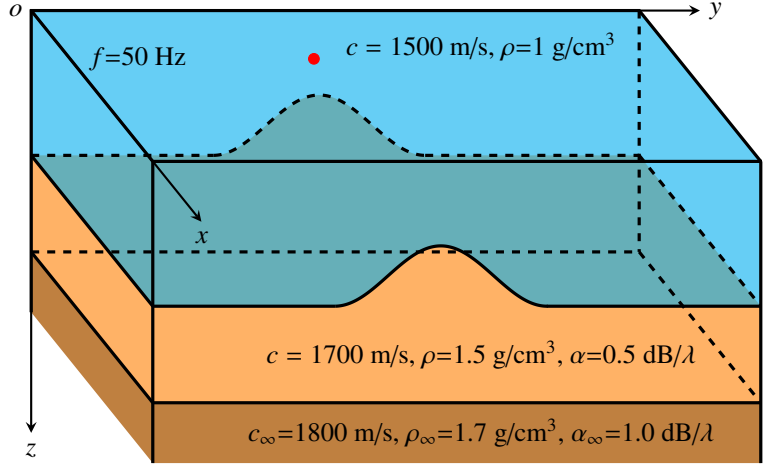
\begin{figure}
	\centering
			\begin{tikzpicture}[node distance=2cm,samples=500,domain=7:10.5,scale=0.8]
		
		\fill[orange,opacity=0.6](2,-2.4)--plot[domain=7:10.5,smooth](\x-2,{-1.9-0.5*cos(103*(\x))})--(12,-2.4)--(14,-4.9)--plot[domain=10.5:7,smooth](\x,{-4.4-0.5*cos(103*(\x))})--(4,-4.9)--cycle;
		\fill[cyan,opacity=0.6] (2,0)--(12,0)--(14,-2.5)--(4,-2.5)--(4,-4.9)--(2,-2.4)--cycle;
		\fill[cyan,opacity=0.6]  (4,-2.5)--(4,-4.9)--(7,-4.9)--plot(\x,{-4.4-0.5*cos(103*(\x))})--(10.5,-4.9)--(14,-4.9)--(14,-2.5)-- cycle;
		\fill[orange,opacity=0.6] (2,-2.4)--(4,-4.9)--(4,-6.5)--(2,-4)--cycle;
		\fill[orange,opacity=0.6]  (4,-4.9)--(7,-4.9)--plot(\x,{-4.4-0.5*cos(103*(\x))})--(10.5,-4.9)--(14,-4.9)--(14,-6.5)--(4,-6.5)--cycle;
		\fill[brown] (2,-4)--(4,-6.5)--(14,-6.5)--(14,-7.5)--(4,-7.5)--(2,-5)--cycle;
		\draw[thick, ->](3,0)--(13,0) node[right]{$y$};
		\draw[thick, ->](2,0)--(2,-7) node[below]{$z$};	    		
		\draw[very thick](1.98,0)--(12.02,0);
		\draw[very thick](4,-2.5)--(14.02,-2.5);
		\draw[very thick](2,0)--(4,-2.5);
		\draw[thick, ->](2,0)--(4.8,-3.5) node[below]{$x$};	
		\draw[very thick](12,0)--(14,-2.5);
		\draw[very thick](2,0.02)--(2,-5);
		\draw[very thick](4,-2.5)--(4,-7.5);
		\draw[dashed, very thick](12,0)--(12,-4);
		\draw[very thick](14,-2.5)--(14,-7.5);
		\draw[dashed, very thick](12,-2.4)--(14,-4.9);
		\draw[dashed, very thick](2,-4)--(12,-4);
		\draw[dashed, very thick](12,-4)--(14,-6.5);
		
		\draw[very thick](2,-4)--(4,-6.5);	
		\draw[very thick](4,-6.5)--(14,-6.5);			
		\filldraw [red] (6.65,-0.8) circle [radius=2.5pt];
		\node at (3.75,-0.8){$f$=50 Hz};
		\node at (1.75,0){$o$};
		
		\draw[very thick](2,-2.4)--(4,-4.9);
		\draw[very thick](4,-4.9)--(7,-4.9);
		\draw[very thick, smooth] plot(\x,{-4.4-0.5*cos(103*(\x))});
		\draw[very thick](10.48,-4.9)--(14,-4.9);
		\draw[dashed, very thick](2,-2.4)--(5,-2.4);
		\draw[dashed,very thick, smooth] plot(\x-2,{-1.9-0.5*cos(103*(\x))});
		\draw[dashed, very thick](8.48,-2.4)--(12,-2.4);
		
		\node at (9.4,-7){$c_{\infty}$=1800 m/s, $\rho_{\infty}$=1.7 g/cm$^3$, $\alpha_{\infty}$=1.0 dB/$\lambda$};					
		\node at (9.4,-5.7){$c=1700$ m/s, $\rho$=1.5 g/cm$^3$, $\alpha$=0.5 dB/$\lambda$};
		\node at (9.4,-0.6){$c=1500$ m/s, $\rho$=1 g/cm$^3$};
		
	\end{tikzpicture}
	\caption{Schematic diagram of the three-dimensional marine environment with an underwater ridge.}
	\label{Figure9}
\end{figure}

A shallow sea with an underwater ridge is one of the commonly used numerical examples in quasi-three-dimensional sound propagation models \cite{Tuhw2022b}. It can be viewed as a two-dimensional structure in the $y$-direction, with no variation in the $x$-direction. Numerical simulations of the underwater ridge waveguide can be used to investigate phenomena such as scattering, reflection, and refraction of sound waves, as well as the influence of seafloor topography on sound propagation paths, transmission losses, and sound energy distribution. In this subsection, we consider an example where there is an underwater ridge. The waveguide environment parameters and source features are shown in Fig.~\ref{Figure9}. The ridge terrain is independent of $x$ and follows the analytical equation in the $y$ direction: 
\begin{equation}
	h(y) = 100 -10\sec^2(\sigma y),\quad \sigma=0.004 \mathrm{m}^{-1}.
\end{equation}
In the simulation process, we consider the first $M=12$ modes excited by the sound source at a sea depth of $H=200$ m.	

\begin{figure}
	\centering
	\includegraphics[width=0.75\textwidth]{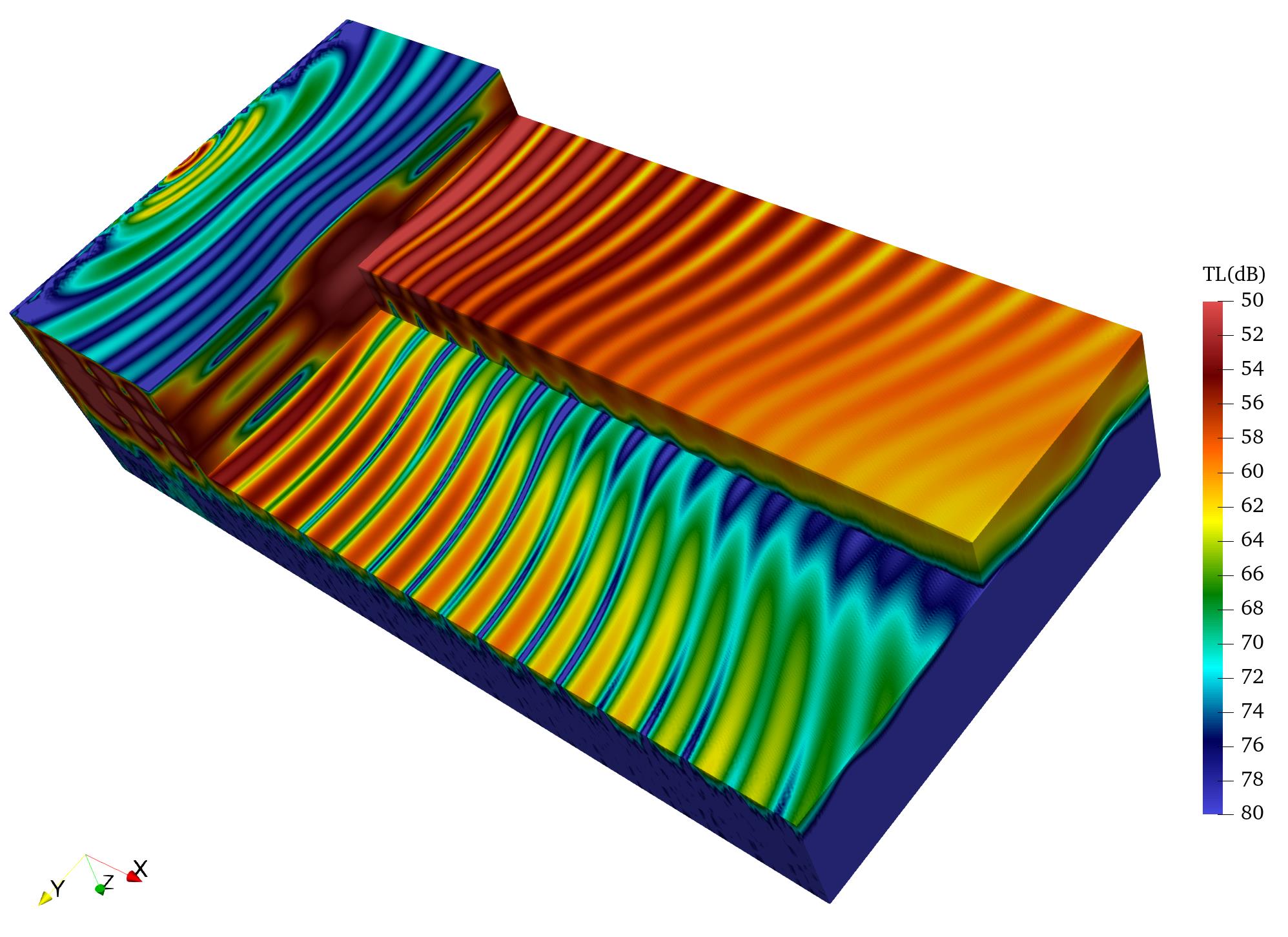}
	\caption{Sound field of the three-dimensional marine environment with an underwater ridge calculated by the spectral model.}
	\label{Figure10}
\end{figure}

\begin{figure}
	\centering
	\subfigure[]{\label{Figure11a}\includegraphics[width=0.49\textwidth]{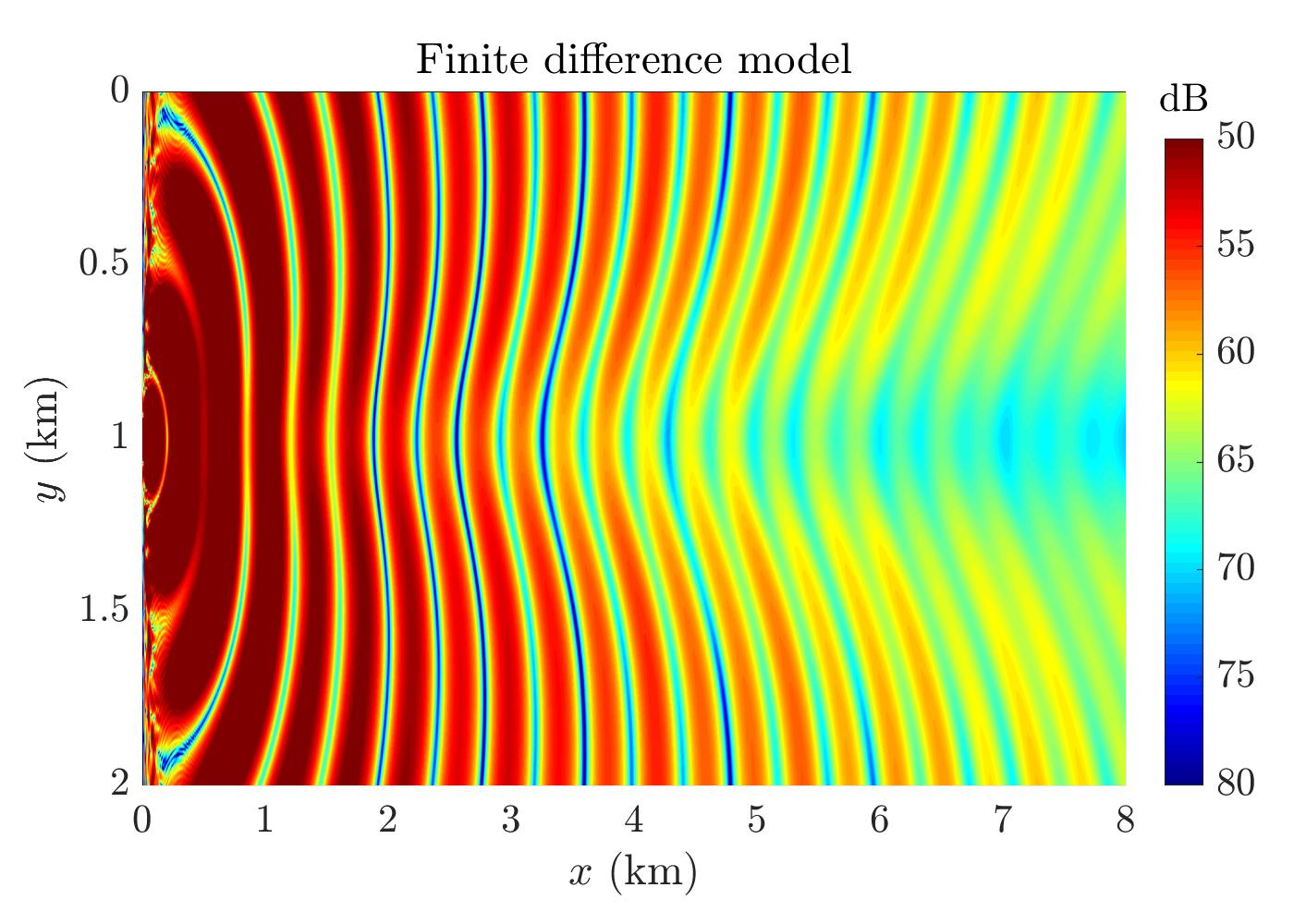}}
	\subfigure[]{\label{Figure11b}\includegraphics[width=0.49\textwidth]{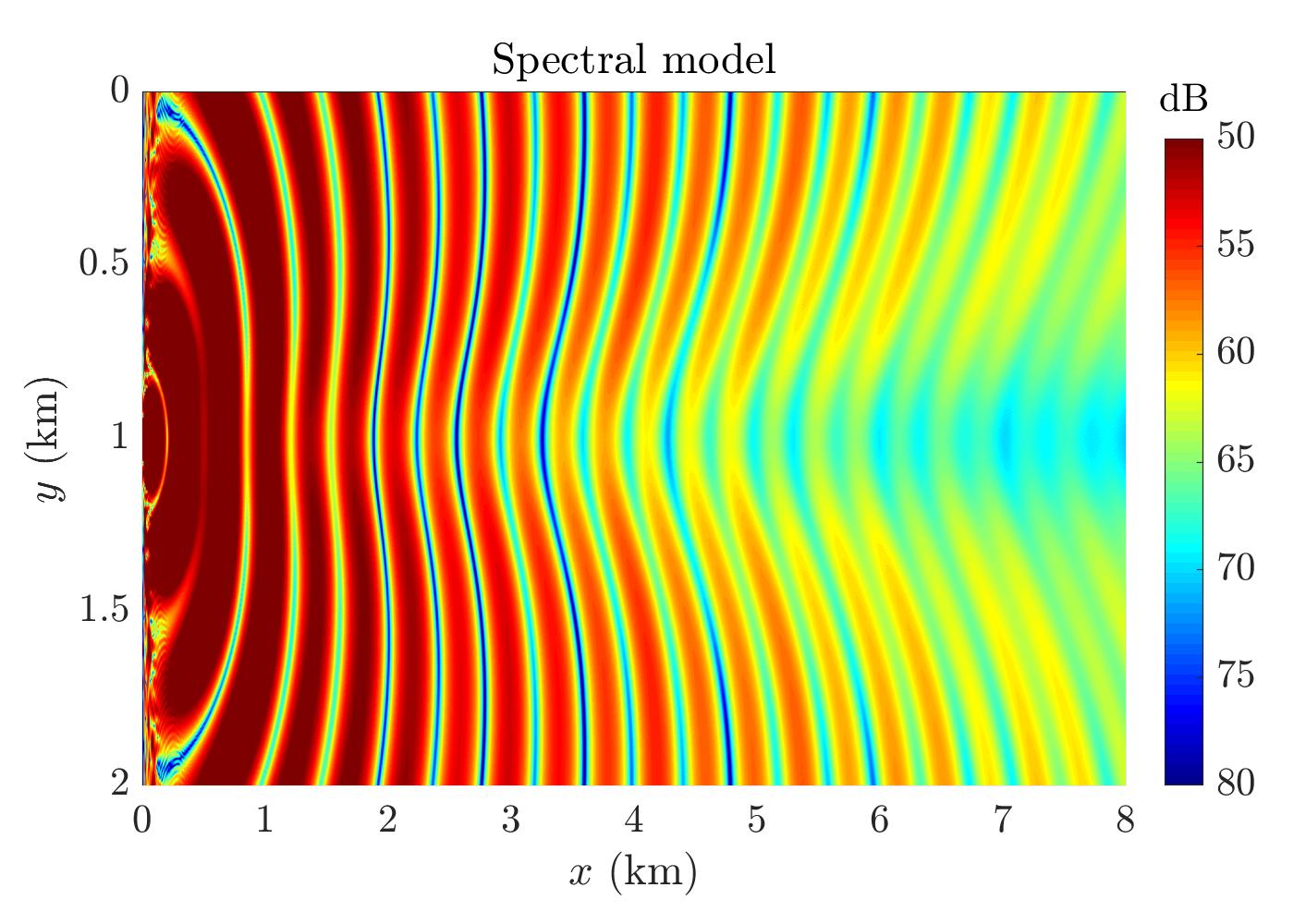}}\\
	\subfigure[]{\label{Figure11c}\includegraphics[width=0.49\textwidth]{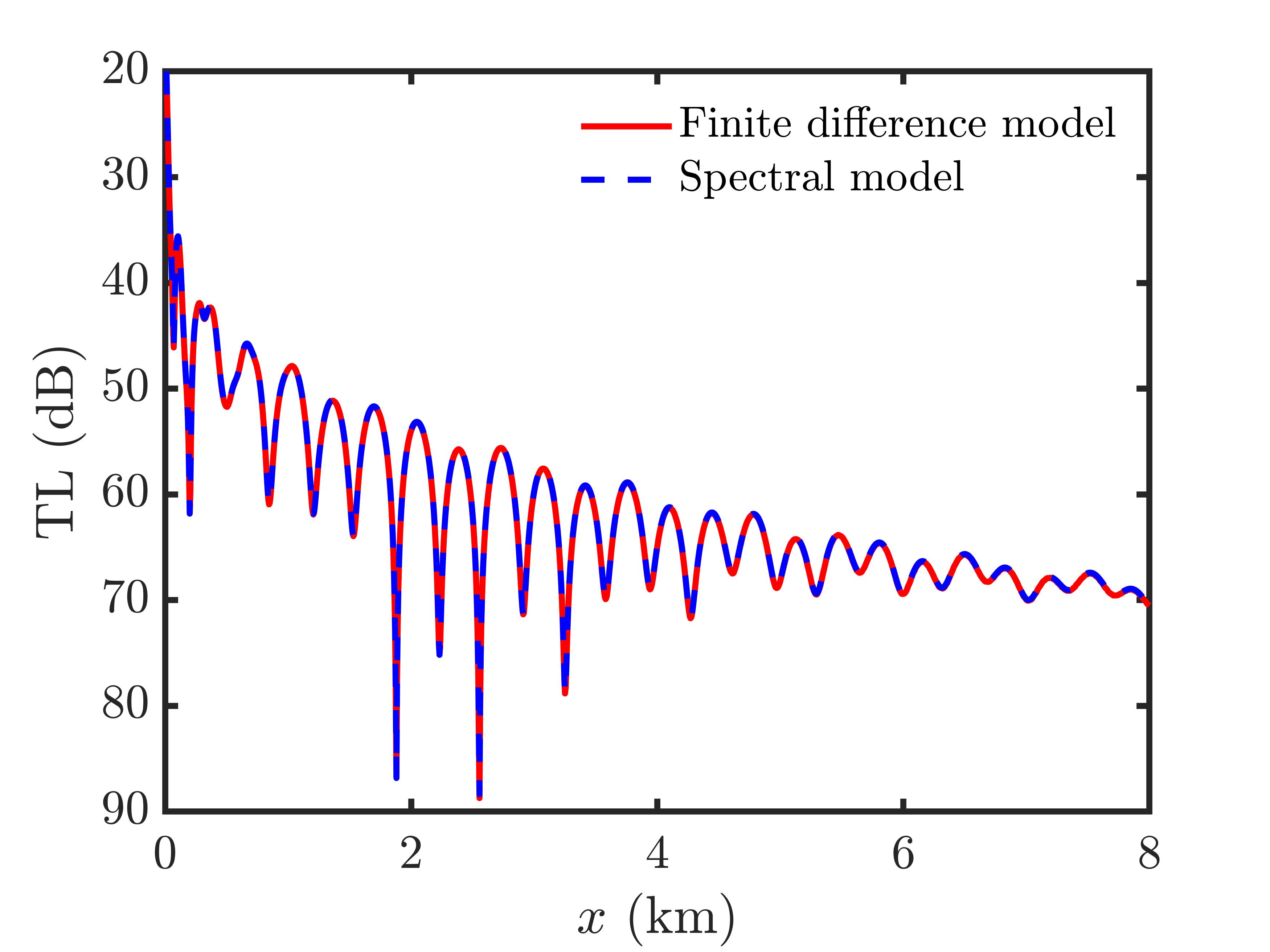}}
	\subfigure[]{\label{Figure11d}\includegraphics[width=0.49\textwidth]{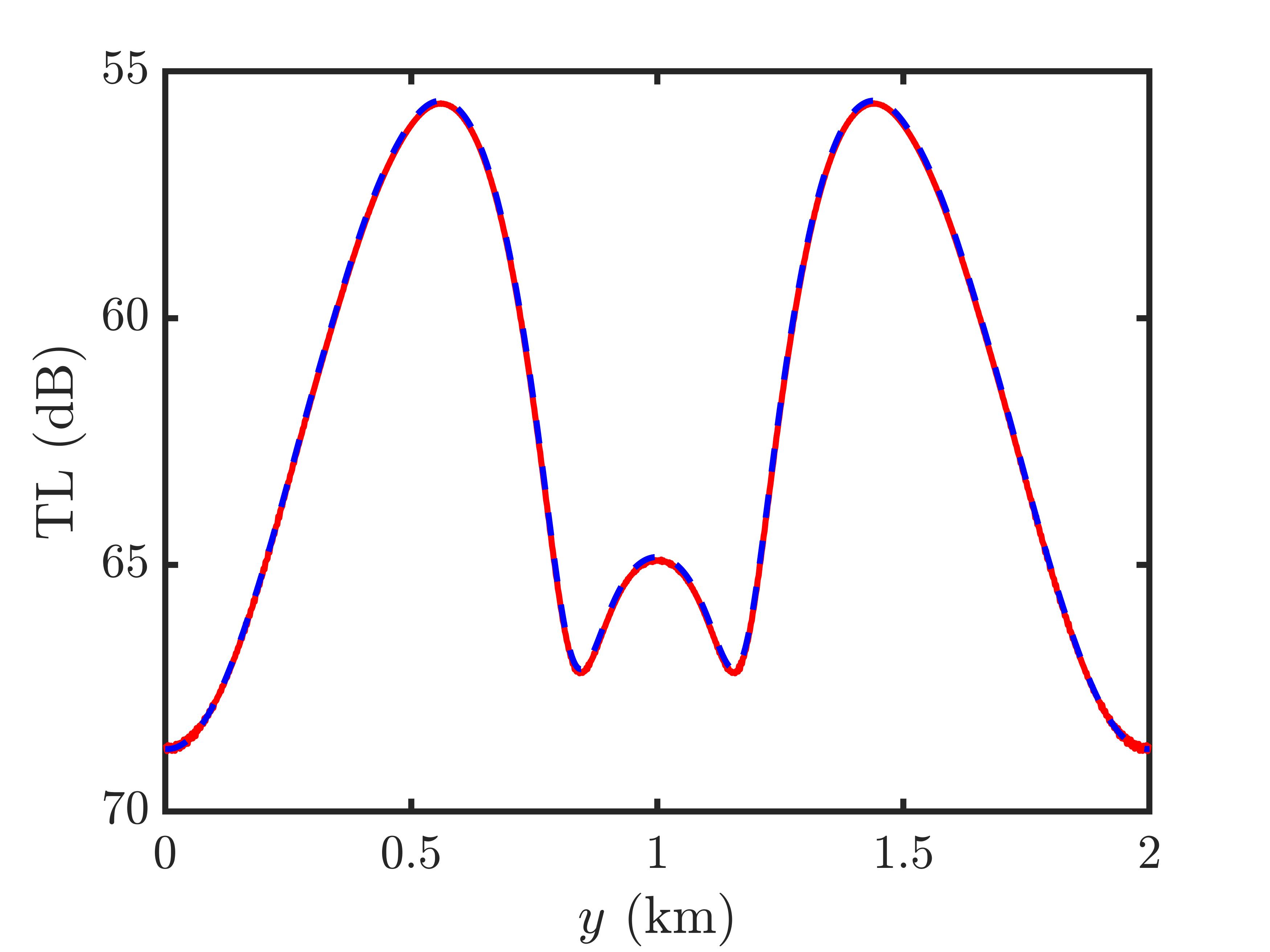}}
	\caption{The sound field slices at $z_\mathrm{r}=90$ m of the underwater ridge waveguide calculated by the finite difference model (a) in Ref.~\cite{Petrov2020a} and the spectral model in this paper (b), both of which use ray-based starters; the TL curves along $y=1$ km (c) and $x=4$ km (d) of the sound field slices.}
\end{figure}

The three-dimensional acoustic field of this configuration simulated by the spectral model developed in this paper is shown in Fig.~\ref{Figure10}. It can be observed from the figure that the ridge causes the sound energy to gradually refract to both sides, and the bending angle increases with the range from the source. The next step is to conduct a more detailed comparison and analysis. Fig.~\ref{Figure11a} and \ref{Figure11b} display the slices of the acoustic field at a depth of 90 m simulated using the finite difference model from Ref.~\cite{Petrov2020a} and the spectral model developed in this paper, respectively. To make a fair comparison, both models employ a ray-based starter with the same parameters. From the figures, it is evident that the results obtained by the spectral model in this paper show good consistency with the finite-difference model. For further confirmation, we also present the TL curves along $y=1$ km and $x=4$ km of the sound field slices in Fig.~\ref{Figure11c} and \ref{Figure11d}. From the figures, it can be observed that, regardless of the $x$- or $y$-direction, the simulation results of the spectral algorithm match well with the finite difference algorithm. The errors between the spectral curves and the finite difference curves are generally within 1 dB. This also indicates that the proposed three-dimensional spectral model in this article yields satisfactory results in addressing quasi-three-dimensional waveguides.

\subsection{Conical seamount}
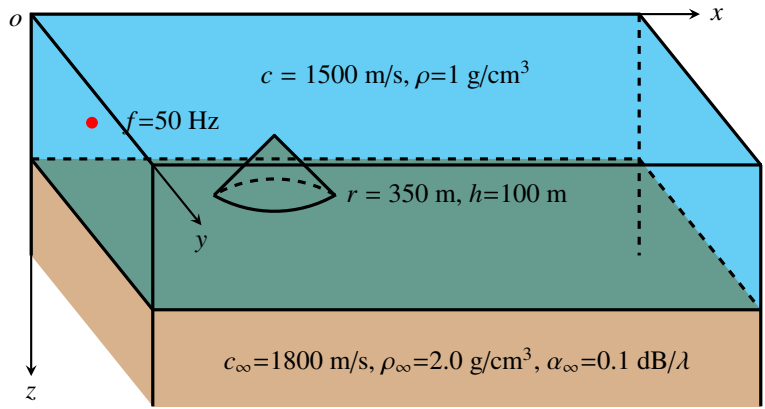
\begin{figure}
	\centering
		\begin{tikzpicture}[node distance=2cm,samples=500,domain=7:10.5,scale=0.8]
		
		\fill[orange,opacity=1](2,-2.4)--(12,-2.4)--(14,-4.9)--(4,-4.9)--cycle;
		\fill[orange,opacity=1]  (5,-3)--(6,-2)--(7,-3)--(5,-3) arc (240:300:2)--cycle;
		\fill[cyan,opacity=0.6] (2,0)--(12,0)--(14,-2.5)--(4,-2.5)--(4,-4.9)--(2,-2.4)--cycle;
		\fill[cyan,opacity=0.6]  (4,-2.5)--(4,-4.9)--(14,-4.9)--(14,-2.5)-- cycle;
		\fill[brown,opacity=0.6] (2,-2.4)--(4,-4.9)--(4,-6.5)--(2,-4)--cycle;
		\fill[brown,opacity=0.6]  (4,-4.9)--(14,-4.9)--(14,-6.5)--(4,-6.5)--cycle;
		
		\draw[thick, ->](2,0)--(13,0) node[right]{$x$};
		\draw[thick, ->](2,0)--(2,-6) node[below]{$z$};	    		
		\draw[very thick](1.98,0)--(12.02,0);
		\draw[very thick](4,-2.5)--(14.02,-2.5);
		\draw[very thick](2,0)--(4,-2.5);
		\draw[thick, ->](2,0)--(4.8,-3.5) node[below]{$y$};	
		\draw[very thick](12,0)--(14,-2.5);
		\draw[very thick](2,0.02)--(2,-4);
		\draw[very thick](4,-2.5)--(4,-6.5);
		\draw[dashed, very thick](12,0)--(12,-4);
		\draw[very thick](14,-2.5)--(14,-6.5);
		\draw[dashed, very thick](12,-2.4)--(14,-4.9);		
		\filldraw [red] (3,-1.8) circle [radius=2.5pt];
		\node at (4.3,-1.8){$f$=50 Hz};
		\node at (1.75,-0.1){$o$};	
		\draw[very thick](2,-2.4)--(4,-4.9);
		\draw[very thick](4,-4.9)--(14,-4.9);
		\draw[dashed, very thick](2,-2.4)--(12,-2.4);
		
		\draw[very thick](5.01,-3)--(6.01,-2);
		\draw[very thick](5.99,-2)--(6.99,-3);
		\draw[very thick](5,-3) arc (240:300.4:2);
		\draw[dashed, very thick](7,-3) arc (60:120:2);
		\node at (9,-3){$r=$ 350 m, $h$=100 m};	
		
		\node at (9,-5.75){$c_{\infty}$=1800 m/s, $\rho_{\infty}$=2.0 g/cm$^3$, $\alpha_{\infty}$=0.1 dB/$\lambda$};					
		\node at (8,-1){$c=1500$ m/s, $\rho$=1 g/cm$^3$};
		
	\end{tikzpicture}
	\caption{Schematic diagram of the three-dimensional marine environment with an underwater conical seamount.}
	\label{Figure12}
\end{figure}

Conical seamounts possess complex terrain shapes and oceanic medium structures, which can cause multiple reflections, scattering, and refractions in sound waveguides. Numerical simulations of conical seamounts enable the study of the principles governing three-dimensional sound propagation and the effects of seafloor topography on sound signal attenuation and deformation \cite{Buckingham1986,Luowy2009,Liuw2021}. Here, we consider a waveguide environment, as shown in Fig.~\ref{Figure12}. The sea depth is $H=250$ m, and below it is a homogeneous acoustic half-space. The sound source is located at (0, 2500, 100) m and excites $M=8$ modes. The summit of the seamount is located at (2000, 2500, 150) m, and the medium inside the mountain is consistent with the acoustic half-space. 

\begin{figure}
	\centering
	\includegraphics[width=0.8\textwidth]{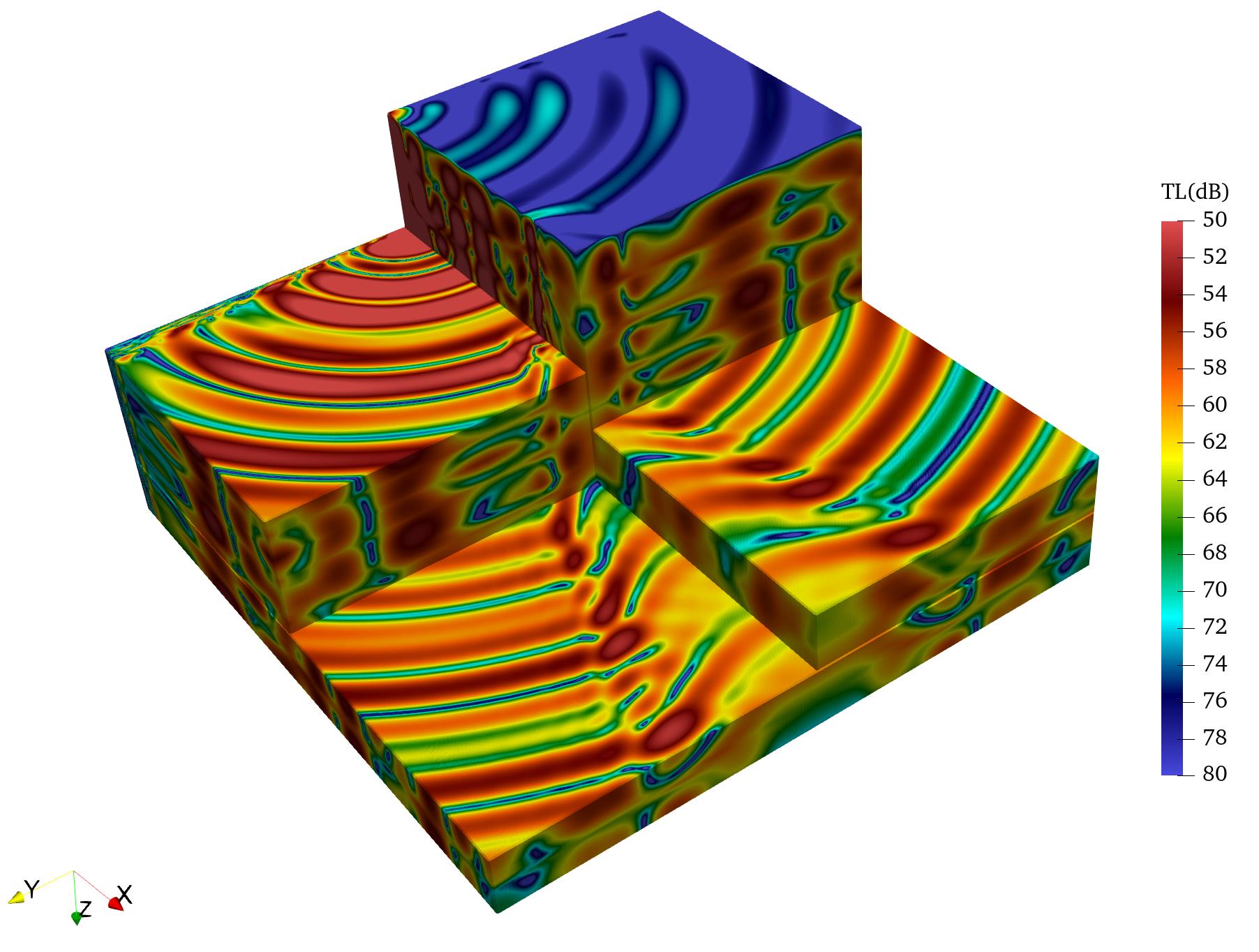}
	\caption{Sound field of the three-dimensional marine environment with an underwater conical seamount calculated by the spectral model.}
	\label{Figure13}
\end{figure}

\begin{figure}
	\centering
	\subfigure[]{\includegraphics[width=0.49\textwidth]{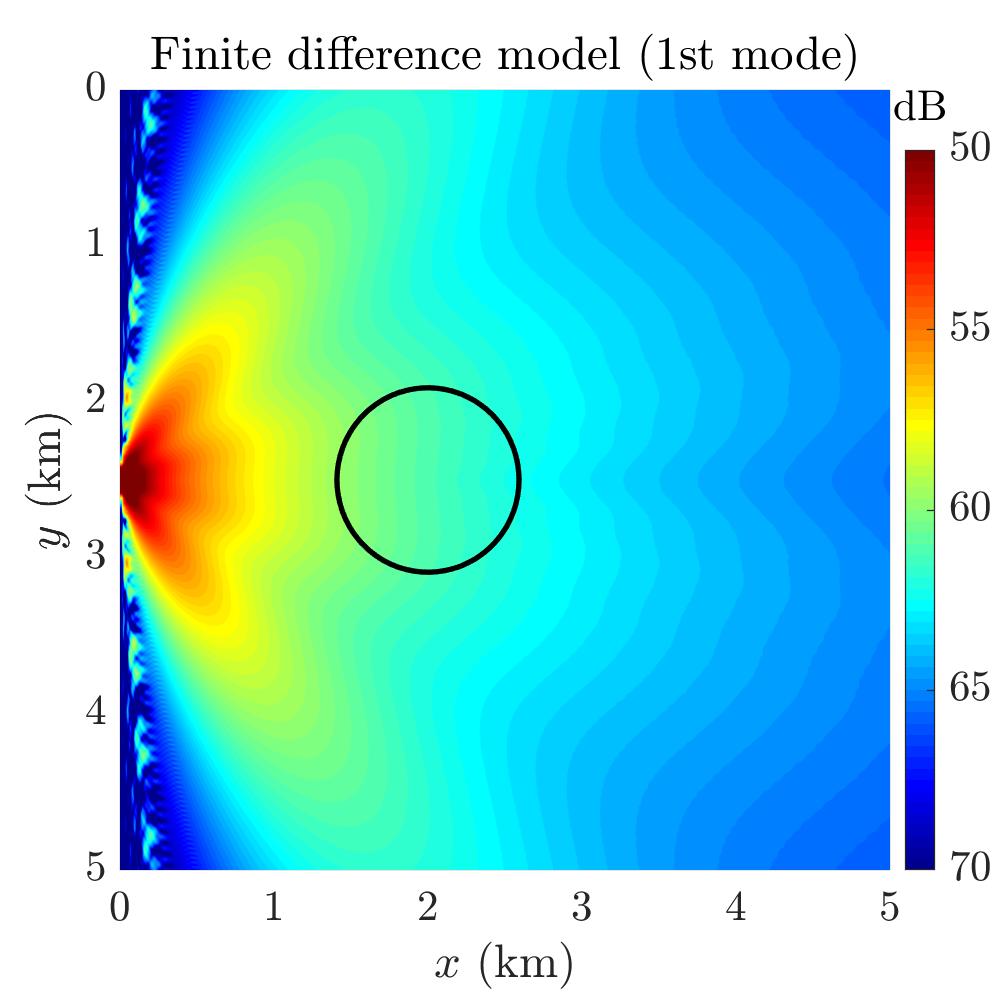}}
	\subfigure[]{\includegraphics[width=0.49\textwidth]{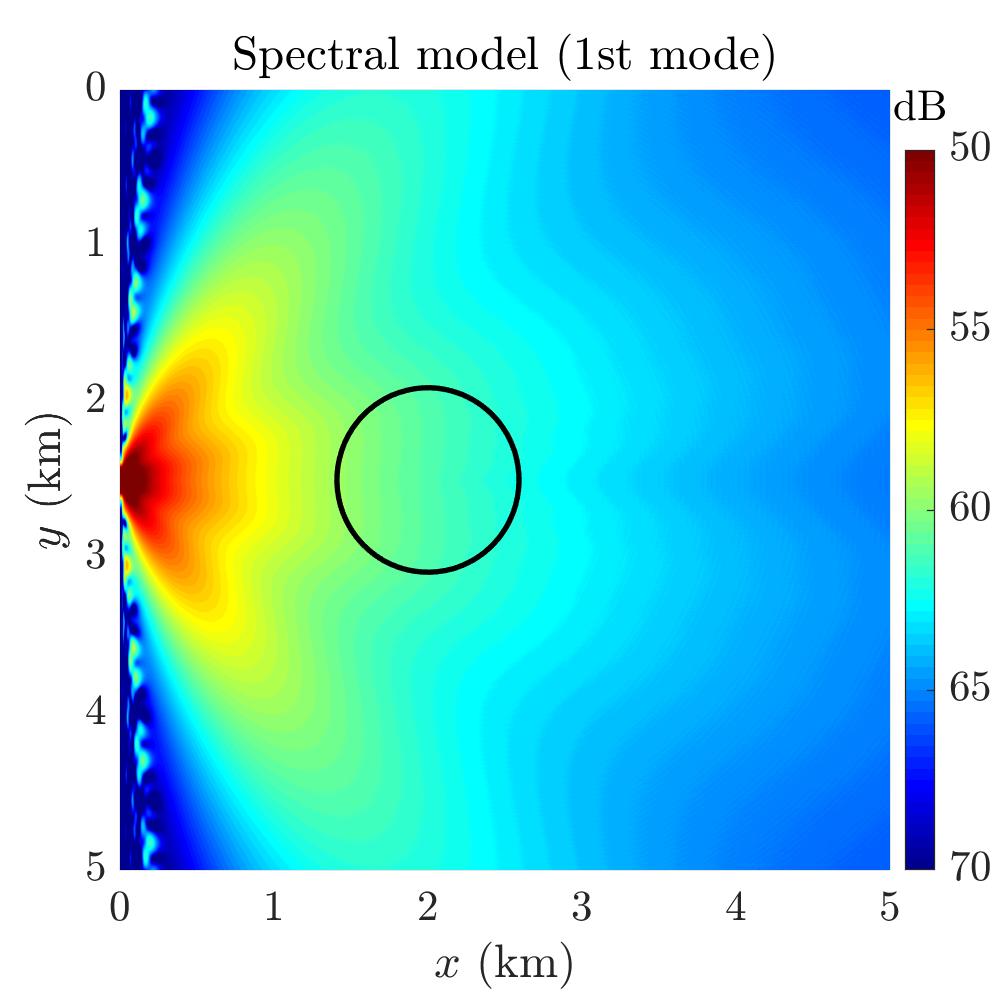}}\\
	\subfigure[]{\includegraphics[width=0.49\textwidth]{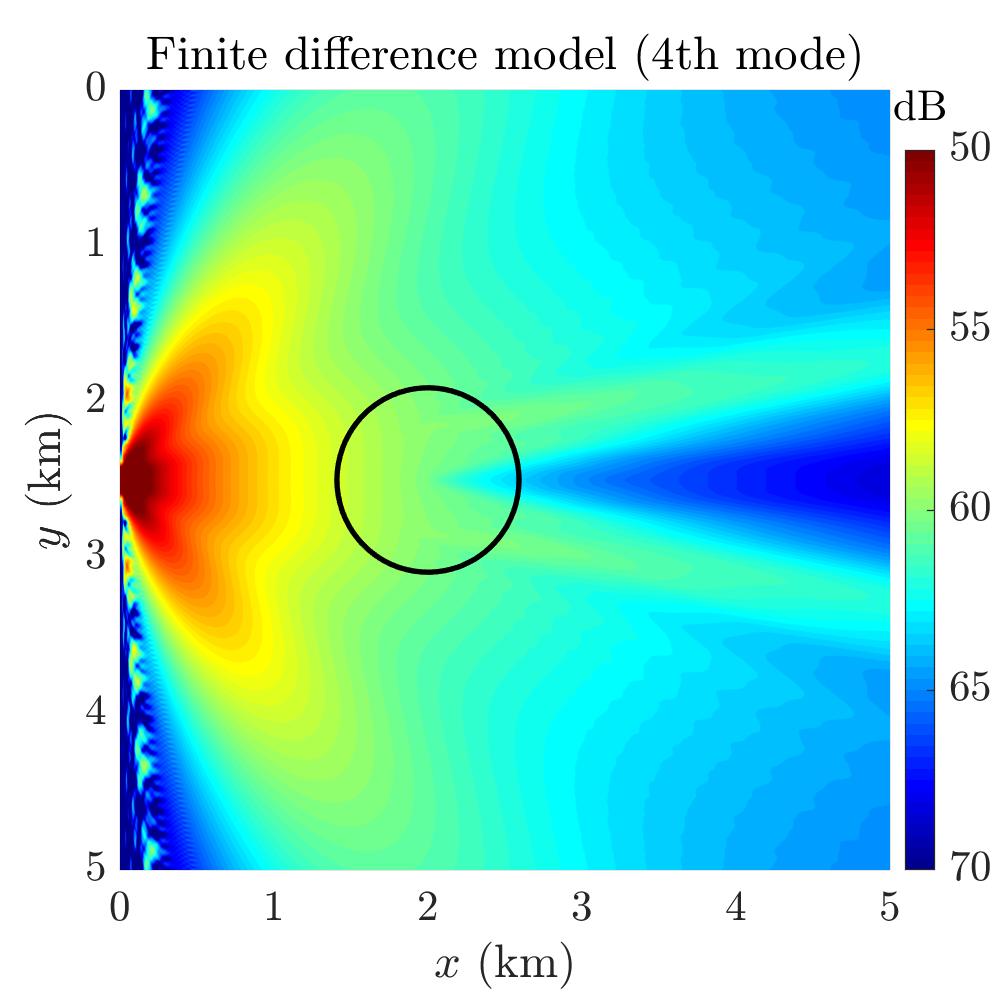}}
	\subfigure[]{\includegraphics[width=0.49\textwidth]{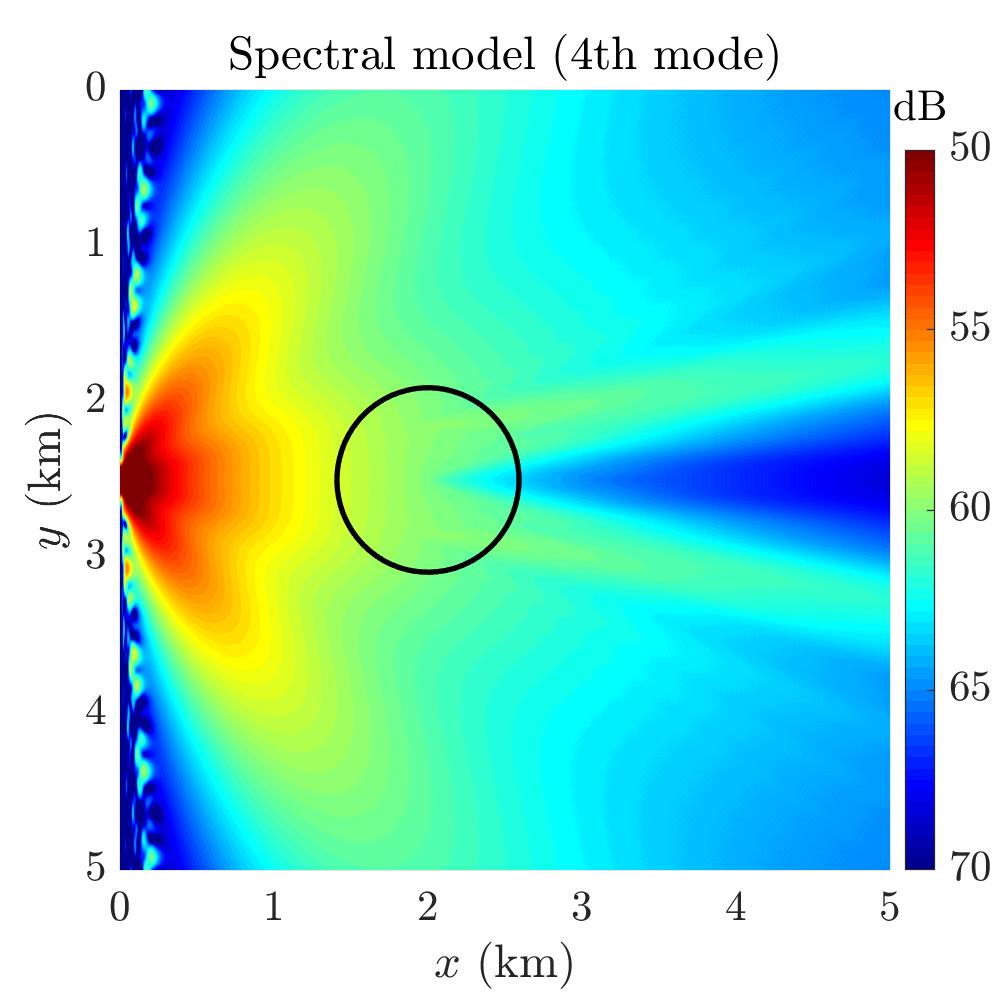}}\\
	\subfigure[]{\includegraphics[width=0.49\textwidth]{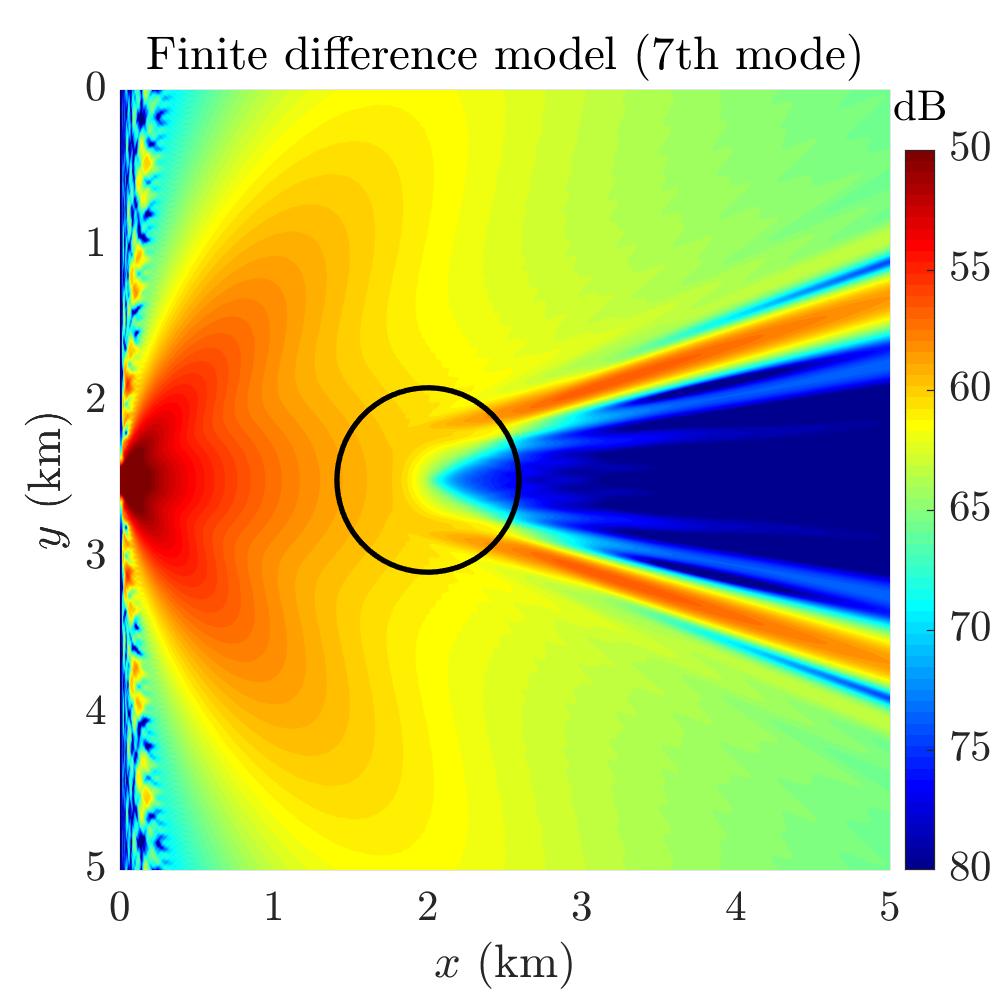}}
	\subfigure[]{\includegraphics[width=0.49\textwidth]{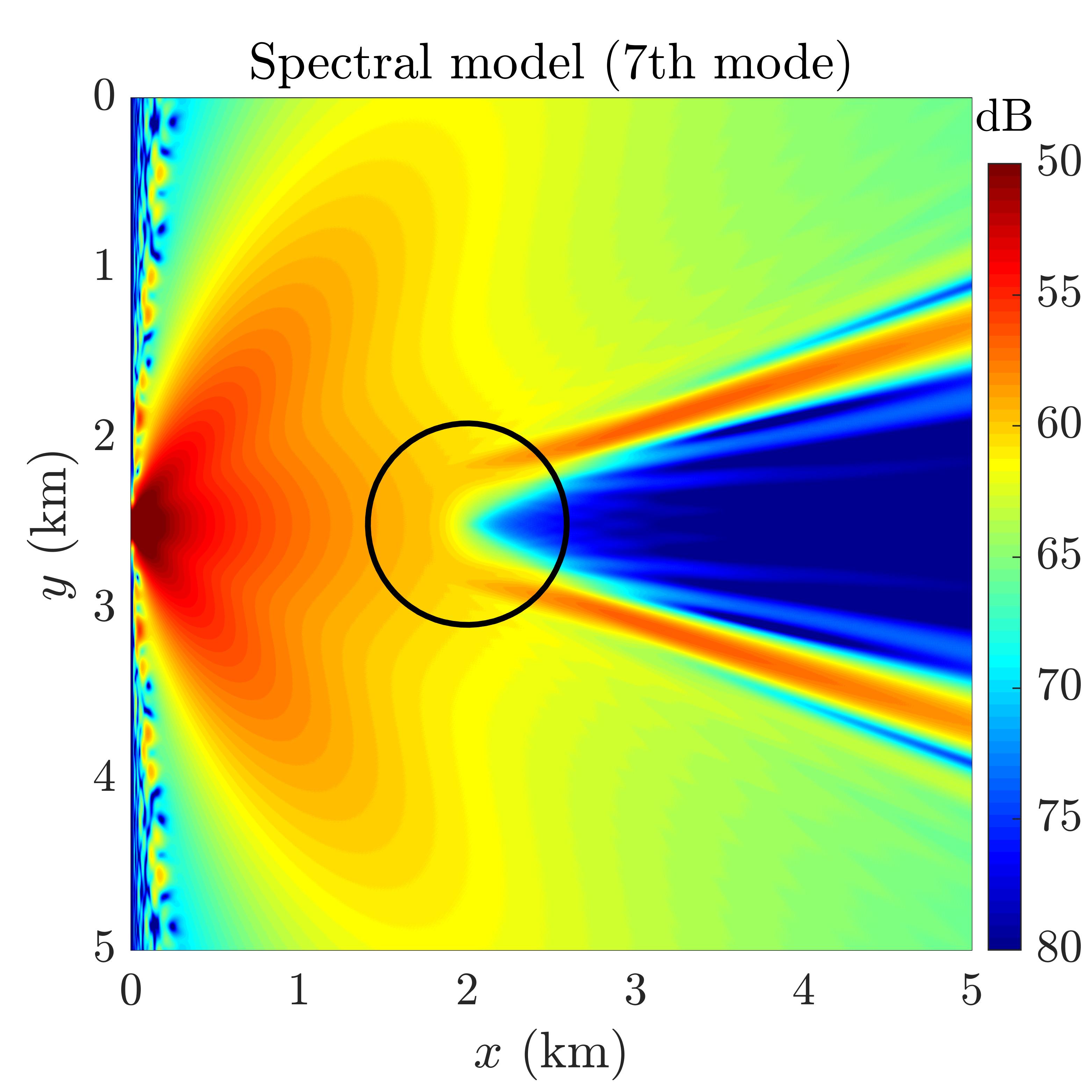}}
	\caption{Horizontal refractive index slices of the conical seamount waveguide on a plane at a depth of $z_\mathrm{r}=100$ m. The position and radius of the seamount are marked by black circles.}
	\label{Figure14}
\end{figure}

\begin{figure}
	\centering
	\subfigure[]{\includegraphics[width=0.49\textwidth]{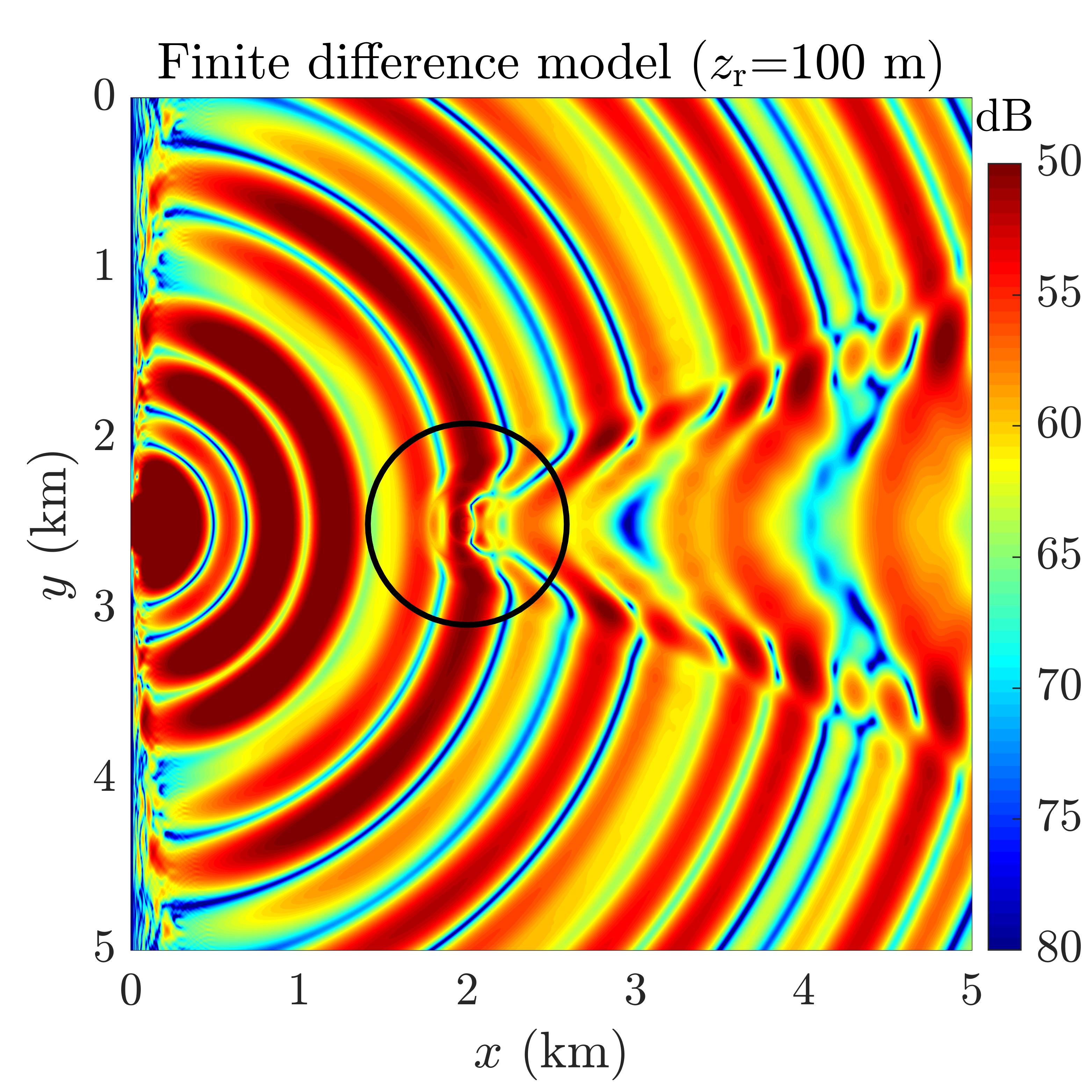}}
	\subfigure[]{\includegraphics[width=0.49\textwidth]{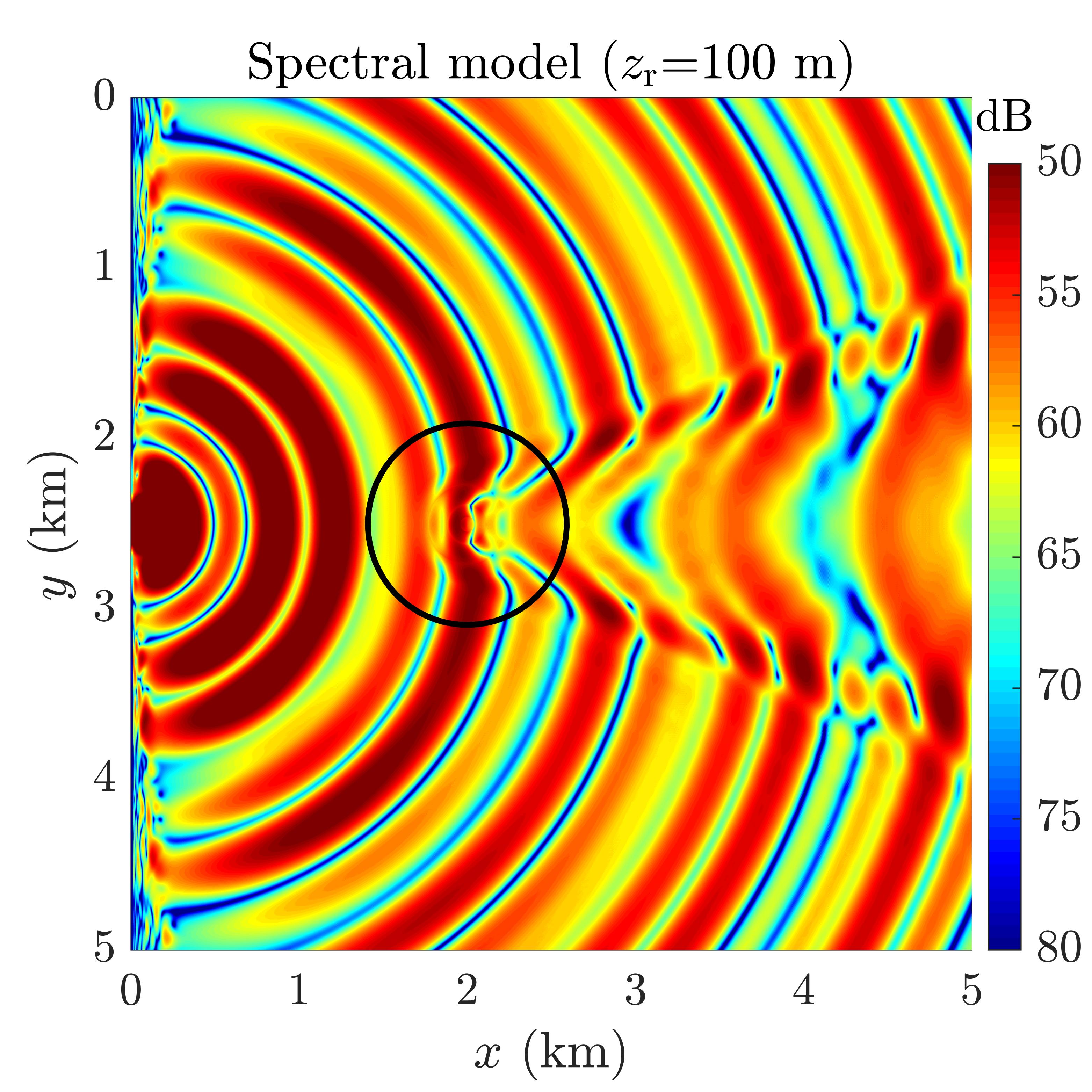}}\\
	\subfigure[]{\includegraphics[width=0.49\textwidth]{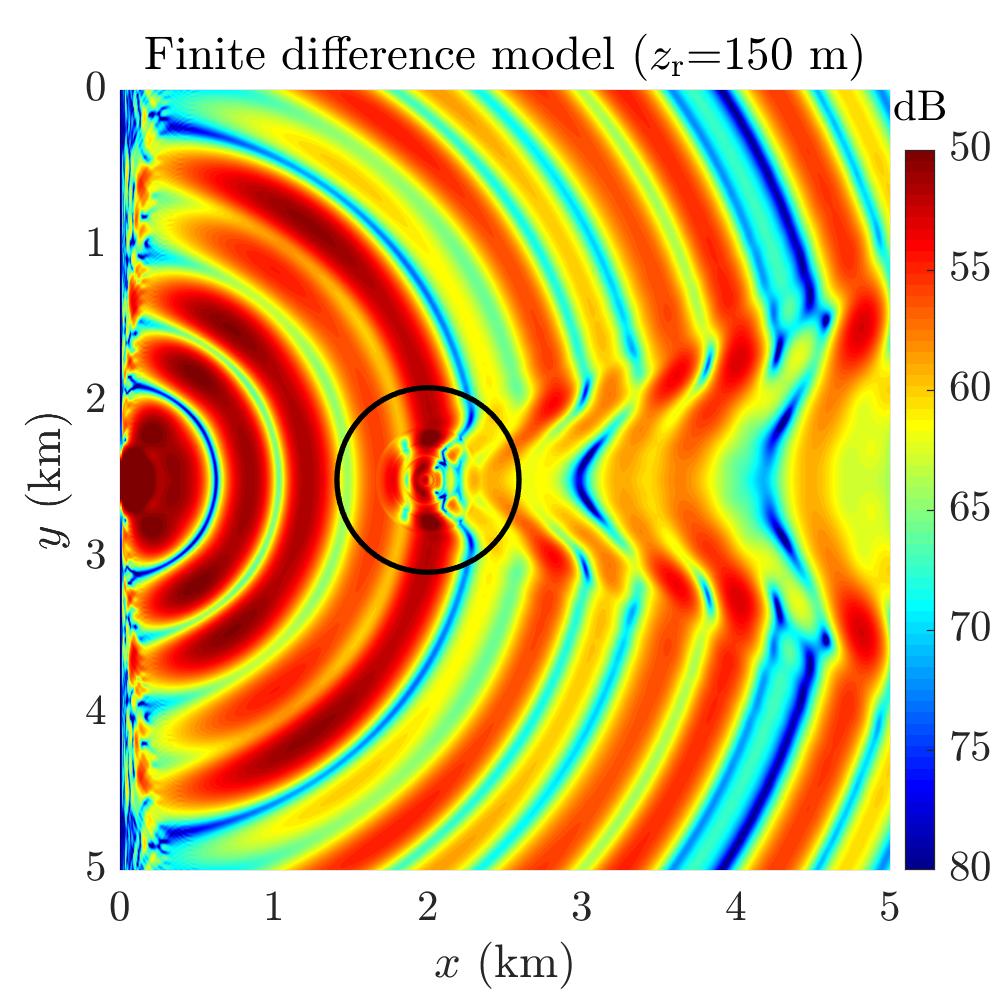}}
	\subfigure[]{\includegraphics[width=0.49\textwidth]{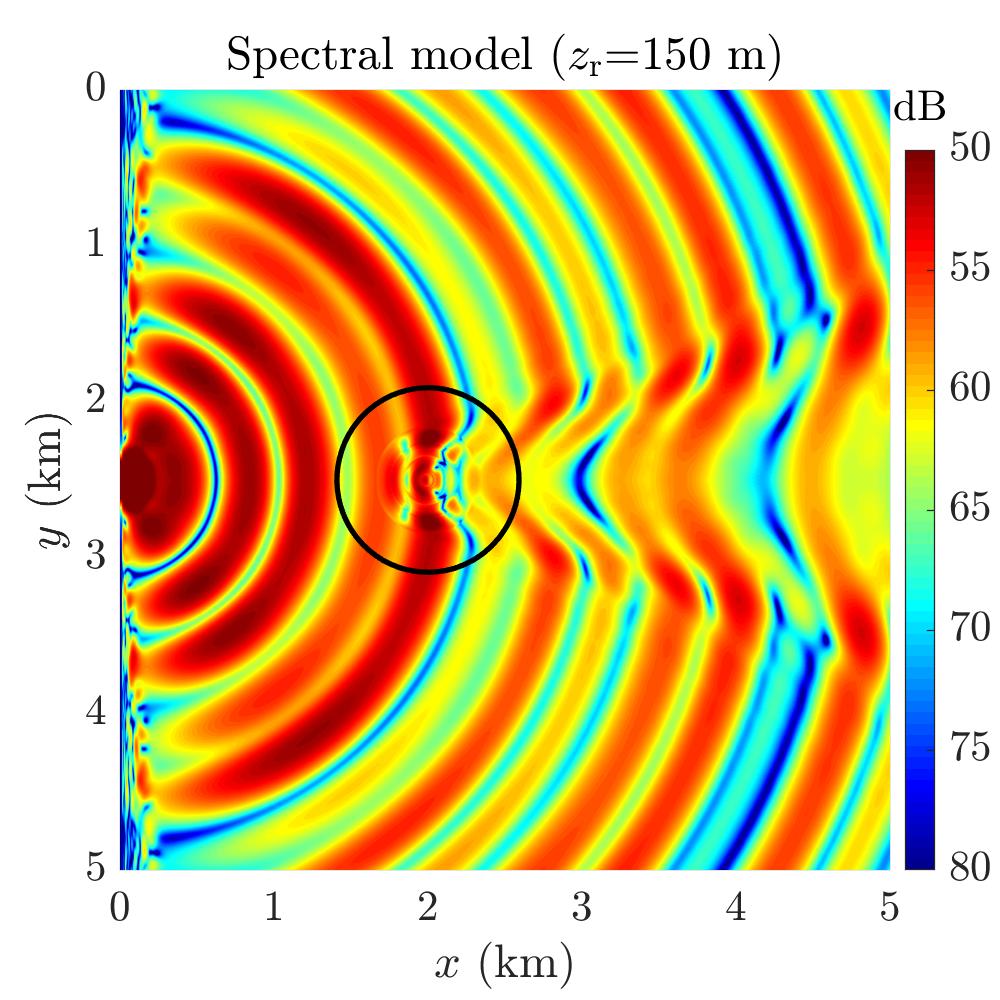}}\\
	\subfigure[]{\includegraphics[width=0.49\textwidth]{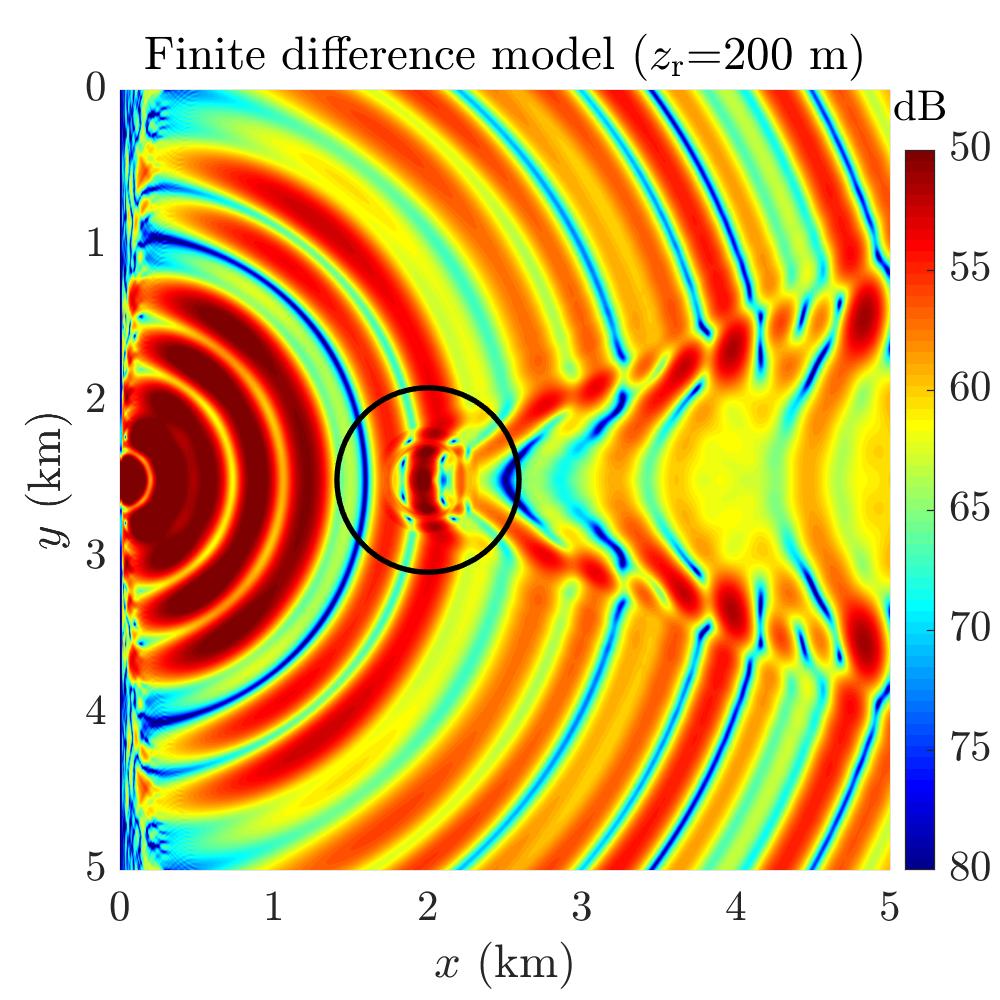}}
	\subfigure[]{\includegraphics[width=0.49\textwidth]{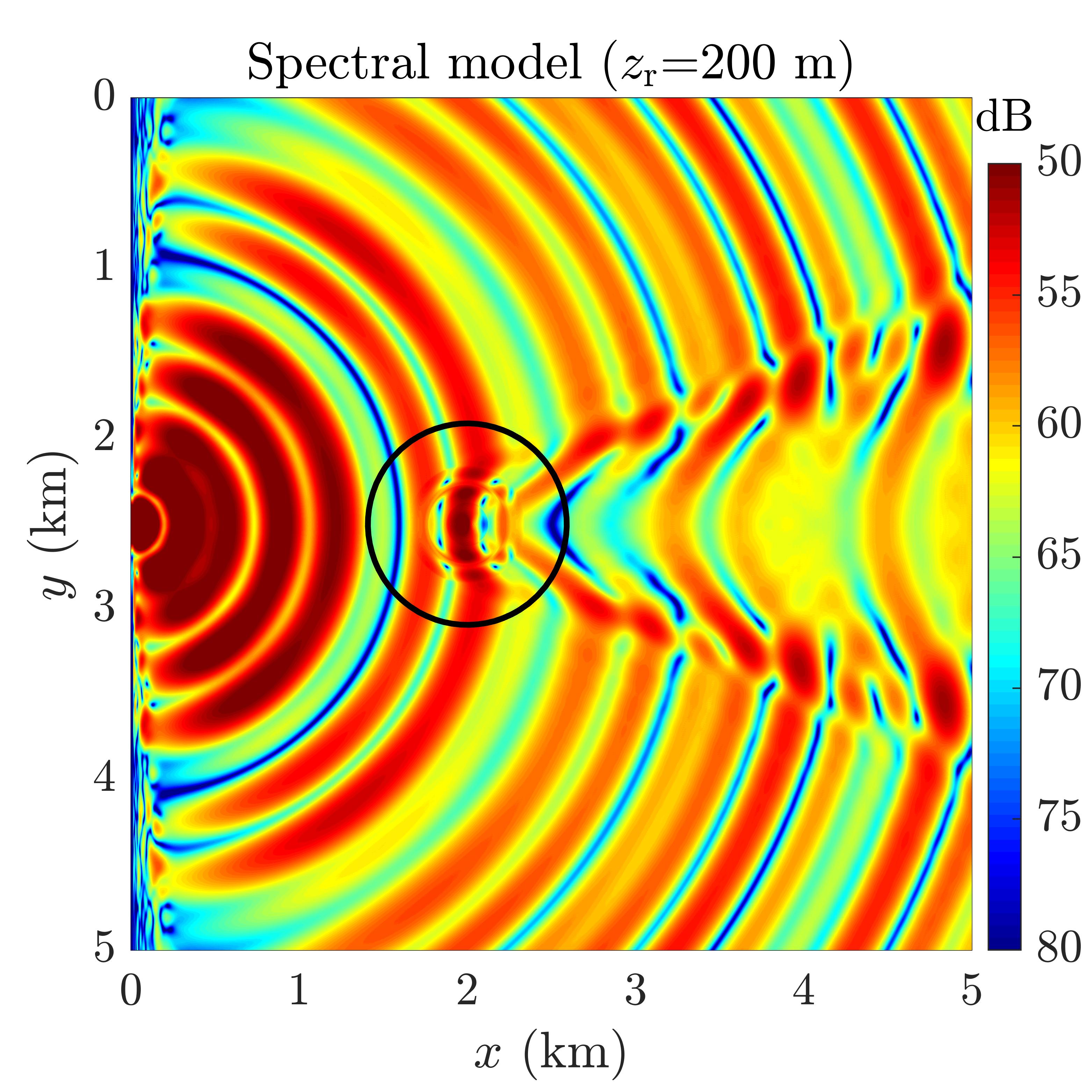}}
	\caption{TL slices of the conical seamount waveguide on a plane at three depths of $z_\mathrm{r}=100$ m, $z_\mathrm{r}=150$ m and $z_\mathrm{r}=200$ m.}
	\label{Figure15}
\end{figure}

\begin{figure}
	\centering
	\subfigure[]{\includegraphics[width=0.49\textwidth]{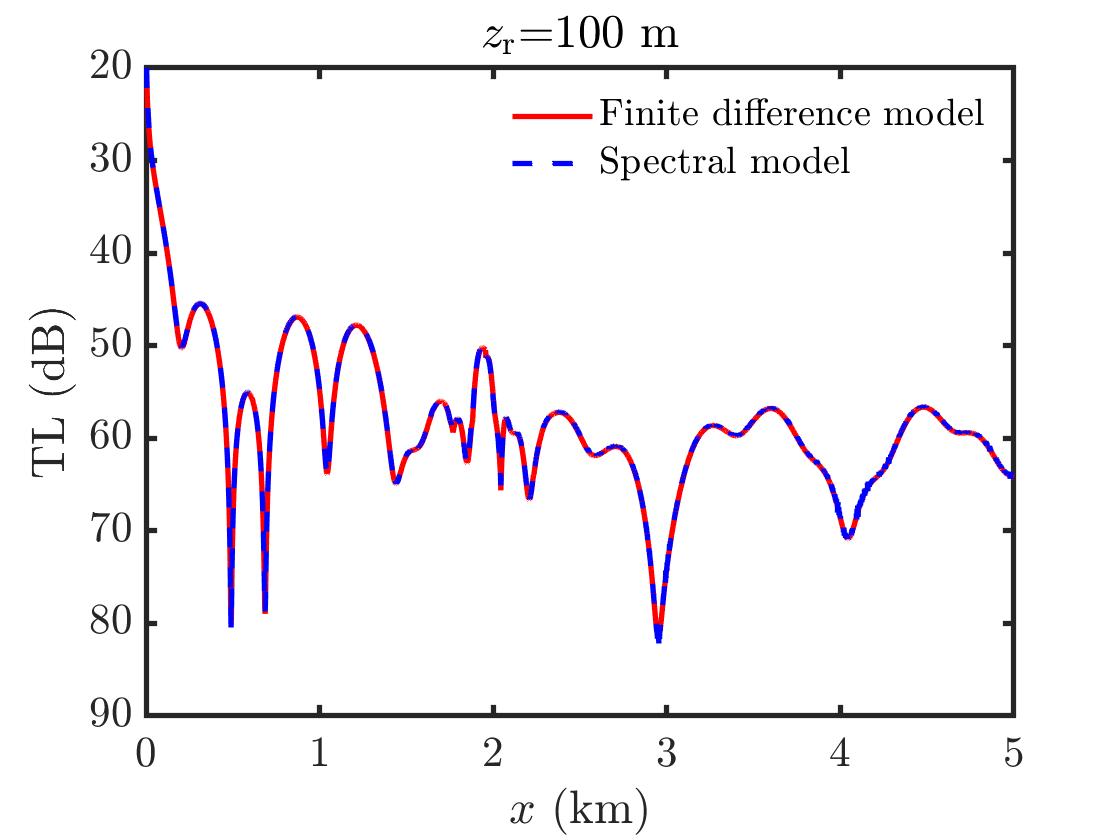}}
	\subfigure[]{\includegraphics[width=0.49\textwidth]{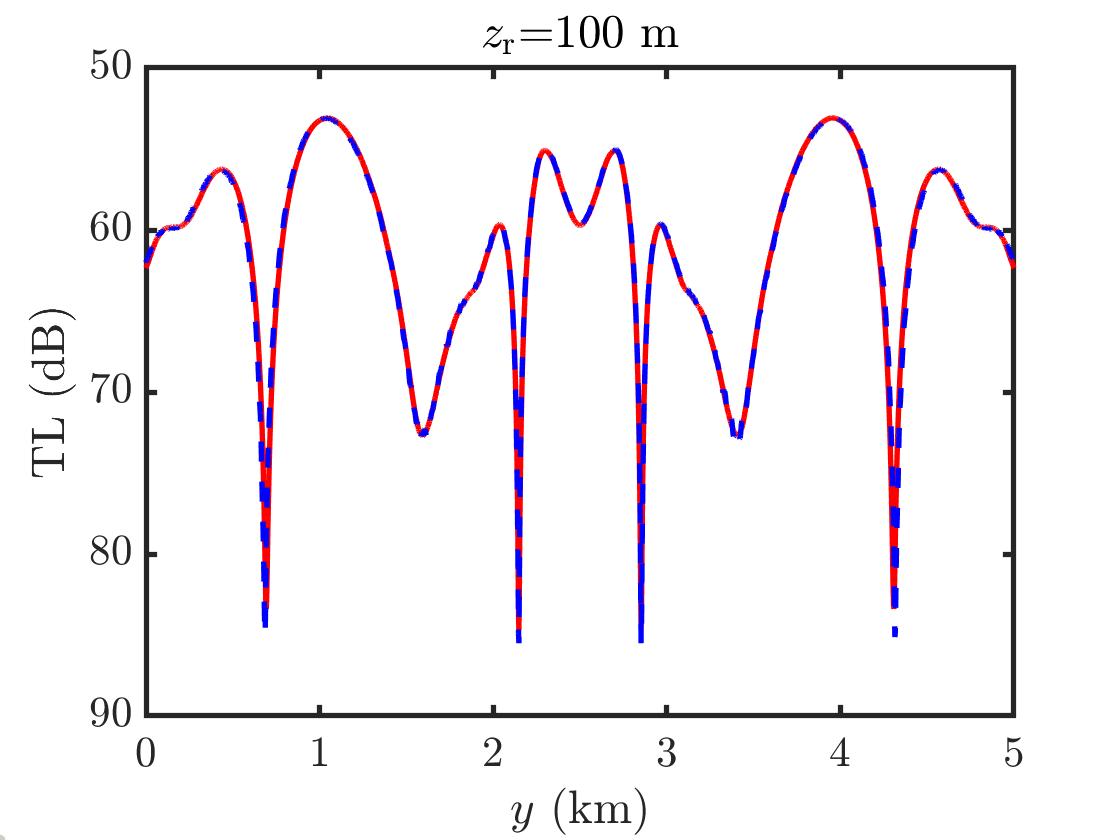}}\\
	\subfigure[]{\includegraphics[width=0.49\textwidth]{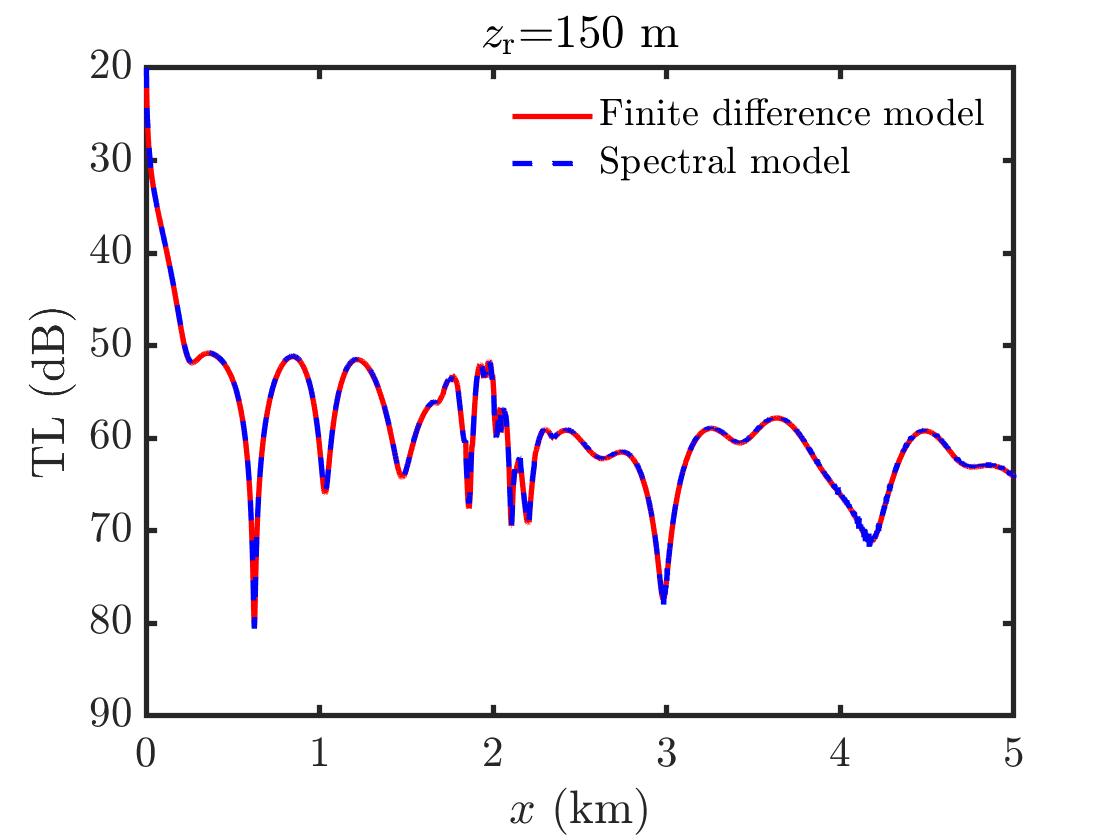}}
	\subfigure[]{\includegraphics[width=0.49\textwidth]{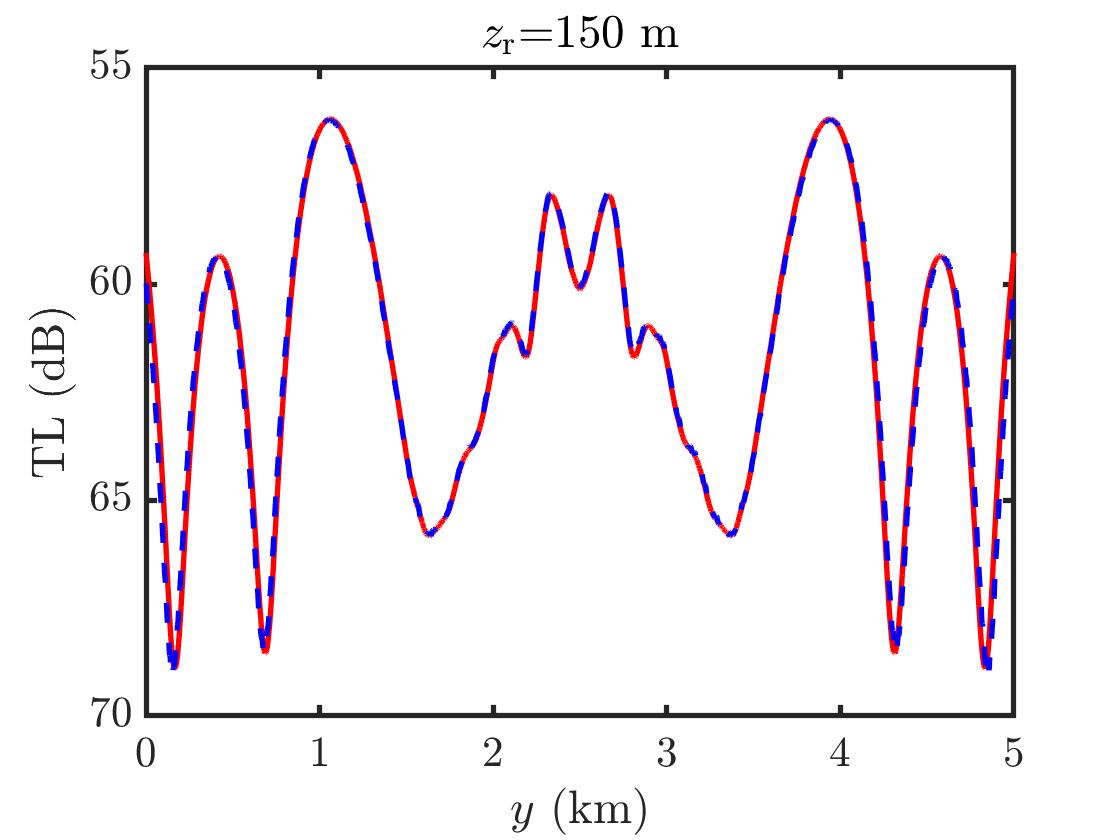}}\\
	\subfigure[]{\includegraphics[width=0.49\textwidth]{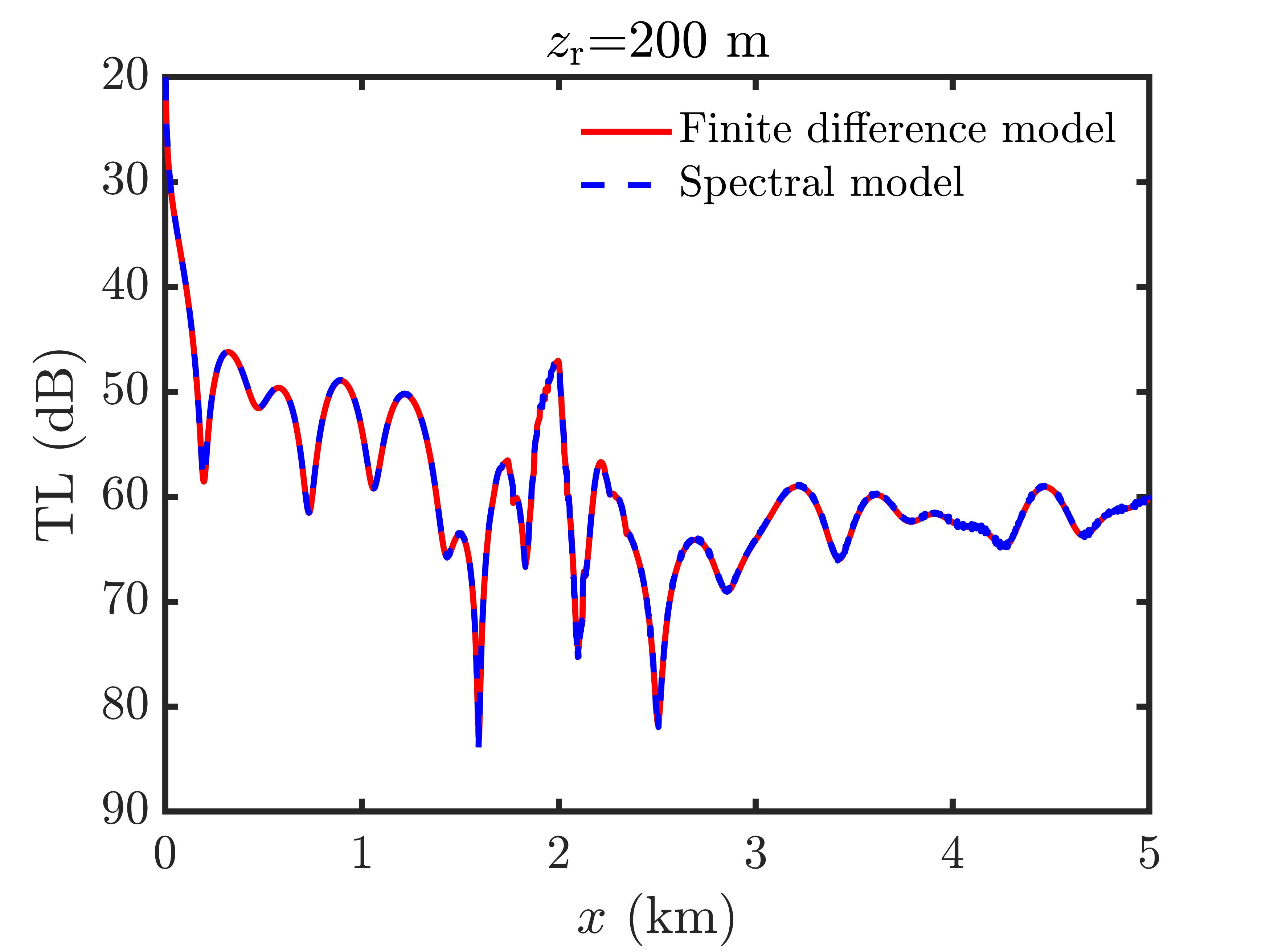}}
	\subfigure[]{\includegraphics[width=0.49\textwidth]{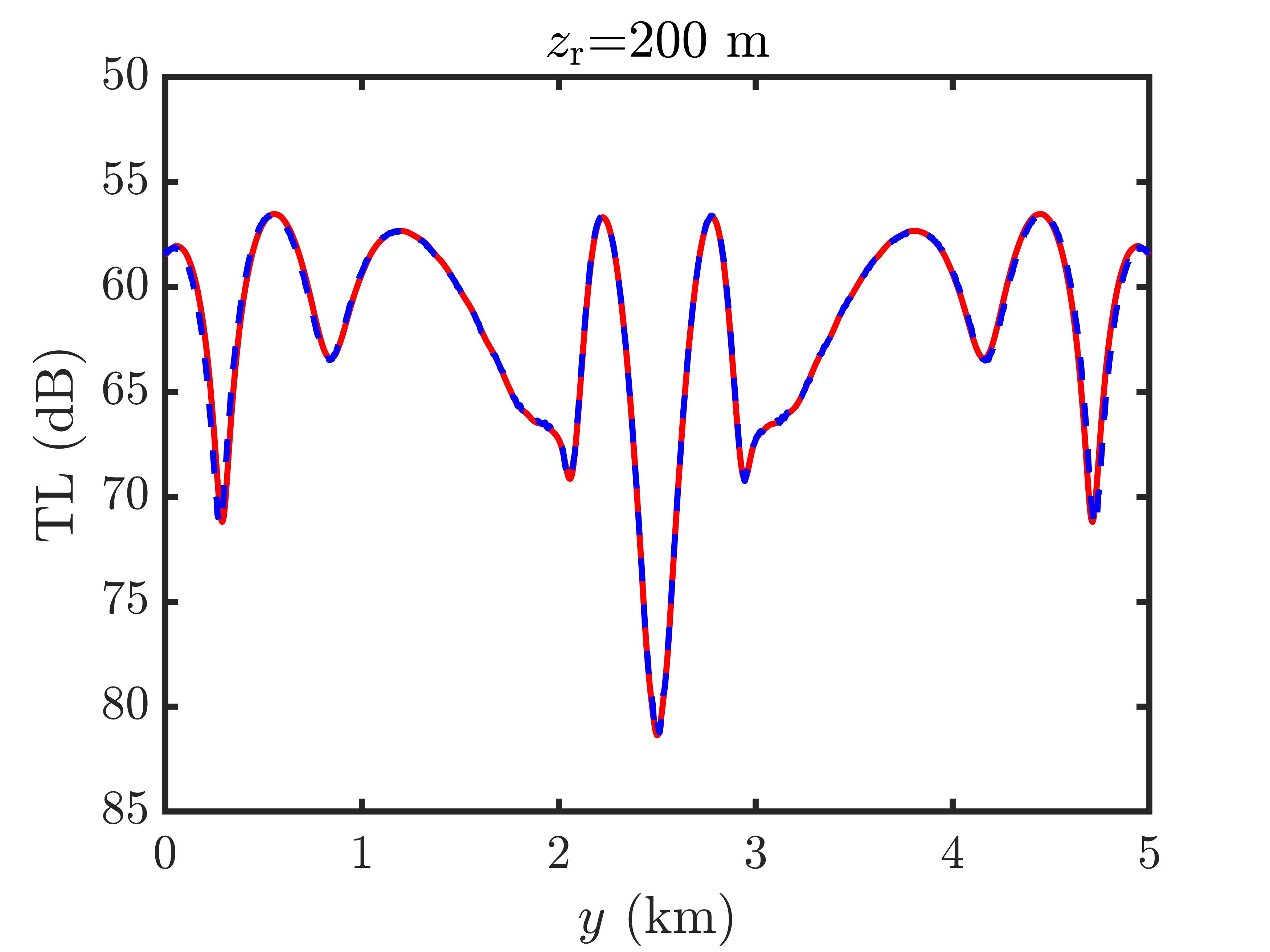}}	
	\caption{The TL curves along $y=2.5$ km [(a), (c), (e)] and $x=2.5$ km [(b), (d), (f)] of the sound field slices at three depths of $z_\mathrm{r}=100$ m, $z_\mathrm{r}=150$ m and $z_\mathrm{r}=200$ m.}
	\label{Figure16}
\end{figure}

Fig.~\ref{Figure13} presents the numerical sound field of this waveguide with a conical seamount. The refraction of sound energy by the seamount disrupts the concentric circular structure of the original sound field and causes noticeable disturbances behind the seamount. The width of the disturbances gradually increases with distance.
Fig.~\ref{Figure14} shows the horizontal refractive index slices at a depth of $z_\mathrm{r}$=100 m calculated using Petrov's finite difference model \cite{Petrov2020a} and the spectral model developed in this study. For the purpose of display, the slices have been normalized according to the calculation of TL. From the figures, it can be seen that the simulation results of both numerical models are consistent for the first-order, fourth-order, and seventh-order HREs. There are only slight differences in the far field. The comparison of the three slices also reveals that the conical seamount has a weaker refraction effect on low-order modes and a stronger refraction effect on high-order modes. The refraction effect of the seamount increases with the increase in mode order. To further compare the level of consistency between the two models, Fig.~\ref{Figure15} provides TL slices at different depths, and Fig.~\ref{Figure16} plots the TL curves along the $x$- and $y$-axes on these slices. From the results in the TL fields, we can see that the spectral model and the finite difference model produce almost identical computational results.	

Overall, the numerical experiments in this section demonstrate that the proposed spectral model in this paper can achieve accurate results in simulating the sound field of three-dimensional gradually varying waveguides. It can effectively simulate the horizontal refractive effects in three-dimensional sound propagation.

\section{Discussion and Summary}
\label{sec6}
In this paper, we developed a new numerical model for simulating fully three-dimensional acoustic waveguides. The three-dimensional model utilizes the theory of ``vertical modes and horizontal parabolic equation'' and calculates the normal modes in the vertical direction while neglecting the coupling effect between modes. By exploiting the orthogonality between local modes, the three-dimensional Helmholtz equation is simplified to a series of two-dimensional Helmholtz equations. The two-dimensional Helmholtz equation serves as the governing equation for the horizontal refraction index, and it is solved using a wide-angle parabolic model based on the split-step Pad\'e approximation. In terms of the starter, we implemented an analytical Greene starter, ray-based starter, and self-starter. Additionally, since the MPEs are unbounded, the numerical solution requires the addition of two PMLs to simulate the free propagation of sound waves at the truncation interfaces.

The main innovation of this paper lies in the introduction of a Chebyshev spectral method to discretize the normal modes and MPEs in the aforementioned theory. Spectral methods are a class of numerical discretization methods that offer high accuracy and fast convergence. It is based on the theory of orthogonal polynomial approximation and the weighted residual principle. In spectral methods, functions are expanded in a chosen set of orthogonal basis functions, and the spectral coefficients are determined by evaluating the residuals at appropriate nodes. Due to the excellent approximation properties of orthogonal basis functions, spectral methods exhibit extremely high accuracy in approximating smooth functions. Furthermore, the convergence rate of spectral methods is typically exponential, which means that as the order of the basis functions increases, the approximation accuracy of the solution improves rapidly.

To solving the local modes, we employ a domain decomposition strategy and use spectral discretization in each layer of the medium. The truncation order of the spectral method used in each layer can be flexibly determined based on the stratification of the medium, ensuring spectral accuracy. The layers are assembled into a block-diagonal global matrix [Eq.~\eqref{eq.47}]. After applying boundary and interface conditions, this large algebraic eigenvalue system is solved in a unified manner. For the waveguides with acoustic half-spaces, traditional root-finding algorithms may exhibit missing root phenomena near the branch cut, and errors in the parameter $\kappa(x,y)$ can further lead to errors in solving the MPEs. We adopt an eigenvalue transformation technique to avoid missing roots and ensure that the obtained $\kappa_m(x,y)$ corresponds to the same-order modes. In solving the MPEs, we employ the spectral method to discretize the $\mathcal{Y}$-operator. To enhance the accuracy of function sampling at GCL nodes in physical space, we adopt a domain decomposition strategy and use four sets of basis functions for spectral discretization in the PMLs and the computational domain separately. The obtained discretized operators are then assembled together [Eq.~\eqref{eq.53}]. After applying the continuity conditions, we proceed in a unified manner with forward stepping. 

In the present study, we propose three sets of numerical experiments to validate the spectral model devised based on the spectral method. The results demonstrate that the spectral model developed in this study can obtain reliable numerical sound fields in three-dimensional variable environments. We compared the wide-angle capability and computational efficiency of three different starters against the analytical example as a benchmark. It was pointed out that in most cases, using a ray-based starter is the most cost-effective option. We also tested the influence of the number of terms in the Pad\'e approximation on the solution and found that, for a ray-based starter, a Pad\'e order of 5 can achieve stable results in most cases. In addition, it is worth mentioning that the spectral algorithm proposed in this paper exhibits good parallelism in both stages mentioned above. The solution of local modes at individual nodes can be naturally parallelized, and the solution of the corresponding MPEs for each mode can also be naturally parallelized. Therefore, the three-dimensional spectral scheme designed in this paper has the potential for further acceleration on high-performance computers.

\section*{Acknowledgments}

This work was supported by the National Key Research and Development Program of China [grant number 2016YFC1401800].

\bibliographystyle{elsarticle-num}

\end{document}